\documentclass[prb,twocolumn,showpacs,preprintnumbers,superscriptaddress,amsmath,amssymb,floatfix]{revtex4}

\usepackage{graphicx}
\usepackage{dcolumn}
\usepackage{bm}
\usepackage[hang,raggedright]{subfigure}
\usepackage{enumerate}
\usepackage{color}
\usepackage{amsmath}
\usepackage{amssymb}
\usepackage{ulem}

\newcommand{\eto}{Er$_{2}$Ti$_{2}$O$_{7}$}
\newcommand{\tto}{Tb$_{2}$Ti$_{2}$O$_{7}$}
\newcommand{\tso}{Tb$_{2}$Sn$_{2}$O$_{7}$}
\newcommand{\pso}{Pr$_{2}$Sn$_{2}$O$_{7}$}
\newcommand{\pio}{Pr$_{2}$Ir$_{2}$O$_{7}$}
\newcommand{\pzo}{Pr$_{2}$Zr$_{2}$O$_{7}$}
\newcommand{\hto}{Ho$_{2}$Ti$_{2}$O$_{7}$}
\newcommand{\dto}{Dy$_{2}$Ti$_{2}$O$_{7}$}

\begin{document}

\title{Towards an Effective Spin Hamiltonian of the Pyrochlore Spin Liquid Tb$_{2}$Ti$_{2}$O$_{7}$}

\author{Hamid R. Molavian}
\affiliation{Department of Physics and Astronomy, University of Waterloo, Waterloo, ON, N2L 3G1, Canada.}
\author{Paul A. McClarty} 
\affiliation{Department of Physics and Astronomy, University of Waterloo, Waterloo, ON, N2L 3G1, Canada.}
\author{Michel J. P. Gingras}
\affiliation{Department of Physics and Astronomy, University of Waterloo, Waterloo, ON, N2L 3G1, Canada.}
\affiliation{Canadian Institute for Advanced Research, 180 Dundas Street West, Suite 1400, Toronto, ON, M5G 1Z8, Canada.}

\date{\today}

\begin{abstract}
\tto\ is a pyrochlore antiferromagnet that has dynamical spins and only short-range correlations even
at $50$ mK $-$ the lowest temperature explored so far $-$ which is much smaller than the scale set by the Curie-Weiss temperature
$\theta_{{\rm CW}} \approx -14$ K. The absence of long-range order in this material is not understood. Recently, virtual crystal
field excitations (VCFEs) have been shown to be significant in \tto, but their effect on spin correlations has not been fully
explored. Building on the work in Phys. Rev. Lett. {\bf 98}, 157204 (2007), we present details of an effective Hamiltonian that
takes into account VCFEs. Previous work found that VCFEs-induced renormalization of the nearest neighbor Ising
exchange leads to spin ice correlations on a single tetrahedron. In this
paper, we construct an effective spin-$1/2$ low-energy theory for \tto\ on the
pyrochlore lattice. We determine semiclassical ground states on a
lattice that allow us to see how the physics of spin ice is connected to the
possible physics of \tto. We observe a shift in the phase boundaries with respect to those of the dipolar spin ice model
as the quantum corrections become more significant. In addition to the familiar classical dipolar spin ice
model phases, we see a stabilization of a $\mathbf{q}=0$ ordered ice phase over a large part of the phase diagram $-$  ferromagnetic
correlations being preferred by quantum corrections in spite of an
antiferromagnetic nearest neighbor exchange in the microscopic
model. Frustration is hence seen to arise from virtual crystal field
excitations over and above the effect of dipolar interactions in spin ice in
inducing ice-like correlations. Our findings imply, more
generally, that quantum effects could be significant in any material related to
spin ices with a crystal field gap of order $100$ K or smaller. 
\end{abstract}


\maketitle

\section{Introduction}

The problem of finding a low energy effective theory from a microscopic theory or directly from experimental
considerations is a ubiquitous one in physics. The purpose is to identify the
relevant degrees of freedom at some energy scale in order to capture the important physics at that scale. Often in condensed matter physics, a large separation of energy scales facilitates
the process of finding an effective theory: for example in the spin ices
\cite{SpinIce, Gingras2, FrusBook} discussed below. When the separation of scales is not large, virtual (quantum mechanical) processes
can become important, as in the Kondo problem in which double occupancy of the
impurity in the Anderson model can be treated as a virtual process that generates
the well-known s-d exchange interaction. \cite{Kondo} One focus of this paper is the construction of such a low energy effective
theory for a highly exotic magnetic material - the \tto\ pyrochlore magnetic material.

A second thread to the present work is frustration, which occurs in magnetism when interactions between
spins cannot be minimized simultaneously. This happens, in the
case of geometric frustration, as a consequence of the topology of the
lattice. As an example, antiferromagnetic isotropic exchange interactions between classical spins on the vertices of the three
dimensional pyrochlore lattice of corner-sharing tetrahedra are frustrated. \cite{Villain,MoessnerChalker,
  MoessnerChalker2, Reimers, Reimers2, GGG} One consequence of this frustration is
an extensive (macroscopic) ground state degeneracy and lack of conventional
long-range order down to arbitrarily low temperatures. Theoretically,
this degeneracy is expected to be lifted, partially, or fully, by other
interactions, \cite{PalmerChalker, Elhajal} perhaps assisted by the presence of thermal or quantum
fluctuations. \cite{MoessnerChalker2, Champion, ChampionHoldsworth} These lessons carry over to real pyrochlore
magnets in which the frustration of the principal spin-spin interaction usually manifests itself in a transition to long-range
order \cite{GTO1,GTO2,GSO,Champion} or a spin glass transition \cite{Gingras3,
  Gardner4} well below the temperature scale set by the interactions $-$ the
Curie-Weiss temperature $\theta_{{\rm CW}}$. In fact, this is a ubiquitous
fingerprint of highly frustrated magnets.

When short-range spin correlations persist down to arbitrarily low temperatures, as in the isotropic exchange
pyrochlore antiferromagnet of Refs.~\onlinecite{MoessnerChalker,MoessnerChalker2}, the system is referred to as a spin liquid or collective
paramagnet. \cite{Villain} Given the large proportion of geometrically frustrated magnetic materials which have been studied
experimentally and which do ultimately exhibit an ordering transition, it does seem that spin liquids are rather rare in two and three
dimensions.\cite{PALee,Levi,Nakatsuji,Mendels,Simonet,Okamoto,Gardner1} One would expect, on general grounds, this scarcity to be particularly
apparent in three dimensional materials where thermal and quantum fluctuations are the most easily quenched.
This paper is concerned with the material \tto\ which is one of the very few three
dimensional spin liquid candidates. \cite{Gardner1} \tto\ is a pyrochlore antiferromagnet that is not magnetically ordered at any temperature above
the lowest explored temperature of 50 mK, \cite{Gardner1,Gardner2,Gardner3,Glass} although the Curie-Weiss temperature,
$\theta_{{\rm CW}}$, is about $-14$ K, that is three hundred times
larger. \cite{Gingras} Despite ten years \cite{Gardner1} of experimental and theoretical
interest in this system, the low energy magnetic properties of this material
are still not currently
understood. \cite{Mirebeau2, Enjalran2,Molavian1}

In this article, we build on earlier work \cite{Molavian1} by presenting further evidence that qualitatively new
physics, in the form of geometrical frustration, is generated via virtual
crystal field excitations (VCFEs) in \tto. The frustration of interactions
coming from high energies is not without precedent in condensed
matter physics: frustrated exchange beyond nearest neighbor and ring exchange
terms arise in small $t/U$ effective theories derived from the Hubbard model
at half-filling. \cite{Hubbard, Delannoy, Delannoy2} In this problem, the higher order terms in the effective model have only a
quantitative effect on the physics which is already captured by the lowest
order terms. \cite{Delannoy} 

In contrast, qualitatively new phenomena have been
proposed to arise by integrating out high energies in a recent work on
Mott systems, \cite{Phillips} and in gauge theories of frustrated magnetic
systems \cite{Gauge1,Gauge2} (which, interestingly,
take as starting points models closely related to the effective
model derived in Sections~\ref{sec:classical} and \ref{sec:quantum} of this paper). The substantial effect of VCFEs on low energy
physics advocated in Ref.~\onlinecite{Molavian1} and in this article is reminiscent of the recent experimentally motivated proposal that
PrAu$_{2}$Si$_{2}$ is a disorder-free spin glass owing to frustration dynamically arising from excited crystal field levels.
\cite{DynamicFrustration} Before launching into the calculations, we first describe some earlier developments relating to \tto\ to
motivate our approach to this problem.

\begin{figure}
\includegraphics[width=0.45\textwidth]{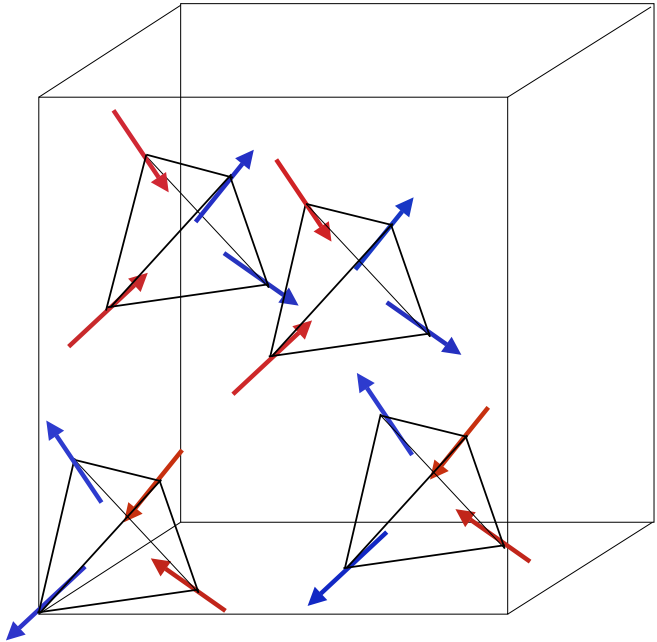}
\caption{\label{fig:UnitCell} (color online). Cubic unit cell of the pyrochlore lattice. The spin configuration shown is the
  ordered LRSI$_{001}$ state of the dipolar spin ice model. \cite{Melko} The spins on each tetrahedron are aligned in the local $[
  111 ]$ direction and satisfy the two-in/two-out ice rule.}
\end{figure}

\subsection{Phenomenology of \tto}

There is one particular property that may be useful for making progress
towards understanding the low energy physics of \tto\ and which is
shared by all the compounds in the R$_{2}$M$_{2}$O$_{7}$ family of compounds
to varying degrees \cite{GGG} (here $R^{3+}$ is a rare earth ion with a
magnetic crystal field ground state and $M^{3+}$ is non-magnetic Ti$^{4+}$ or Sn$^{4+}$). It is the smallness of the energy scale due to interactions, $V$, compared with the
crystal field splitting, $\Delta$, between the single ion ground state doublet and the first (lowest) excited states. The interactions are typically of the order of $0.1$ K or smaller while the lowest crystal field
splitting is of the order of tens or hundreds of Kelvin.\cite{Rosenkranz, Gingras, Mirebeau} This means that the ground state wavefunction and low
energy excitations mainly ``live'' in the Hilbert space spanned by the ground state crystal field states on all lattice sites. As we
shall see in detail later on, the interactions, V, admix excited crystal field wavefunctions into the ground state doublet and these quantum
corrections are weighted by $\langle V\rangle/\Delta$. \cite{Vbracket}  For the spin ices, \hto\ and \dto, for which $\Delta$ is of the order of $300$ K,\hspace{1pt}\cite{Rosenkranz} the effect of excited crystal
field levels can be ignored to a very good approximation and the angular momenta can then be treated as classical Ising spins.
\cite{SpinIce, denHertog, Gingras2} In common with the spin ices, \tto\ has a crystal field ground state that can be described in terms of Ising
spins.\cite{Gingras} But, the (classical) dipolar spin ice model (DSIM) which has, through various studies demonstrated its veracity in comparisons to the
spin ices, \cite{denHertog, Bramwell, Yavorskii} is not a good model for \tto.  

\begin{figure}
\includegraphics[width=0.5\textwidth]{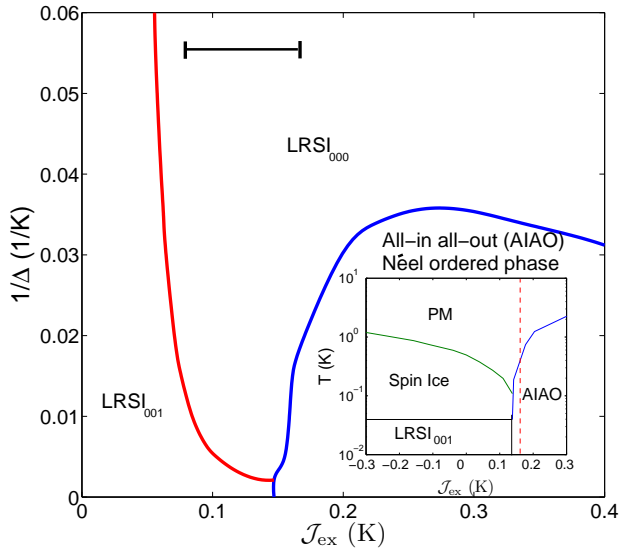}
\caption{\label{fig:PhaseDiagram} (color online). Semiclassical ground state phases for the cubic unit cell model with Ewald summed
  dipole-dipole interactions as the crystal field gap, $\Delta$, and the bare exchange coupling, $\mathcal{J}_{\rm ex}$, are
  varied. The horizontal bar indicates a value for $1/\Delta$ ($\Delta=18$ K) and a range of $\mathcal{J}_{\rm ex}$ that are consistent with
  experimental results on \tto. \cite{Gingras,Mirebeau} The inset is the phase diagram of the dipolar spin ice model
  \cite{denHertog, Melko} adopted for \tto\ with $\mathcal{D}=0.0315$ K with a vertical dotted line showing an estimated
  $\mathcal{J}_{\rm ex}=1/6$ K coupling for \tto. \cite{Gingras}}
\end{figure}

An estimate of the antiferromagnetic exchange coupling in \tto\
\nolinebreak\hspace{1pt}~\cite{Gingras} puts this compound close to the phase boundary of the DSIM between the paramagnetic spin
ice state (or lower temperature long-range ordered spin ice phase) and the
four sublattice long-range N\'{e}el antiferromagnetic phase (see inset to Fig.~\ref{fig:PhaseDiagram}). \cite{denHertog, Melko, Melko2} None of these states adequately describes \tto. The
long-ranged ordered phases can be ruled out on the grounds that no Bragg peaks are observed in the diffuse neutron scattering
pattern. \cite{Gardner2,Gardner3} A comparison with spin ice phenomenology is a little more
subtle. One of the main features of the spin ice state is that it harbors a large residual entropy as deduced by integrating the heat
capacity downwards from high temperatures. \cite{Ramirez} Whereas, similarly
to what has been observed in spin ices, \cite{SpinIce,Ramirez} there is a broad bump in the specific
heat $C_{\rm V}$ between $1$ K and $2$ K as the temperature is lowered, at
present it remains difficult to determine whether there is a residual entropy in the
collective paramagnetic state of \tto. \cite{Gingras, Hamaguchi} The study in
Ref.~\onlinecite{Hamaguchi} finds a slightly different heat capacity to the
one in Ref.~\onlinecite{Gingras} and claims no evidence of residual entropy in
\tto\ owing to almost a complete recovery of the full entropy of the doublet-doublet crystal
field levels (see also Ref.~\onlinecite{Ke} for a similar finding). Instead it reports that there is a sharp feature in the heat capacity at about $300$ mK indicating the onset of a glassy
state. Glassiness has also been observed in the susceptibility measurements of
Ref.~\onlinecite{Luo}. Finally, the diffuse paramagnetic neutron scattering pattern \cite{Gardner1,Gardner2, Gardner3, Yasui} of \tto\ differs drastically from
the experimental spin ice pattern (which has been reproduced by Monte Carlo simulations of the DSIM
\cite{Bramwell} and its improvements \cite{Yavorskii}). This strongly suggests
that the Ising nature of the localized moments is not an appropriate
description for the magnetism in \tto, as noted in Ref.~\onlinecite{Enjalran}.

Some important insight into the microscopic nature of \tto\ is provided by a mean field theory for classical
spins with only a finite Ising anisotropy. \cite{Enjalran} Specifically,
Ref.~\onlinecite{Enjalran} finds that a toy model in which spins, subject to a
finite anisotropy and interacting via isotropic exchange and dipole-dipole interactions, captures the main features of the
experimental paramagnetic diffuse neutron scattering pattern in \tto. \cite{Gardner1} The
results of Ref.~\onlinecite{Enjalran} lead one to suspect that the weaker anisotropy
of the spins in \tto, in contrast to those in the spin ices, can be attributed
to the fact that because the ground to first excited
crystal field gap is much smaller in \tto, the effect of excited crystal field states cannot be ignored. The effects of VCFEs can be studied, albeit incompletely, within the random phase approximation (RPA). A computation
of the RPA diffuse neutron scattering intensity in the paramagnetic regime using the full crystal field level structure and wavefunctions \cite{Kao} leads to
results that are in good qualitative agreement with experiment, \cite{Gardner2} adding weight to the idea that one of the effects of VCFEs in \tto\ is to
decrease the Ising anisotropy of the spins.

Having identified VCFEs as an important contribution to the physics of \tto, we look for a way of examining the
effect of VCFEs on the ground state of perhaps the simplest minimal model for \tto. An approach that is well-suited to this
problem is an effective Hamiltonian formalism. The low energy theory that is obtained within this formalism inhabits a product of
two dimensional Hilbert spaces $-$ one for each magnetic site $-$ spanned by the ground state crystal field doublet. So, the effective
theory can be written in terms of (pseudo) spins one-half. Neglecting VCFEs,
the effective Hamiltonian is simply the theory obtained by projecting onto the ground state crystal field doublet on each magnetic
ion which, as we shall see, is the DSIM of interacting
(classical) Ising spins i.e. a model in which transverse spin fluctuations are
absent. \cite{Gingras2} The separation of energy scales to which we have
alluded then allows us to develop a perturbation series in the parameter
$\langle V\rangle/\Delta$ \hspace{1pt}\cite{Vbracket} where the zeroth order term is
the DSIM \hspace{1pt} \cite{Gingras2} and higher order terms explicitly incorporate the effect of VCFEs in terms of operators
acting within the projected Hilbert space. The procedure can be written schematically as
\begin{align*} H({\mathbf{J}}) & = H_{\rm cf} + V  \\ & \xrightarrow[{\rm
      perturbation \hspace{3pt}
    theory}]{{\rm projection}} H_{\rm eff} ({\mathbf{S}_{\rm eff}})   \end{align*}
where the bare microscopic Hamiltonian $H$, depending on magnetic moments
    ${\mathbf{J}}$ through the crystal field $H_{\rm cf}$ and interactions $V$, is
    used to derive an effective Hamiltonian $H_{\rm eff}$ in terms of
    pseudospins $1/2$, ${\mathbf{S}_{\rm eff}}$.

One advantage of this approach is that, by decreasing $\langle V\rangle/\Delta$, we can smoothly connect our results to
the physics of spin ice. \cite{SpinIce, Gingras2, FrusBook} A second more
practical advantage is that, since the dimensionality of the relevant Hilbert space is reduced,
exact diagonalization calculations on finite size clusters (albeit small
clusters), series expansion techniques and the linked cluster method may
become tractable. \cite{SeriesExpansion} 

\begin{figure}
\includegraphics[width=0.5\textwidth]{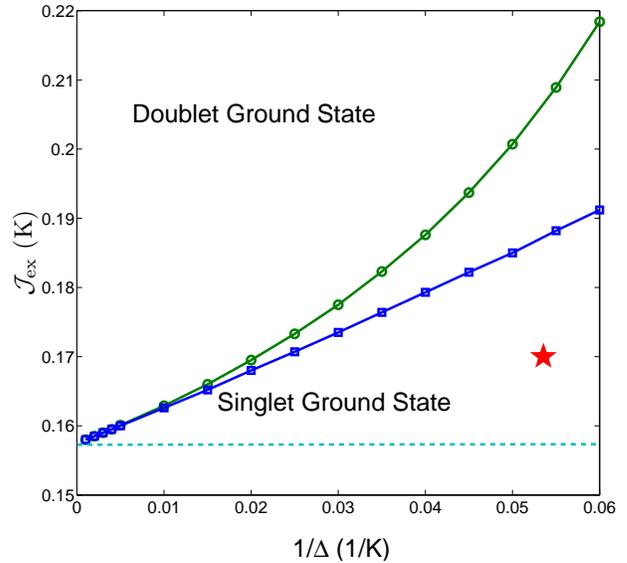}
\caption{\label{fig:GSdeg} (color online). Figure showing exact
  diagonalization of a minimal Hamiltonian, $H=H_{\rm cf}+V$, on a single tetrahedron. The ground state
  degeneracy is shown for different values of the ground-to-first
  excited crystal field gap $\Delta$, and the bare exchange coupling
  $\mathcal{J}_{\rm ex}$ for fixed dipolar strength $\mathcal{D}=0.0315$ K
  relevant to \tto. There are two regions: one with a singlet ground state, the other with a
  doubly degenerate ground state. The boundary between the two regions is marked for the two models considered. For the effective
  Hamiltonian the boundary is marked by circles and for the four crystal field
  state microscopic model on a single tetrahedron described in the main text (based on
  the crystal field Hamiltonian Eq.~(\ref{eqn:truncate})), the boundary is
  traced out by squares. For the estimated parameters $(\mathcal{J}_{\rm ex},
  D, \Delta)$ for
  \tto, indicated by a star, the boundaries agree to within ten percent. The horizontal dashed line shows the phase boundary of the classical
  part ($1/\Delta=0$) of the effective Hamiltonian between the all-in/all-out
  doublet configurations and sextet (degenerate two-in/two-out) ``spin ice'' ground states. Within the classical
  description, \tto\ would be in the doublet all-in/all-out state (i.e. above
  the horizontal dashed line). However, when VCFEs are included, the phase
  boundary is shifted towards larger (i.e. more antiferromagnetic)  values of
  the bare exchange in such a way that \tto\ ``finds itself'' below the boundary in a singlet
  ground state. The singlet arises because of fluctuations that lift the
  sixfold degeneracy of classical two-in/two-out configurations on a single tetrahedron.}
\end{figure}

A comparison has previously been made \cite{Molavian1} between the effective
Hamiltonian  to lowest order in quantum corrections, $\langle \mathcal{J}_{\rm
ex}\rangle/\Delta$, with the crystal
field gap $\Delta$ as a free parameter and the ``high energy'' microscopic (bare) model from which it was obtained. This involved an exact
diagonalization of the two models on a single tetrahedron to determine the ground state as a function of $\Delta$ and the exchange
coupling. \cite{Molavian1} The result is shown in Fig.~\ref{fig:GSdeg}. The
ground state degeneracies largely coincide over the range of
parameters explored, which includes the estimated exchange coupling of
\tto. Most importantly, in the singlet region of the phase
diagram, the ground state of the exact bare microscopic model is a nondegenerate superposition of states each
satisfying the spin ice constraint. In contrast, for the classical dipolar ice model with
the same exchange coupling, on a single tetrahedron and on a lattice, the ground state is
a doubly degenerate all-in/all-out state (see Fig. \ref{fig:AIAO}). That the full quantum problem favors spin ice-like
correlations at the single tetrahedron level was shown to arise from a
renormalization of the Ising exchange in the effective anisotropic spin-$1/2$
Hamiltonian when VCFEs are
included. \cite{Molavian1}  Finally, it was found that the level structure from exact diagonalization of the original model on a
single tetrahedron is sufficient to reproduce the main semi-quantitative
features of the experimental diffuse neutron scattering pattern for \tto. \cite{Molavian1}

The renormalization of the effective nearest neighbor Ising exchange by VCFEs such that spin ice correlations are energetically
preferred over a larger range of the bare exchange couplings than would be the case without quantum corrections shows clearly that
quantum effects can have a significant effect on the nature of the correlations in \tto. However, owing to the presence of a
long-range dipole-dipole interaction and the fact that VCFEs in themselves generate interactions beyond nearest neighbor, it was
not clear on the basis of earlier work \cite{Molavian1} whether VCFEs would have a significant, or even the same qualitative
effect on the \tto\ correlations when considering the full lattice. That is the main problem that we resolve in this work.

\subsection{Scope of the paper}

In this article, we present a more detailed derivation of the effective Hamiltonian for \tto\ than was possible in the earlier
work \cite{Molavian1} owing to lack of space. We also take some initial steps beyond the single tetrahedron approximation by
calculating the ground states of the effective model assuming that the
effective $S_{\rm eff}=1/2$ spins are classical spins of fixed length (large
$S$ approximation). Our main result is shown in Fig.~\ref{fig:PhaseDiagram} which is discussed more fully in
Section~\ref{sec:16sbl}. The plot shows the semiclassical phase diagram of the
effective model on a cubic unit cell with periodic boundary conditions as a
function of the gap $\Delta$ and the isotropic exchange coupling
$\mathcal{J}_{\rm ex}$ in the
microscopic model. When $1/\Delta=0$, all quantum
corrections are suppressed and we recover the limit of the dipolar spin ice
model (DSIM) with two phases - a state with the spin ice rule satisfied on each
tetrahedron and ordering wavevector $001$ (LRSI$_{001}$) and a
four-in/four-out Ising state (AIAO) for more antiferromagnetic $\mathcal{J}_{\rm ex}$. Compared to the dipolar spin
ice model ground states, the effective model contains one other phase $-$ a
$\mathbf{q}=0$ long range ordered spin ice phase (LRSI$_{000}$). Also, the
magnetic moments in the LRSI$_{000}$ and LRSI$_{001}$ phases are canted
away from the local Ising directions as $\Delta$ decreases. The region over
which the LRSI$_{000}$ is the ground state forms a wedge, broadening out to
lower $\Delta$ until it is the only phase found within the explored range of
$\mathcal{J}_{\rm ex}$ at the expense of the antiferromagnetic AIAO phase. There
are two main physical mechanisms (contributions) to the stabilization of the LRSI$_{000}$ state
across the phase diagram. The first is that the effective nearest neighbor Ising
coupling becomes more ferromagnetic in character as $\Delta$
decreases. However, it does eventually change sign as $\mathcal{J}_{\rm ex}$ increases
over the entire range of $\Delta$ studied. So the second reason for the
spreading of a spin ice state across the phase diagram as $\Delta$
decreases is due to beyond nearest neighbor interactions that arise purely
from effective VCFEs and which monotonically increase in strength as $\Delta$
decreases.

The outline of the paper is as follows. In Section \ref{sec:Heff}, we introduce some notation
and describe the microscopic (bare) model for \tto\ from which the effective model is derived. With this in hand, we formulate our approach in more detail than in this introduction. Section \ref{sec:classical} discusses the form and properties of the lowest order
(classical dipolar spin ice) term in the effective Hamiltonian. In Section \ref{sec:quantum}, the quantum corrections to this model are
enumerated to lowest order in $\langle V\rangle/\Delta$ and we study how the
longitudinal (Ising) exchange coupling in the dipolar spin ice model (DSIM) is
renormalized to this order. Having obtained the effective Hamiltonian for \tto\ to lowest order in the $1/\Delta$, we
treat the effective $S=1/2$ spins as classical spins and present, in Section
\ref{sec:ground}, the resulting semiclassical ground states. This study of the ground states allows us to see how the effect of VCFEs is connected to the physics of
spin ice and also clearly shows that spin ice correlations are present even though the bare microscopic exchange coupling
$\mathcal{J}_{\rm ex}$ is
antiferromagnetic. 

\begin{quotation}
{\it In other words, geometric frustration in the model (Eqs.~(\ref{eqn:bare}),(\ref{eqn:HCF}) and  (\ref{eqn:interactions})) of \tto\
  emerges from quantum virtual crystal field excitations (VCFEs) and many-body physics.}
\end{quotation}

This is the main result of our paper. We discuss these
results, in Section~\ref{sec:discussion}, in the light of experiments on \tto\
and describe some possible further applications of the effective Hamiltonian
that we derive for \tto. Finally, we provide in
Appendix \ref{sec:general}, details of the effective Hamiltonian method as a background to the main application to
\tto\ described in the remainder of the paper. Appendix~\ref{sec:calc} contains further details behind the calculations presented
in Section~\ref{sec:quantum} and Appendix~\ref{sec:CFP} gives some data used
to convert between crystal field parameters for different rare earth
pyrochlore titanates using a point charge approximation.

We note here that while our specific focus is on the \tto\ pyrochlore magnet, the formalism that we employ below could be
straightforwardly used to construct effective low energy theories for many other frustrated rare earth systems where the excited
crystal field levels have a somewhat larger energy scale than the microscopic interactions.

\section{Effective Hamiltonian}
\label{sec:Heff}

\subsection{Microscopic (Bare) Model}
\label{sec:model}
 
The microscopic or bare Hamiltonian for the magnetic Tb$^{3+}$ ions in \tto\ is given by
\begin{equation}
H = H_{\rm cf} + V
\label{eqn:bare}
\end{equation}
where $H_{\rm cf}$ is the crystal field Hamiltonian and $V$ are the interactions between the ions. In the remainder of this
section we explain the form of both terms in some detail.

The magnetic Tb$^{3+}$ ions in \tto\ are arranged on the sites of a pyrochlore lattice. The pyrochlore lattice consists of corner-shared
tetrahedra which can otherwise be thought of as a face-centered cubic (fcc) lattice with primitive translation vectors
$\mathbf{R}_{A}$ for $A=1,2,3$ and a basis of four ions $\mathbf{r}^{a}$ ($a=1,\ldots,4$). We follow the same labeling of
the four sublattice basis vectors as in Ref.~\onlinecite{Enjalran}. It is useful to introduce a coordinate system on each of the
four sublattices with local $\mathbf{\hat{z}}^{a}$ unit vector along the local
cubic $[ 111 ]$ direction. The sublattice basis
vectors and local Cartesian $\mathbf{\hat{x}}^{a}$, $\mathbf{\hat{y}}^{a}$ and $\mathbf{\hat{z}}^{a}$  directions are given in Table \ref{tab:lattice}. Below, we also make use of rotation
matrices $u^{a}_{\alpha\beta}$ (the elements of which are contained in
Table~\ref{tab:lattice}) which achieve a passive transformation that takes the local sublattice coordinate system for
sublattice $a$ into the global Cartesian laboratory axes. 
 
\begin{table*}
\caption{\label{tab:lattice} Basis of four magnetic ions on a pyrochlore indexed by position vectors $\mathbf{r}^{a}$. The local
  $[ 111 ]$ direction on each sublattice is $\mathbf{z}^{a}$. The edge length
  of the cubic unit cell is $a$. The rotation matrix $u^{a}_{\alpha\beta}$
  takes the form $\left( \mathbf{x}^{a},\mathbf{y}^{a},\mathbf{z}^{a}
  \right)_{\alpha\beta}^{T}$ in which the vector components are placed in the
  matrix columns. In the main text, we make use of vectors
  $\mathbf{\hat{n}}^{x}=(1,0,0)$,$\mathbf{\hat{n}}^{y}=(0,1,0)$,$\mathbf{\hat{n}}^{z}=(0,0,1)$
  in the laboratory coordinate system. }
\begin{ruledtabular}
\begin{tabular}{lcccc}
Sublattice & $\mathbf{r}^{a}$ & $\mathbf{x}^{a}$ & $\mathbf{y}^{a}$ & $\mathbf{z}^{a}$  \\
\hline
\vspace{.05mm} \\
$1$ & $(a/4)(0,0,0)$ & $(1/\sqrt{6})(-1,-1,2)$  & $(1/\sqrt{2})(1,-1,0)$   & $(1/\sqrt{3})(1,1,1)$   \\
$2$ & $(a/4)(1,1,0)$ & $(1/\sqrt{6})(1,1,2)$    & $(1/\sqrt{2})(-1,1,0)$   & $(1/\sqrt{3})(-1,-1,1)$ \\ 
$3$ & $(a/4)(1,0,1)$ & $(1/\sqrt{6})(1,-1,-2)$  & $(1/\sqrt{2})(-1,-1,0)$  & $(1/\sqrt{3})(-1,1,-1)$ \\
$4$ & $(a/4)(0,1,1)$ & $(1/\sqrt{6})(-1,1,-2)$  & $(1/\sqrt{2})(1,1,0)$    & $(1/\sqrt{3})(1,-1,-1)$ \\
\end{tabular}
\end{ruledtabular}
\end{table*}

Spin-orbit coupling within the relevant localized $4f$ levels of the Tb$^{3+}$ ions leaves total angular momentum $\mathbf{J}$ as
a good quantum number with ${\rm J}=6$. The local environment about each Tb$^{3+}$ ion is responsible for breaking the $2{\rm
  J}+1$ degeneracy. Its effect can be computed from a crystal field Hamiltonian, $H_{{\rm cf}}$, which is constrained by
symmetry to take the form \cite{Gingras, Rosenkranz,Mirebeau}
\begin{multline} H_{{\rm cf}} = \sum_{ i,a} B_{2}^{0}O_{2}^{0}(i,a) + B_{4}^{0}O_{4}^{0}(i,a) +
  B_{4}^{3}O_{4}^{3}(i,a) \\ + B_{6}^{0}O_{6}^{0}(i,a) + B_{6}^{3}O_{6}^{3}(i,a) + B_{6}^{6}O_{6}^{6}(i,a).
\label{eqn:HCF}
\end{multline}
The magnetic ions are labeled by an fcc site $i$ and a sublattice index $a$. Expressions for the operators ${O_{l}^{m}}$ in
terms of the local angular momentum components can be found, for example, in Hutchings. \cite{Hutchings} The crystal field in
\tto\ has been studied in  Refs.~\onlinecite{Gingras} and
\onlinecite{Mirebeau} resulting in somewhat differing estimates for the parameters
${B_{l}^{m}}$. In the following, all quantitative results that we present for \tto\ were obtained using crystal field parameters
for \hto, obtained from inelastic neutron scattering in
Ref.~\onlinecite{Rosenkranz}, which have been rescaled to the \tto\ parameters according to
\begin{equation}  (B_{l}^{m})_{{\rm Tb}} = \left(\frac{(S_{l})_{{\rm Tb}}}{(S_{l})_{{\rm Ho}}}\right)\left( \frac{\langle r^{m} \rangle_{{\rm
    Tb}}}{\langle r^{m} \rangle_{{\rm Ho}}}\right)(B_{l}^{m})_{{\rm Ho}}. \label{eqn:convert} \end{equation} 
Here, the $S_{l}$ are Stevens factors. \cite{Stevens} These and the radial
expectation values $\langle r^{m} \rangle$ for the rare earth ions
\cite{Freeman} can be found in Appendix~\ref{sec:CFP}. We have checked that using
the crystal field parameters of Ref.~\onlinecite{Mirebeau} instead leads to results that are in fairly
close quantitative agreement with those obtained using the rescaled parameters
from Eq.~(\ref{eqn:convert}). 

\begin{figure}
\includegraphics[width=0.25\textwidth]{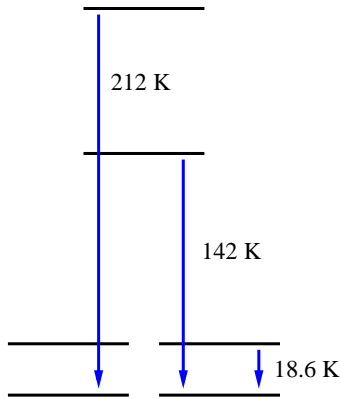}
\caption{\label{fig:spectrum} (color online). Figure indicating the four
  lowest levels of the crystal field spectrum (not to scale). The splitting
  between the ground state doublet and the first excited state is called $\Delta$. The ground state and the first excited
  state are doublets. \cite{Gingras} The two other excited states are singlets. \cite{Gingras}}
\end{figure}

The crystal field
Hamiltonian, $H_{\rm cf}$, can be diagonalized numerically exactly; the eigenvalues are $E_{n}$ and the eigenstates $| n \rangle$ for
$n=1,\ldots,13$, which we implicitly arrange in order of increasing energy. One finds a level structure that includes a ground
state and a first excited state that are both doubly degenerate. \cite{Gingras,Mirebeau} The splitting, $\Delta$, between the
ground and first excited states is about $18.6$ K, \cite{Gingras, Mirebeau} which is much smaller than the corresponding gap in the
spin ices (for example, the gap in Ho$_{2}$Ti$_{2}$O$_{7}$ is about $230$ K \hspace{1pt}\cite{Rosenkranz}). It is the smallness of
this value of $\Delta$ compared to $V$ for \tto\ and the possibility of
admixing between the ground state and excited state crystal field levels
that are at the root of all the phenomenology that we explore in the rest of
this paper. Fig.~\ref{fig:spectrum} shows the level structure of the crystal
field spectrum for the four lowest levels determined on the basis of an exact diagonalization of Eq.~(\ref{eqn:HCF}).

We emphasize two features of this spectrum that will be important later on. First of all, let us write down the time reversal
  properties of the eigenstates, $|n \rangle$. Let $|n \rangle$ be written as a linear combination of the eigenstates of
  $\mathbf{J}$, denoted $|{\rm J}, M\rangle$, 
\[ |n \rangle = \sum_{M} c^{M}_{n} |{\rm J}, M\rangle. \]
Time reversal invariance requires that the coefficients are related to one another by $c^{M}_{n} =
(-)^{J-M}c^{-M}_{n}$. \cite{TimeReversal} Secondly, it is possible to interpret the non-interacting single ion angular momenta as Ising-like at low
energies, as was done in Ref.~\onlinecite{denHertog}. This is because, at sufficiently low energies, thermal occupation of excited
crystal field levels is negligible and one can focus on the ground state doublet. The ground state doublet states, $|1\rangle$ and
$| 2\rangle$, have
\begin{equation}  \langle 1| \widetilde{{\rm J}^{z}}|1 \rangle = - \langle 2 | \widetilde{{\rm J}^{z}}| 2 \rangle \equiv \langle
\widetilde{{\rm J}^{z}} \rangle 
\label{eqn:ME}
\end{equation} 
as the only nonvanishing matrix elements, where the tilde indicates that the $z$ axis is taken along the local $[
111 ]$ direction appropriate to each magnetic ion (see Table \ref{tab:lattice}). So, this doublet considered on its own has
nonzero angular momentum expectation values only along one axis with vanishing transition matrix elements $\langle
1|\widetilde{{\rm J}}^{\pm}|2\rangle =0$.

The interactions between the angular momenta, $V\equiv H_{{\rm ex}}+H_{{\rm dd}}$, are taken to be nearest neighbor isotropic exchange
$H_{{\rm ex}}$ and dipole-dipole interactions, $H_{{\rm dd}}$:
\begin{align}
 H_{\rm ex} & = \mathcal{J}_{{\rm ex}}\sum_{\langle (i,a),(j,b)\rangle}
 \mathbf{J}_{i,a}\cdot\mathbf{J}_{j,b} \nonumber \\
 H_{\rm dd} & = \mathcal{D}r_{{\rm nn}}^{3}\sum_{{\rm pairs}}
 \frac{\mathbf{J}_{i,a}\cdot\mathbf{J}_{j,b}}{|\mathbf{R}^{ab}_{ij}|^{3}} - 3
 \frac{(\mathbf{J}_{i,a}\cdot\mathbf{R}^{ab}_{ij})(\mathbf{J}_{j,b}\cdot\mathbf{R}^{{\rm
 ab}}_{ij})}{|\mathbf{R}^{ab}_{ij}|^{5}}.  
\label{eqn:interactions}
\end{align}
The notation $\mathbf{R}^{ab}_{ij}$ is short for $\mathbf{R}_{i}^{a}-\mathbf{R}_{j}^{b}$ with
$\mathbf{R}_{i}^{a}=\mathbf{R}_{i}+\mathbf{r}^{a}$ and $r_{{\rm nn}}=3.59\AA = a\sqrt{2}/4$ (where $a$ is the edge length of the
conventional cubic unit cell) is the distance between
neighboring magnetic ions. \cite{Gingras} Here, we employ the convention that $\mathcal{J}_{{\rm ex}}>0$ is antiferromagnetic and
$\mathcal{J}_{{\rm ex}}<0$ is ferromagnetic. This is the
simplest Hamiltonian consistent with the nonvanishing Tb$^{3+}$ dipole-dipole coupling,
$\mathcal{D}=(\mu_{0}/4\pi)(g_{J}\mu_{B})^{2}/r_{{\rm nn}}^{3}=0.0315$ K with the Land\'e factor, 
 $g_{J}=3/2$ and with the negative Curie-Weiss
temperature $\theta_{{\rm CW}}=-14$ K. \cite{Gingras} The exchange coupling $\mathcal{J}_{{\rm ex}}$ has been estimated from
$\theta_{{\rm CW}}$ for \tto\ and $\theta_{{\rm CW}}$ for the diluted compound (Y$_{0.98}$Tb$_{0.02}$)$_{2}$Ti$_{2}$O$_{7}$
\hspace{1pt}\cite{Gingras} to be about $0.17$ K, while a fit in Ref.~\onlinecite{Mirebeau} gives a value for $\mathcal{J}_{{\rm
    ex}}=0.083$ K that is significantly less antiferromagnetic. \cite{Estimate}

In summary, our bare microscopic model for \tto\ consists of three terms: the
crystal field Hamiltonian $H_{\rm cf}$, an isotropic
exchange $H_{\rm ex}$ with an antiferromagnetic coupling and a dipole-dipole interaction, $H_{\rm dd}$. \cite{Interactions} An extension of the
present work could include (i) bare exchange couplings beyond nearest neighbors, (ii) anisotropic nearest neighbor exchange as
described in Ref.~\onlinecite{ETOPaper} and (iii) direct or virtual (phonon-mediated) multipolar interactions. \cite{Multipole}

\subsection{Route to an effective Hamiltonian}
\label{sec:route}

If we were able to ignore the excited crystal field levels in \tto, the angular momenta could be treated as classical Ising
spins \cite{denHertog,Gingras2} because the only nonvanishing matrix elements of the angular momentum are those in Eq.
(\ref{eqn:ME}). \cite{Gingras2} However, for reasons outlined in the Introduction, this is not a good approximation for this material. The
interactions between the angular momenta induce VCFEs that admix excited crystal field wavefunctions into the space spanned by the
non-interacting crystal field doublets with the consequence that the magnetic
moments behave much less anisotropically than one would expect on the basis of
the $[ 111 ]$ Ising-like ground state crystal field doublet. These quantum fluctuations can
be treated perturbatively because there is a small dimensionless parameter
$\langle V\rangle/\Delta$, where $\langle V \rangle\sim
O({\rm max}(\mathcal{J}_{{\rm ex}},\mathcal{D}))$. To lowest order in such a perturbation theory, and in a low energy effective model, the spins
should be perfectly Ising-like and hence we recover the DSIM. We now proceed to make these ideas more concrete.

Because we seek a Hamiltonian operating within a low energy subspace, we need a projection operator onto the
non-interacting single ion crystal field ground states. For a single ion at the site specified by indices $i,a$, the projection is accomplished by
\[ \mathcal{P}(i,a) =  |1_{i,a} \rangle\langle 1_{i,a} | + |2_{i,a} \rangle\langle 2_{i,a}|. \]
This operator satisfies the conditions $\mathcal{P}^{2}(i,a)=\mathcal{P}(i,a)$ and Hermiticity. With moments on all the sites of
the lattice, the projector is $\mathcal{P}\equiv \prod_{i,a}\mathcal{P}(i,a)$. The subspace of the full Hilbert space selected by the
projector will be called the model space, $\mathfrak{M}\equiv \prod_{\otimes (i,a)}\mathfrak{M}_{i,a}$, from now on. The Hilbert space
$\mathfrak{M}_{i,a}$ is defined as the space spanned by states $|1_{i,a} \rangle$ and $|2_{i,a} \rangle$ on site $(i,a)$.

The spin-spin interaction 
\begin{equation} V\equiv H_{{\rm ex}}+H_{{\rm dd}} \end{equation} 
is to be treated as a perturbation. Because the perturbation $V$ is
``small'' compared to the difference between the ground and first excited crystal
field energies $\Delta$, $H_{{\rm cf}}\equiv H_{0}$, we expect that on a crystal of $N$ sites, the $2^{N}$ lowest energy eigenstates of $H$ lie mainly within $\mathfrak{M}$ because the
admixing of excited crystal field wavefunctions into the model space is a small effect. Our effective
Hamiltonian will be defined in such a way that its eigenstates live entirely within $\mathfrak{M}$ while its eigenvalues exactly
correspond to the $2^{N}$ lowest energy eigenvalues of the exact Hamiltonian, $H$. The $2^{N}$ lowest energy eigenstates of $H$ mainly
lie within $\mathfrak{M}$ in the sense that the rotation of exact states out of the model space is determined by the relatively small
perturbation $\langle V\rangle/\Delta$.

In practice, the exact eigenvalues can be approximated by carrying out perturbation theory in the construction of the effective
Hamiltonian $H_{\rm eff}$. After some work, that is briefly laid out in Appendix~\ref{sec:general}, one finds that the effective
Hamiltonian can be written as \cite{Book} 
\begin{equation} H_{\rm eff} = \mathcal{P}H_{0}\mathcal{P} + \mathcal{P} V \mathcal{P} + \mathcal{P}V\mathcal{R}V\mathcal{P} +
  \ldots       \label{Heff} 
\end{equation} 
The operator $\mathcal{R}$ $-$ the resolvent operator$-$ is given by 
\begin{equation} \mathcal{R} = \sum_{| P\rangle \notin \mathfrak{M}} \frac{|P \rangle\langle P|}{E_{g}-E_{P}}
 \label{eqn:resolvent} \end{equation}
where, for a finite crystal of $N$ sites, $E_{g}$ is $N$ times the energy of the degenerate ground state crystal field levels
$E_{0}$. The numerator of each term in the resolvent is a projector onto a space orthogonal to $\mathfrak{M}$ $-$ a product
of crystal field operators $|P \rangle\langle P| \equiv \prod_{\otimes} |n\rangle\langle n|$ where the product is taken over all sites of
the lattice with at least one such operator having $n>2$ (i.e. belonging to the group of excited crystal field states); this is the
meaning of the notation $| P \rangle \notin \mathfrak{M}$ in the summation index of Eq. (\ref{eqn:resolvent}). The third term on the
right-hand-side of Eq. (\ref{Heff}) is the lowest order term in the perturbation series to include the effects of crystal
field states outside the model space. This term is therefore the lowest order contribution of the VCFEs that we have referred to
above.

Equation (\ref{Heff}) makes no reference to a particular model. In Sections \ref{sec:classical} and \ref{sec:quantum}, we
develop the terms in the effective Hamiltonian for the model $H = H_{0}+V\equiv H_{\rm cf}+ H_{\rm dd}+H_{\rm ex}$ of \tto\
described in Section \ref{sec:model}. Section \ref{sec:classical} is devoted to the lowest order, or classical, term $\mathcal{P}H
\mathcal{P}$. Section \ref{sec:quantum} enumerates the lowest order terms generated by VCFEs, relating each underlying class of
terms that originate from $\mathcal{P}H\mathcal{R}H\mathcal{P}$ to specific
virtual excitation channels. Higher order corrections than $\mathcal{P}H\mathcal{R}H\mathcal{P}$ are computationally difficult to determine mainly
because of the presence of the long-range dipole interactions $H_{\rm dd}$. See Ref.~\onlinecite{Bergman} for a model on a pyrochlore for which
degenerate perturbation theory can be carried out to much higher order than is done is this work.

To spare readers the details of
this rather technical derivation if they so choose, we include a short summary
(Section \ref{sec:summary}) of the form of the low energy model for \tto.
Finally, in Section~\ref{sec:results}, we summarize some results that have
been obtained from the effective Hamiltonian which have already appeared in
the literature. \cite{Molavian1, Molavian2} All in all, we shall see that the
DSIM couplings are renormalized by VCFEs and that effective anisotropic
spin-spin couplings appear in addition to the Ising interactions of the DSIM. 
In other words, the effective theory allows for fluctuations of the moments
perpendicular to the local $\mathbf{z}^{a}$ axes.
We shall study the variation of the
effective couplings in $H_{\rm eff}$ as $\mathcal{J}_{\rm ex}$ is varied. This information will be useful in the interpretation of the semiclassical ground
states of the effective model (Section~\ref{sec:ground}) and hence in assessing the effects of VCFEs on the physics of \tto.

\section{Classical part of $H_{\rm eff}$}
\label{sec:classical}

\subsection{$[111]$ Ising model for \tto}

In this subsection, we consider the (lowest order) term $\mathcal{P}H\mathcal{P}$ in Eq. (\ref{Heff}). The effective Hamiltonian derived from $H$
for \tto\ can be rendered in the form of a spin one-half model by rewriting the model space operators in Eq. (\ref{Heff}) in
terms of Pauli matrices. This is possible because the model space, in our case, is a direct product of two dimensional Hilbert
spaces spanned by the ground state crystal field doublet. The correspondence between Pauli matrices and operators on the crystal
field ground state:
\begin{align}
\tilde{\sigma}^{x} = |1\rangle\langle 2 | & + |2 \rangle\langle 1| \\
\tilde{\sigma}^{y} = -i(|1\rangle\langle 2 | & - |2 \rangle\langle 1|) \\
\tilde{\sigma}^{z} = |1 \rangle\langle 1 | & - | 2 \rangle\langle 2|
\label{eqn:Pauli} 
\end{align}
together with the unit operator $\mathbb{I} =|1 \rangle\langle 1 | + | 2
\rangle\langle 2|$. Note, however, that despite the fact they do satisfy
the commutation rules 
\[ [\tilde{\sigma}^{\alpha}, \tilde{\sigma}^{\beta}] = 2 i \epsilon_{\alpha
    \beta \gamma} \tilde{\sigma}^{\gamma}, \] where $\epsilon_{\alpha \beta \gamma}$ is the Levi-Civita symbol, the
$\tilde{\sigma}^{\alpha}$ do not swap sign under time reversal so they are not true angular momentum operators. For this reason,
we shall call them pseudospins or effective spins. It is helpful for later sections to give their properties under $\mathfrak{T}$,
the time reversal transformation:
\begin{align} 
\mathfrak{T}: \tilde{\sigma}^{x} & \rightarrow \tilde{\sigma}^{x} \\
\mathfrak{T}: \tilde{\sigma}^{y} & \rightarrow \tilde{\sigma}^{y} \\
\mathfrak{T}: \tilde{\sigma}^{z} & \rightarrow -\tilde{\sigma}^{z} 
\label{eqn:time}
\end{align}
because $\mathfrak{T}:| 1 (2)\rangle\rightarrow | 2 (1)\rangle$.

If we apply the projector to the full Hamiltonian to obtain $\mathcal{P}H\mathcal{P}$, we find that the crystal
field part $H_{{\rm cf}}$ becomes $E_{0}\sum_{i,a}\mathbb{I}_{i,a}$ with $E_{1}=E_{2}\equiv E_{0}$. From now on, we omit
this constant energy shift. To project the interaction part $\mathcal{P}V\mathcal{P}$, we write the angular momentum components in the local coordinate
system with local $\mathbf{\hat{z}}^{a}$ axes in the directions given in Table \ref{tab:lattice}:
$J^{\alpha}_{i,a}=u^{a}_{\alpha\beta}\widetilde{J}^{\beta}_{i,a}$. All operator components that refer to the local coordinate systems
are labeled with a tilde. Also, when it is not important to distinguish different sublattices, we abbreviate $(i,a)$
with the site index $I$. We add further numerical subscripts to $I$ to label different sites. With this notation, the projector acting on $\tilde{J}^{z}_{I}$ gives 
\[ \langle\widetilde{J}^{z}_{I}\rangle (|1_{I}\rangle\langle 1_{I} | - |2_{I}\rangle\langle 2_{I}|) =
\langle\widetilde{J}^{z}_{I}\rangle\tilde{\sigma}_{I}^{z}  \]
where $\langle\widetilde{J}^{z}_{I}\rangle = \langle 1| \widetilde{J}^{z}|1 \rangle$.
Owing to $\langle 1|\widetilde{J}^{\pm}|2\rangle=0$,  all matrix elements of the other angular momentum components vanish. So, the
isotropic exchange $H_{\rm ex}$ becomes
\begin{equation} 
\mathcal{P} H_{\rm ex}\mathcal{P} = \mathcal{J}_{{\rm classical}}\sum_{\langle I_{1},I_{2}\rangle} (\mathbf{\hat{z}}^{a}\cdot \mathbf{\hat{z}}^{b}) \tilde{\sigma}_{I_{1}}^{z} \tilde{\sigma}_{I_{2}}^{z} 
\label{eqn:Jclassical}
\end{equation}
and the dipole-dipole interaction becomes
\begin{multline} 
\mathcal{P} H_{\rm dd}\mathcal{P} = \mathcal{D}_{{\rm classical}} r_{{\rm nn}}^{3} \\ \times\frac{1}{2}\sum_{(i,a;j,b)} \left( \frac{(\mathbf{\hat{z}}^{a}\cdot
  \mathbf{\hat{z}}^{b}) }{|\mathbf{R}^{ab}_{ij}|^{3}} \right. \left. - 3
\frac{(\mathbf{\hat{z}}^{a} \cdot\mathbf{R}^{ab}_{ij})(\mathbf{\hat{z}}^{b}
  \cdot\mathbf{R}^{ab}_{ij})}{|\mathbf{R}^{ab}_{ij}|^{5}} \right) \tilde{\sigma}_{i,a}^{z} \tilde{\sigma}_{i,b}^{z}.
\label{eqn:Dclassical}
\end{multline}
The renormalized, or effective, exchange and dipole-dipole couplings are, respectively, $\mathcal{J}_{{\rm classical}} = \mathcal{J}_{{\rm
    ex}}\langle \widetilde{J}^{z} \rangle^{2}$ and $\mathcal{D}_{{\rm classical}} = \mathcal{D}\langle \widetilde{J}^{z} \rangle^{2}$. 
$H_{\rm DSM}=\mathcal{P}(H_{\rm ex}+H_{\rm dd})\mathcal{P}$ is the celebrated DSIM. \cite{SpinIce, Gingras2, FrusBook,denHertog,
    Melko,Fennell,Yavorskii} It is a classical (local Ising) model because 
all the terms mutually commute as they solely consist of
    $\tilde{\sigma}^{z}_{i,a}$ operators. This model exhibits two different
    ground states depending on the ratio of the exchange to the dipolar coupling; these are shown in the inset of Fig.~\ref{fig:PhaseDiagram}. When $\mathcal{J}_{{\rm ex}}/\mathcal{D}>4.525$, the ground state has ordering wavevector $\mathbf{q}=0$ with the spins on a single tetrahedron in the
    $|1\rangle$ state or the $|2\rangle$ state $-$ the all-in/all-out phase. \cite{denHertog}  When $\mathcal{J}_{{\rm
    ex}}/\mathcal{D}<4.525$,  the ordering wavevector of the ground state is $(0,0,2\pi/a)$ and each tetrahedron has spins
    satisfying the two-in/two-out ice rule; we refer to this state as the
    LRSI$_{001}$ phase, \cite{SpinIce, Melko, Melko2} with one of the domains
    shown in Fig. \ref{fig:UnitCell}. \cite{Transition} Above a nonzero critical temperature,
    the LRSI$_{001}$ phase gives way \cite{SpinIce} to a spin ice state with no
    conventional long-range order (Fig.~\ref{fig:PhaseDiagram}). 

Formally speaking, the spin ice state is a collective paramagnetic state \cite{Villain} $-$ a
    classical spin liquid of sorts.  That the DSIM has proved to be a good
    model for spin ice materials is largely
    due to the substantial gap $\Delta$ between the crystal field ground state doublet and first excited state which results in a
    roughly $1/\Delta$ suppression of VCFEs. \cite{SpinIceFNN}  This model is not a good description for \tto. Indeed, if we consider
    the estimated couplings given in Section \ref{sec:model}, we find $\mathcal{J}_{{\rm ex}}/\mathcal{D}\sim 5.4$ (recalling
    $\mathcal{J}_{\rm ex}\approx 0.17$ K and $\mathcal{D}\approx 0.0315$ K as stated in Section~\ref{sec:model}), which would put
    \tto\ in the all-in/all-out 
    phase with a critical temperature into this phase from the paramagnetic
    phase at $T_{c}\sim 0.5$ K (see vertical dashed line in the inset to Fig.~\ref{fig:PhaseDiagram}). \cite{denHertog} This is in
    contradiction with neutron scattering experiments which find no magnetic Bragg peaks in zero field. \cite{SpinIceFNN2} If we allow
    for inaccuracies in the estimate of $\mathcal{J}_{\rm ex}$,
    \cite{Mirebeau,SpinIceFNN2} such that a classical, dipolar spin ice state is implied by the coupling, we find various
     properties of spin ices that are not compatible with those of \tto. Some of these conflicting properties $-$ the diffuse
    neutron scattering pattern and differing spin anisotropies $-$ were
    discussed in the Introduction. Therefore, in the next
    section, we investigate what happens when $\Delta$ is small enough that the lowest order fluctuation term
    $\mathcal{P}H\mathcal{R}H\mathcal{P}$ in Eq.~(\ref{Heff}) becomes important.

\subsection{Exchange convention}
\label{sec:convention}

In Eq.~(\ref{eqn:interactions}), we use the opposite sign convention for the exchange coupling to the one used in
Refs.~\onlinecite{Melko}, \onlinecite{denHertog} and \onlinecite{Melko2}. \cite{Transition} The convention in these works is to include a
minus sign in front of the exchange coupling in contrast to our Eq. (\ref{eqn:interactions}). In this article, in the global
coordinate system, antiferromagnetic corresponds to $\mathcal{J}_{\rm ex}>0$. 

A warning must be made regarding the convention within the local coordinate system. In rotating to the local system, geometrical
factors appear in front of the couplings. For example, as shown in Section~\ref{sec:classical}, the local Ising exchange part of the coupling $\mathcal{J}_{\rm
  ex}\mathbf{J}_{i,a}\cdot\mathbf{J}_{j,b}$ is $\mathcal{J}_{\rm ex}(\mathbf{\hat{z}}^{a}\cdot \mathbf{\hat{z}}^{b})
\tilde{J}_{i,a}\cdot \tilde{J}_{j,b}^{z}$ where $(\mathbf{\hat{z}}^{a}\cdot \mathbf{\hat{z}}^{b})=-1/3$ arises from the fact that
the local $\mathbf{\hat{z}}$ axes are not collinear. In the following pages, we adopt the simplifying scheme of absorbing the geometrical factors
into the couplings. In doing so, it will be useful to describe how to go from the sign of the local effective Ising coupling, $J^{zz}$, in
front of $J^{zz}\tilde{\sigma}^{z}\tilde{\sigma}^{z}$ to the type of order that is energetically favored by the coupling. Thus,
when the local coupling is said to be ferromagnetic, $J^{zz}$ is negative and the Ising components of the spins prefer to lie in
an all-in/all-out configuration. When, instead, $J^{zz}$ is positive, it is said to be antiferromagnetic and the local Ising components are
frustrated, leading to a spin ice configuration on each tetrahedron.

\section{Lowest order quantum fluctuations}
\label{sec:quantum}

\subsection{General Considerations}

In this section we present a derivation of the quantum terms $\mathcal{P}V\mathcal{R}V\mathcal{P}$ in the effective
Hamiltonian. We refer those readers interested only in the results of this
technical derivation to Section~\ref{sec:summary}. 
We begin by introducing a little more notation to describe the structure of the term
$\mathcal{P}V\mathcal{R}V\mathcal{P}$ which we shall refer to as $H^{(2)}_{\rm eff}$.  We write the interaction term $V$ in the form 
\begin{align} V  = & \sum_{I_{1},I_{2}}\sum_{\alpha,\beta} \mathcal{K}_{I_{1}I_{2}}^{\alpha,\beta}
 J_{I_{1}}^{\alpha}J_{I_{2}}^{\beta} \nonumber \\  = & \sum_{I_{1},I_{2}}\sum_{\alpha,\beta}
 \mathcal{\tilde{K}}_{I_{1}I_{2}}^{\alpha,\beta}\widetilde{J}^{\alpha}_{I_{1}} \widetilde{J}^{\beta}_{I_{2}}   \label{eqn:V} \end{align}
where, in the second line, we have absorbed the rotation matrices $u^{a}_{\alpha\beta}$ into the definition of
 $\mathcal{\tilde{K}}$. When the spins interact via nearest neighbor
 isotropic exchange and long-ranged dipole-dipole interactions as in
 Eq.~(\ref{eqn:interactions}), we have for $\mathcal{K}$:
\begin{multline} \mathcal{K}_{(i,a),(j,b)}^{\alpha,\beta} = \frac{1}{2}\mathcal{J}_{\rm ex}\delta_{\mathbf{R}^{ab}_{ij}, r_{{\rm nn}}}
    (\mathbf{n}^{\alpha}\cdot \mathbf{n}^{\beta}) \\ +  \frac{1}{2}\mathcal{D}r_{{\rm nn}}^{3}\left( \frac{(\mathbf{n}^{\alpha}\cdot
      \mathbf{n}^{\beta})}{|\mathbf{R}^{ab}_{ij}|^{3}} - 3
      \frac{(\mathbf{n}^{\alpha}\cdot\mathbf{R}^{ab}_{ij})(\mathbf{n}^{\beta}\cdot\mathbf{R}^{ab}_{ij})}{|\mathbf{R}^{ab}_{ij}|^{5}}
      \right). 
\label{Eqn:Kint}
\end{multline} 
with unit vectors $\mathbf{n}^{\alpha}$ for $\alpha=x,y,z$ in the laboratory
$x,y,z$ directions respectively (see Table~\ref{tab:lattice}). The prefactors of one-half cure the double counting of pairs in
Eq. (\ref{eqn:V}). 

The model space $\mathfrak{M}$ basis states are products of ground state doublet states $|1_{I}\rangle$ and
$|2_{I}\rangle$ over all lattice sites $I$ while excited crystal field states are denoted
$|W_{I}\rangle$ for $W=3,\ldots,13$ on each site $I$. 
The state $| P \rangle$ in Eq. (\ref{eqn:resolvent}) is a direct product of crystal field states on different sites
with the condition that at least one of the states in $|P\rangle$ lies outside the ground state crystal field doublet; in other
words, at least one Tb$^{3+}$ ion must be virtually excited in a state $|n\rangle$ with $n\geq 3$. With this notation in hand, we write the quantum term
$H_{\rm eff}^{(2)}\equiv \mathcal{P}V\mathcal{R}V\mathcal{P}$ as 
\begin{equation}
 \sum_{I_{1}, ..,I_{4}}\sum_{\alpha,\beta,\gamma,\delta} \mathcal{P} \left( \mathcal{K}_{I_{1}I_{2}}^{\alpha,\beta}
 J_{I_{1}}^{\alpha}J_{I_{2}}^{\beta}\right)\hspace{1pt} \mathcal{R} \hspace{1pt}\left(\hspace{1pt}
 \mathcal{K}_{I_{3}I_{4}}^{\gamma,\delta} J_{I_{3}}^{\gamma}J_{I_{4}}^{\delta}\right) \mathcal{P}.
\label{eqn:expanded} 
\end{equation}

There are a few observations that we can make from Eq. (\ref{eqn:expanded}) that identify classes of nonvanishing
terms. Suppose we choose magnetic sites $I_{p}$ on the pyrochlore lattice for $p=1,2,3,4$ in Eq.(\ref{eqn:expanded}). Then, when
we evaluate Eq.(\ref{eqn:expanded}) for all other sites, we obtain unit operators $|1_{I_{m}}\rangle\langle
1_{I_{m}}|+|2_{I_{m}}\rangle\langle 2_{I_{m}}|$ for all sites $I_{m}$ with $m\neq 1,2,3,4$. This follows because the resolvent operator $\mathcal{R}$
and projection operators $\mathcal{P}$ are diagonal on each site. In the following, we do not write out all these unit
operators explicitly. A second observation is that when we consider a term with magnetic sites $I_{p}$ ($p=1,2,3,4$) all different, we
find that such a term vanishes. The reason for this is that the resolvent and angular momentum operators are sandwiched by projectors into the
model space. That way, a virtual excitation induced, for example, by $J_{I_{3}}$ in the $\mathcal{K}_{I_{3}I_{4}}$ bilinear
operator must be ``de-excited'' by an angular momentum operator in the other bilinear operator $\mathcal{K}_{I_{1}I_{2}}$,
(with $I_{1}=I_{3}$, for example). If all $I_{p}$ are different there can be
no virtual excitations and, because the resolvent operator is orthogonal to
the model space states, such terms must vanish.

\begin{figure}
\centering
\subfigure[\hspace{1pt} \label{fig:caseA} An example of Case A with $I_{2}=I_{4}$ and $I_{1}\neq I_{3}$. Only the ion on site $I_{2}$ is virtually
  excited. The two pairs of sites $I_{1}$, $I_{2}$ and $I_{2}$, $I_{3}$ are
  not restricted to be nearest neighbors because the bare Hamiltonian has
  long-ranged dipole-dipole interactions. ]{\includegraphics[width=7cm,clip]{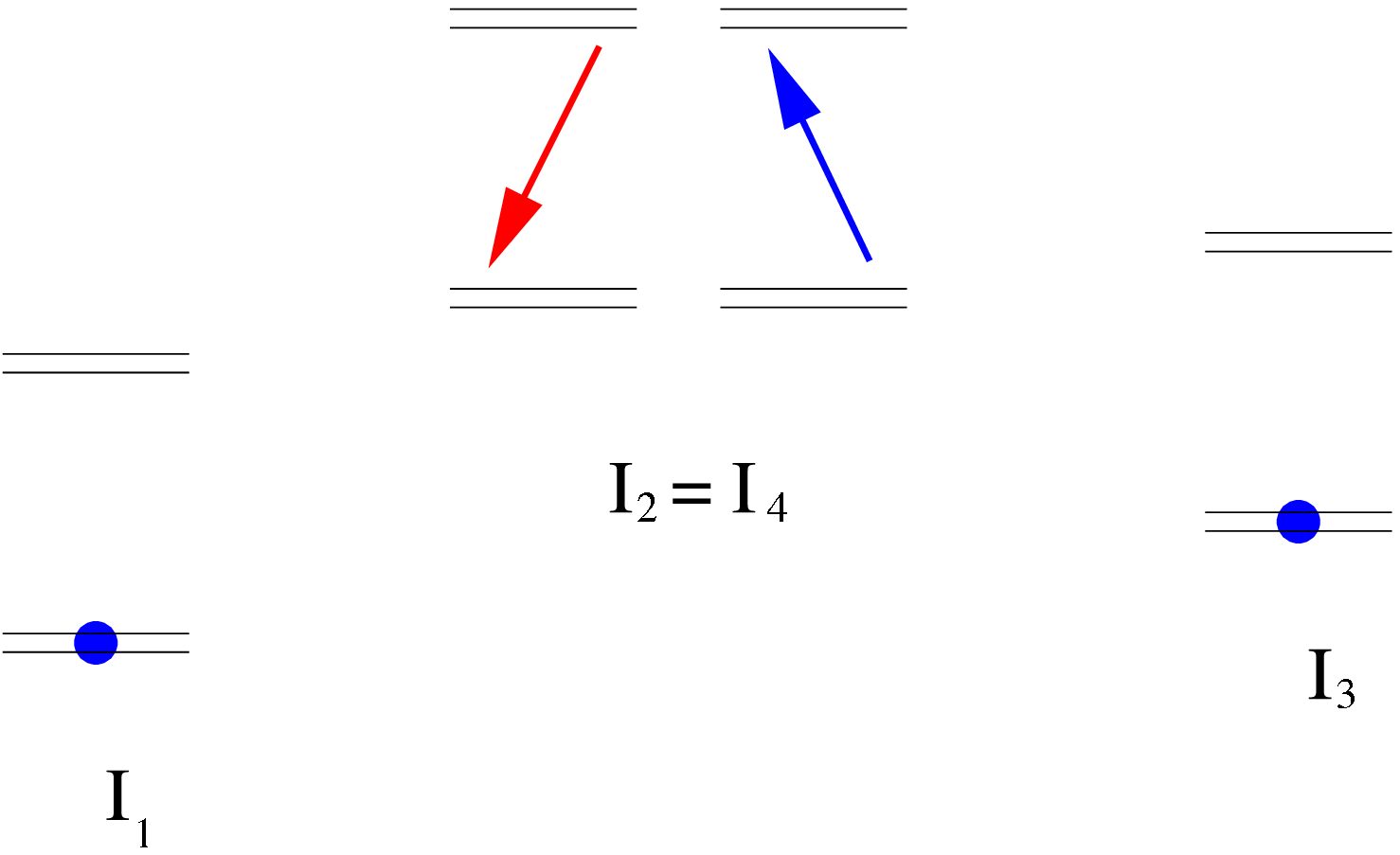}}
\\ $\left. \right.$ \\
\subfigure[\hspace{1pt} \label{fig:caseB} Case B with $I_{1}=I_{3}$ and $I_{2}=I_{4}$ with virtual excitations only on site
  $I_{2}$. There is no virtual excitation on
  site $I_{1}$. $I_{1}$ and $I_{2}$ need not be n- earest neighbors because
  $\mathcal{K}$ includes an intera- ction between dipole moments.]{\includegraphics[width=7cm,clip]{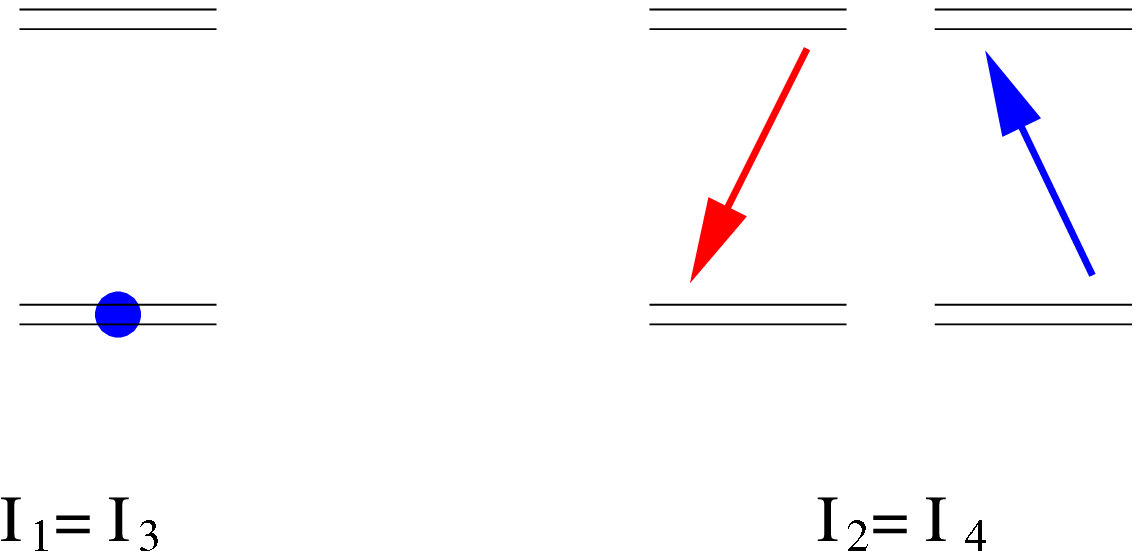}}
\\ $\left. \right.$ \\
\subfigure[\hspace{1pt} \label{fig:caseC} Case C with $I_{1}=I_{3}$ and $I_{2}=I_{4}$. Ions on
  both sites are virtually excited. Similarly to Cases A and B, $I_{1}$ and $I_{2}$ need not be
  nearest neighbors.]{\includegraphics[width=7cm,clip]{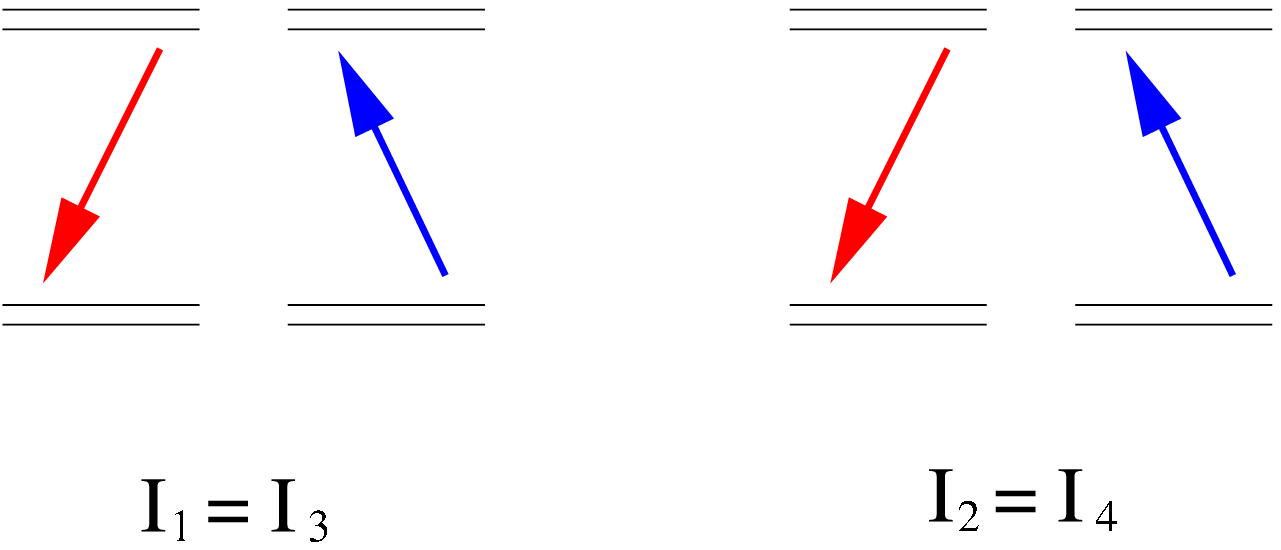}}
\caption{(color online). Figure illustrating the virtual excitations distinguishing three classes of terms in the effective Hamiltonian which are
  enumerated and described in the main text. The arrows show virtual excitations and de-excitations within the lowest-lying
  pair of crystal field doublets belonging to the ion on the labeled
  site. Sites with a blue circle over the ground state doublet indicate that the ion on
  that site remains in its original state within the ground state crystal field doublet.}
\label{fig:cases}
\end{figure}

Having found those terms that must always vanish, we now divide all the
potentially nonvanishing terms into three classes that we shall analyze in turn
in the next three subsections.
\begin{enumerate}[{CASE} A]
\item The first class of terms has two groups $(I_{1}, I_{2})$ and
  $(I_{3},I_{4})$ of sites, with exactly one site in the first group in common
  with a site in the second group. In this case, the ion on the common site
  must be virtually excited and de-excited, and the other two ions
  remain in their ground doublets. This is because the resolvent operator demands that there be some virtual excitations and that the
  projectors require that any virtual excitation must be de-excited. So, only when two angular momentum operators (one in each $V$
  operator of $\mathcal{P}V\mathcal{R}V\mathcal{P}$) belong to a given site can that site be virtually excited.
\item This class of terms has identical pairs $(I_{1}, I_{2})$ and $(I_{3},
  I_{4})$ regardless of label ordering, but with only {\it one
  single} ion ($I_{1}$ or $I_{2}$) that is virtually excited.  
\item Finally, we shall consider the case where $(I_{1}, I_{2})$ and $(I_{3},
  I_{4})$ are identical pairs and where {\it both} ions are virtually excited.
\end{enumerate}
The virtual excitations belonging to each of these three cases are illustrated in Fig.~\ref{fig:cases}.

It will be convenient, while considering the possibilities enumerated above, to make use of the following explicit decomposition
of the quantum term $H_{\rm eff}^{(2)}$:
\begin{eqnarray}
\mathcal{P}V\mathcal{R}V\mathcal{P} & = \mathcal{P}H_{\rm ex}\mathcal{R}H_{\rm
  ex}\mathcal{P} + \left( \mathcal{P}H_{\rm
  ex}\mathcal{R}H_{\rm dd}\mathcal{P} \right. \nonumber
\\ & \left. + \mathcal{P}H_{\rm dd}\mathcal{R}H_{\rm ex}\mathcal{P} \right) + \mathcal{P}H_{\rm dd}\mathcal{R}H_{\rm dd}\mathcal{P}.
\label{eqn:decomp}
\end{eqnarray}
We refer to $\mathcal{P}H_{\rm ex}\mathcal{R}H_{\rm ex}\mathcal{P}$ as the
exchange-exchange part, $\left( \mathcal{P}H_{\rm
  ex}\mathcal{R}H_{\rm dd}\mathcal{P} + \mathcal{P}H_{\rm dd}\mathcal{R}H_{\rm
  ex}\mathcal{P}\right)$ as the exchange-dipole part and $\mathcal{P}H_{\rm
  dd}\mathcal{R}H_{\rm dd}\mathcal{P}$ as the dipole-dipole part.

\subsection{Case A}
\label{sec:caseone}

We will show that the situation in Case A  described above leads to (i) effective Hamiltonian bilinear interactions between the
local $z$ components of the spins and also to (ii) three-body interactions of the form
$\tilde{\sigma}_{I_{1}}^{z}\tilde{\sigma}_{I_{2}}^{\alpha}\tilde{\sigma}_{I_{3}}^{z}$
where $\alpha=x$ or $y$, but not $z$. 

We write the bilinear operator on the left-hand-side of Eq.~(\ref{eqn:expanded}) as $\mathcal{\tilde{K}}_{I_{1}I_{2}}^{\alpha,\beta}
\widetilde{J}^{\alpha}_{I_{1}}\widetilde{J}^{\beta}_{I_{2}}$ and the other bilinear as $\mathcal{\tilde{K}}_{I_{2}I_{3}}^{\alpha,\beta}
\widetilde{J}^{\alpha}_{I_{2}}\widetilde{J}^{\beta}_{I_{3}}$ with $I_{1}\neq I_{3}$ ($I_{2}=I_{4}$, see Fig.~\ref{fig:caseA}) with all angular momentum
components referred to the local coordinate system. As we discussed above, the contribution of all the
other sites gives identity operators for each site. Omitting these unit operators, we are left with
\begin{multline}
\sum_{\alpha,\beta,\rho,\sigma} \sum_{{m_{p}}} \sum_{W} \mathcal{P}( m_{1},m_{2},m_{3} )  \tilde{K}_{I_{1}I_{2}}^{\alpha\beta}\widetilde{J}_{I_{1}}^{\alpha}
  \widetilde{J}_{I_{2}}^{\beta} \\ \times \frac{|m_{4,I_{1}},W_{I_{2}},m_{3,I_{3}}\rangle\langle m_{4,I_{1}}, W_{I_{2}},
  m_{3,I_{3}}|}{E_{0} - E_{W}} \\ \times \tilde{K}_{I_{2}I_{3}}^{\rho\sigma}\widetilde{J}_{I_{2}}^{\rho} \widetilde{J}_{I_{3}}^{\sigma}
  \mathcal{P}( m_{4},m_{5},m_{6} )
\label{eqn:caseone}
\end{multline}
with
\begin{align}  \mathcal{P}( m_{1},m_{2},m_{3} ) & \equiv |m_{1,I_{1}},m_{2,I_{2}},m_{3,I_{3}}\rangle\langle
m_{1,I_{1}},m_{2,I_{2}},m_{3,I_{3}} | \nonumber   \\
 \mathcal{P}( m_{4},m_{5},m_{6} ) & \equiv |m_{4,I_{1}},m_{5,I_{2}},m_{6,I_{3}}\rangle\langle
m_{4,I_{1}},m_{5,I_{2}},m_{6,I_{3}} |. 
\end{align}
$E_{W}$ is the energy of an excited crystal field state on a single ion.
Here, the angular momentum components are expressed in their respective local coordinate systems with local $\mathbf{z}$
axes given in Table \ref{tab:lattice}, the rotation matrices
having been absorbed implicitly into
$\tilde{K}_{I_{1}I_{2}}^{\alpha\beta}$. The integers ${m_{p}}$ run over $1$ and $2$ with the states lying  within
$\mathfrak{M}_{I}$ on their respective sites. We factor out the part for site $I_{1}$: $\sum_{m_{1},m_{4}}|m_{1}\rangle\langle m_{1}|
\widetilde{J}_{I_{1}}^{\sigma}|m_{4}\rangle\langle m_{4} |$. Recalling the property, Eq. (\ref{eqn:ME}) and the mapping in
Eq. (\ref{eqn:Pauli}), we obtain $\langle \widetilde{J}^{z}\rangle\tilde{\sigma}^{z}_{I_{1}}$. We reach the same result for the sum
over states $m_{3}$ and $m_{6}$ on site $I_{3}$. Equation (\ref{eqn:caseone}) then simplifies to
\begin{multline}
\sum_{\beta,\rho} \tilde{K}_{I_{1}I_{2}}^{z \beta} \tilde{K}_{I_{2}I_{3}}^{\rho z} \langle \widetilde{J}^{z}\rangle^{2} \tilde{\sigma}_{I_{1}}^{z}
\tilde{\sigma}_{I_{3}}^{z} \\ \times  \left( \sum_{m_{2}, m_{5}} \sum_{W}   |m_{2}\rangle\langle m_{5}| \frac{ \langle m_{2} | \widetilde{J}_{I_{2}^{\beta}}|W\rangle\langle W |
\widetilde{J}_{I_{2}}^{\rho}| m_{5}\rangle }{E_{0}-E_{W}}  \right)
\label{eqn:caseone2}
\end{multline}
in the local coordinate system where we have dropped the $I_{2}$ site labels from the state vectors enclosed by brackets.
After summing over the excited states $|W\rangle$, and rendering the sum of operators in terms of Pauli matrices, we find
an Ising interaction term $\tilde{\sigma}_{I_{1}}^{z}\tilde{\sigma}_{I_{3}}^{z}$ and three-body operators
$\tilde{\sigma}_{I_{1}}^{z}\tilde{\sigma}_{I_{2}}^{\alpha} \tilde{\sigma}_{I_{3}}^{z}$ with $\alpha=x,y$ and $\alpha\neq z$ since,
if $\alpha$ were to equal $z$, the three-body term would violate time reversal invariance. The sum over virtual excited state and
the subsequent rendering in terms of Pauli operators is discussed in some more detail in Appendix \ref{sec:calc}

We have reduced the most general three ion terms in $H_{\rm eff}^{(2)}$ (case A) to interactions between pseudospins one-half but we have not
made any assumptions yet about the form of the interactions
$\mathcal{K}_{I_{1}I_{2}}^{\alpha\beta}$. In the following, we shall consider the four terms of Eq.~(\ref{eqn:decomp})
in turn within Case A. These terms determine the spatial range of the
resulting effective interactions between the pseudospins and are obtained by distinguishing the exchange and dipolar parts
of $\tilde{K}$ in Eq. (\ref{eqn:caseone2}).

\begin{figure}
\includegraphics[width=0.3\textwidth]{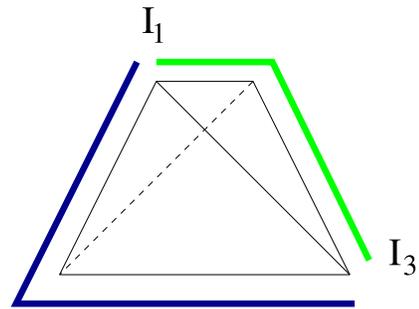}
\caption{\label{fig:nnIsing} (color online). In deriving $H_{\rm eff}^{(2)}
  =\mathcal{P}V\mathcal{R}V\mathcal{P}$, where $V$ is a sum of bilinear interactions, we consider a single
  pairwise interaction in the right-hand interaction $V$ on sites $I_{2}$ and $I_{3}$ and a pairwise interaction in the left-hand
  $V$ between $I_{1}$ and $I_{2}$. This choice of terms in  $H_{\rm
  eff}^{(2)}$ is referred to as Case A in the main text. Because $I_{1}\neq I_{3}$, and because the
  operators on these sites are sandwiched between $\mathbf{P}$ projectors, the only virtually excited site is $I_{2}$ while the
  other two sites remain in their (noninteracting) crystal field ground state. As we show in Appendix~\ref{sec:calc}, one obtains
  effective $\tilde{\sigma}^{z}$ operators on site $I_{1}$ and $I_{3}$. There are two possibilities for the effective operator on
  site $I_{2}$ after calculation: it could be a unit operator leaving an Ising coupling between sites $I_{1}$ and $I_{3}$, or it
  could give rise to a transverse operator $\tilde{\sigma}^{x}_{I_{2}}$ or
  $\tilde{\sigma}^{y}_{I_{2}}$, generating an effective three-body term
  connecting sites $I_{1}$, $I_{2}$ and $I_{3}$. The figure shows a single tetrahedron in a pyrochlore lattice with the two ways of
  joining sites $I_{1}$ and $I_{3}$. The total effective Ising exchange for
  this Case A between ions $I_{1}$ and $I_{3}$, $J_{I_{1}I_{2}}^{zz}$ is the sum of the
  Ising exchange terms corresponding to each path in the figure.}
\end{figure}

\begin{figure}
\includegraphics[width=0.4\textwidth]{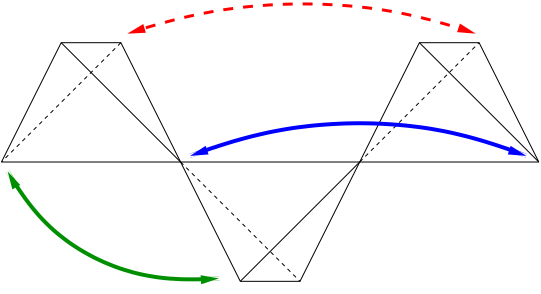}
\caption{\label{fig:fnnIsing} (color online). Part of a chain of tetrahedra in
  a pyrochlore lattice. For case A (with only exchange interactions in the
  bare Hamiltonian), the quantum
  part, $H_{\rm eff}^{(2)}=\mathcal{P}V\mathcal{R}V\mathcal{P}$, of the
  effective Hamiltonian has nonvanishing couplings for interactions between
  the sites connected by the thick lines. The (green) curve on the
  left-hand-side indicates that Ising exchange acts between second nearest neighbors. The (blue) curve on the right-hand-side indicates
  that those third nearest neighbors lying along chains of sites are coupled
  by Ising exchange.  The red dashed curve indicates a second type of third nearest
  neighbor for which no effective coupling is generated as explained in the
  main text. Three-body interactions also
  arise that couple all three ions along each of the paths joined in the
  figure. These further neighbor couplings receive a further contribution when
  the dipolar coupling is switched on. Also, in the presence of the
  dipolar interaction, there are off-diagonal effective couplings of the form
  $J_{I_{1}I_{2}}^{\mu\nu}\tilde{\sigma}_{I_{1}}^{\mu}\tilde{\sigma}_{I_{2}}^{\nu}$
  with $\mu \neq z$, $\nu\neq z$ in addition to an Ising coupling that renormalizes the classical (Ising, first order) DSIM $\mathcal{P}V\mathcal{P}$ term.}
\end{figure}

\begin{figure}
\includegraphics[width=0.5\textwidth]{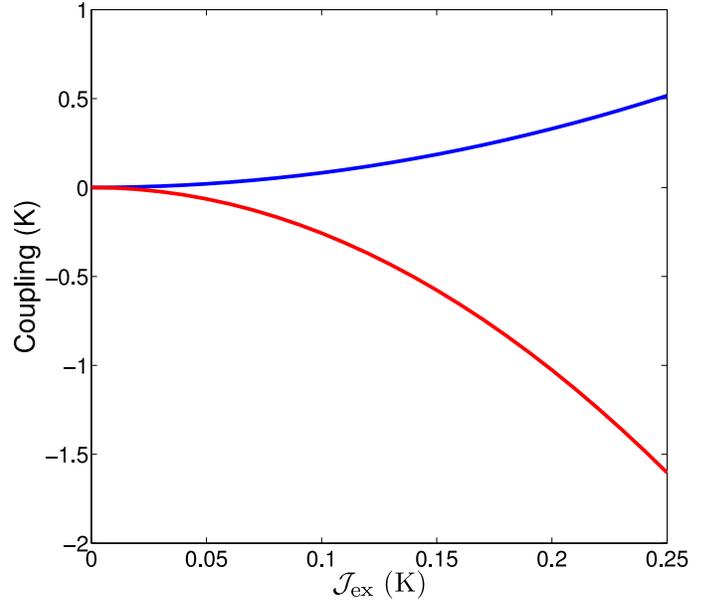}
\caption{\label{fig:fnnCoupling} (color online). Further neighbor Ising interactions generated by quantum fluctuations when
  $\mathcal{D}=0$. The upper curve is the second nearest neighbor Ising coupling (between spins with sublattice labels $1$ and
  $3$). It is antiferromagnetic in nature. The lower curve is the ferromagnetic third nearest neighbor interaction (between two
  spins with sublattice label $1$ $-$ see straight right-hand blue line in Fig.~\ref{fig:fnnIsing}). When $\mathcal{D}\neq 0$,
  these couplings are renormalized by other contributions from Class A and Class C terms. Since couplings are referred to
  operators in the local coordinate system, they contain geometrical factors from the rotation matrices (see
  Sections~\ref{sec:model} and \ref{sec:convention}).}
\end{figure}

\underline{Exchange-exchange part} The exchange-exchange part (referring to the first term of Eq. (\ref{eqn:decomp}))
$-$ which is nothing more than Eq. (\ref{eqn:caseone2}) for $\tilde{K}$ with $\mathcal{D}=0$ $-$ is
a short-range, but not strictly nearest neighbor, effective interaction. If $I_{1}$, $I_{2}$ and $I_{3}$ all lie on the same tetrahedron in the lattice, then the Ising
interaction acts between nearest neighbors and, for given $I_{1}$ and $I_{3}$, there are two choices for the position of the
``mediating ion'' $I_{2}$ as shown in Fig. \ref{fig:nnIsing}. The thick lines in this figure join the $I_{1}$ and $I_{3}$ ions via
two different choices for the ion $I_{2}$ and generate, together with three-body interactions connecting $I_{1}$, $I_{2}$ and $I_{3}$,
 a nearest neighbor effective Ising interaction between $I_{1}$ and $I_{3}$. As we
describe in more detail later on, this Ising interaction renormalizes the
$\mathcal{J}_{{\rm classical}}$ exchange, defined in Eq. (\ref{eqn:Jclassical}). Within
a single tetrahedron, the three-body interactions couple all three pseudospins along each of the paths in Fig.
\ref{fig:nnIsing}. There are two other exchange-exchange pseudospin terms
arising from Eq. (\ref{eqn:caseone2}) $-$ those for which the interaction extends
further than a single tetrahedron. The ions at the endpoints of the lines on the left-hand-side and right-hand-side of the Fig. \ref{fig:fnnIsing} are coupled,
respectively, by effective Ising-like second nearest neighbor interactions and
effective third nearest neighbor Ising exchange. The ions along each line -
those at the endpoints and the one at the center of each line - interact also via effective three-body interactions. The
dependence of the effective couplings
for second and third nearest neighbor Ising interactions, on the bare exchange coupling $\mathcal{J}_{{\rm ex}}$ is shown in
Fig. \ref{fig:fnnCoupling} (the dipole-dipole coupling $\mathcal{D}$ being set equal to
zero). The second nearest neighbor effective coupling is antiferromagnetic, and the
third nearest neighbor effective coupling is ferromagnetic when the bare
exchange coupling is antiferromagnetic ($\mathcal{J}_{\rm ex}> 0$). See Section~\ref{sec:convention} for the convention on the
exchange that we use in this paper. We note that there are two distinct types of
third nearest neighbors on the pyrochlore lattice (see, for example Fig. $1$
of Ref.~\onlinecite{Wills} or Fig. $2$ in Ref.~\onlinecite{DelMaestro}) and that only one type $-$ those connected by one
mediating ion, or two edges, along the lattice $-$ appear in $H_{\rm
  eff}^{(2)}$. Third nearest neighbours of the other type have sites that are connected via
three edges along the lattice (as shown by the dotted line of Fig. \ref{fig:fnnIsing}) and hence couplings between ions on these sites
would require two mediating ions to appear in the effective Hamiltonian. But,
to the (second) order of perturbation theory that we are considering, there is a
maximum of one mediating ion so such couplings do not appear in $H_{\rm eff}^{(2)}$.

Having discussed the exchange-exchange part, we switch on the dipolar
interaction. In doing so, the interactions become anisotropic even in the bare
Hamiltonian. Nevertheless, in Case A, the types of couplings that arise in the
presence of the dipolar coupling are the same as those arising in the
exchange-exchange case, the only difference being in the range over which the
couplings act.

\underline{Exchange-dipole part}  The exchange-dipole part of the Class A interactions consists of those terms in Eq. (\ref{eqn:caseone2}) with $\mathcal{D}_{\rm ex}=0$
between, say $I_{1}$ and $I_{2}$ and $\mathcal{J}_{\rm ex}=0$ between $I_{2}$ and $I_{3}$. There are then long-range Ising-like
interactions between $I_{1}$ and $I_{3}$ although the exchange Hamiltonian constrains two sites, 
$I_{1}$ and $I_{2}$, in this case, to be nearest neighbors. The bare
microscopic dipole interaction acts between 
sites $I_{2}$ and $I_{3}$ and decays as
$|\mathbf{R}_{I_{2}I_{3}}|^{-3}$. Overall the Case B effective interactions decay as
$|\mathbf{R}_{I_{1}I_{3}}|^{-3}$ at long distances just as the bare dipole interactions do on their own. The same is true, by symmetry, of the Class A
interactions belonging to the dipole-exchange term in Eq.~(\ref{eqn:decomp}). There are also three-body interactions
originating from the exchange-dipole and dipole-exchange parts of Eq.~(\ref{eqn:decomp}) where two of the spins operators must lie
on nearest neighbor sites. 

\underline{Dipole-dipole part} Finally, the terms in Class A belonging to the dipole-dipole term of Eq. (\ref{eqn:decomp}) ($\mathcal{J}_{\rm ex}=0$ in Eq.
(\ref{eqn:caseone2})) have a range that is a product of two dipole interactions
so the overall two-body interaction between sites $I_{1}$ and $I_{3}$ decays
as a function of $I_{2}$ as
$|\mathbf{R}_{I_{1}I_{2}}|^{-3}|\mathbf{R}_{I_{2}I_{3}}|^{-3}$ before summing
over. 

In summary, we have
learned in this section, in the discussion following Eq.~(\ref{eqn:decomp}), and in Appendix \ref{sec:calc},
that the effective pseudospin operator on  the ``connecting site'' $I_{2}$
involved in the virtual excitation process can be
a unit operator so that the resulting nontrivial effective interaction is a bilinear of Ising pseudospin operators between $I_{1}$ and $I_{3}$. In this circumstance,  there are contributions to this interaction for $I_{2}$
positions at arbitrarily large distances from $I_{1}$ and $I_{3}$. Together with two body Ising interactions, there are also three spin
long-range interactions in this group of terms following from the argument given above. There are no constraints on the positions of
the coupled sites $I_{1}$, $I_{2}$ and $I_{3}$ in the lattice.

\subsection{Case B}
\label{sec:casetwo}

Now, referring to Eq.~(\ref{eqn:expanded}), we impose the constraint $I_{3}=I_{1}$ and $I_{4}=I_{2}$ and, without loss of generality,
suppose that only the ion on site $I_{2}$ is virtually excited (see Fig.~\ref{fig:caseB}). After computing Eq.~(\ref{eqn:expanded}) with the
aforementioned constraint, we obtain a result that is proportional to the unit operator. That is, this case gives rise to a
constant shift in energy. As a first step, we write out Eq.~(\ref{eqn:V}) with the unit operators on all sites but $I_{1}$ and
$I_{2}$ factored out:
\begin{multline}
\sum_{\alpha,\beta,\rho,\sigma} \sum_{{m_{p}}} \sum_{W} \mathcal{P}( m_{1},m_{2} )  \tilde{K}_{I_{1}I_{2}}^{\alpha\beta}\widetilde{J}_{I_{1}}^{\alpha}
  \widetilde{J}_{I_{2}}^{\beta} \\ \times \frac{|m_{3},W \rangle\langle m_{3}, W |}{E_{0} - E_{W}}  
  \tilde{K}_{I_{1}I_{2}}^{\rho\sigma}\widetilde{J}_{I_{1}}^{\rho} \widetilde{J}_{I_{2}}^{\sigma} \mathcal{P}( m_{4},m_{5} ),
\end{multline}
where the notation $\mathcal{P}( m_{1},m_{2} )$ stands for operator $|m_{1, I_{1}},m_{2, I_{2}} \rangle\langle m_{1,I_{1}},m_{2,
  I_{2}} |$. 
Because only the ion on site $I_{2}$ is virtually excited, the operator $\widetilde{J}_{I_{1}}^{\alpha}$ acts
  entirely on states within $\mathfrak{M}_{I_{1}}$ so, referring to the matrix elements in Eq. (\ref{eqn:ME}) we obtain, for ion $I_{1}$, \begin{multline}
 |1_{I_{1}} \rangle\langle 1_{I_{1}}| \langle 1_{I_{1}}| \widetilde{J}_{I_{1}}^{z}|1_{I_{1}} \rangle^{2} + |2_{I_{1}} \rangle\langle
  2_{I_{1}} | \langle 2_{I_{1}} | \widetilde{J}_{I_{1}}^{z} |2_{I_{1}} \rangle^{2} \\ = \langle
  \widetilde{J}_{I_{1}}^{z}\rangle^{2}\mathbb{I}_{I_{1}}  
\end{multline}
$-$ the unit operator acting on $I_{1}$, which we can omit in the following. When the sum is performed over excited crystal field states on
  ion $I_{2}$, the resulting operators map to the unit operator $\mathbb{I}_{I_{2}}$ and Pauli matrices
  $\tilde{\sigma}^{x}_{I_{2}}$ and $\tilde{\sigma}^{y}_{I_{2}}$ $-$ all the
  resulting operators being consistent with time reversal. This
  calculation is performed along the lines described in Appendix
  \ref{sec:calc}. 

It would thus seem that, together with the constant energy shift, there are
  effective nontrivial single-site transverse field operators in the effective
  Hamiltonian of the form $\tilde{\sigma}^{x}$ and $\tilde{\sigma}^{y}$. However, these effective transverse field terms cancel on
  any given lattice site when one sums the contributions to these single-site operators coming from the bare microscopic pairwise
  interactions in $V$. Without going into the details of the sum over different contributions to site $I_{2}$ we see that there
  must be such a cancellation because neither the original model nor the effective Hamiltonian formalism distinguishes particular
  directions (as opposed to particular axes) on individual sites. When the Tb$^{3+}$ ions are randomly diluted with nonmagnetic ions, as in
  (Tb$_{p}$Y$_{1-p}$)$_{2}$Ti$_{2}$O$_{7}$,\cite{Keren} there is no longer perfect cancellation of the effective single site
  operators. These effective fields have the effect of splitting the degenerate $|1\rangle$, $|2\rangle$ doublet on each ion for
  which the cancellation does not occur. So, in fact, the low energy effective theory of the diluted compound
  (Tb$_{p}$Y$_{1-p}$)$_{2}$Ti$_{2}$O$_{7}$ would be somewhat different than that of the pure \tto\ by admitting effective random
  transverse fields. The possible generation of effective random transverse fields generated by dilution in
  (Tb$_{p}$Y$_{1-p}$)$_{2}$Ti$_{2}$O$_{7}$ had previously been proposed in Ref.~\onlinecite{Duijn}. In the remainder of the article, we shall assume that the magnetic ions are not diluted.

\subsection{Case C}
\label{sec:casethree}

The class of terms where the two ions $I_{1}$ and $I_{2}$ are both virtually
excited (see Fig.~\ref{fig:caseC}) is the most complicated of the three cases A,B and C in the sense that all
two body terms consistent with time reversal and the lattice symmetries can
and do arise. Because the dipolar coupling is nonzero, long range effective
interactions appear in $H_{\rm eff}$. The calculation of the types of terms
and their couplings in Case C is most easily accomplished by the projection method given
in Appendix~\ref{sec:calc}. In this section then, we give only the results of
this calculation $-$ the means of calculation having been outlined in the
discussion of Section~\ref{sec:caseone} and in Appendix~\ref{sec:calc}. As with the terms
in Case B, the net single-site ``fields'', $\tilde{\sigma}^{x}$ and $\tilde{\sigma}^{y}$ cancel, leaving the Ising interaction
$\tilde{\sigma}^{z}_{I_{1}}\tilde{\sigma}^{z}_{I_{2}}$, and the transverse exchange interactions
$\tilde{\sigma}^{\alpha}_{I_{1}}\tilde{\sigma}^{\beta}_{I_{2}}$ where each of
$\alpha$ and $\beta$ can be $x$ and $y$. 

We make the observation here, that is discussed in more detail in Appendix~\ref{sec:calc}, that the transverse exchange
interactions can appear in the effective Hamiltonian from VCFEs involving only the ground state doublet and first excited states
because there are nonvanishing $\tilde{J}^{x}$, $\tilde{J}^{y}$ and $\tilde{J}^{z}$ matrix elements between these states. Hence,
the appearance of these effective transverse exchange interactions in $H_{\rm
  eff}^{(2)}$  is strongly tied to the specific form of the \tto\ single ion
crystal field wavefunctions. 

Because the bare exchange interaction vanishes if $I_{1}$ and $I_{2}$
are not nearest neighbors, even if one of the bilinear interactions in
Eq. (\ref{eqn:decomp}) is a dipole-dipole interaction, the cutoff coming from
the bare exchange ensures vanishing of the effective interaction beyond
nearest neighbor for Case C.
This means that the only interactions in Case C extending beyond nearest
neighbors come from
the dipole-dipole part of Eq.~(\ref{eqn:decomp}) and, because there are only
two ions involved in Case C (see Fig.~\ref{fig:caseC}), the interaction falls
off as the square of the dipole-dipole interaction: $1/|\mathbf{R}_{I_{1}I_{2}}|^{6}$.

\subsection{Treatment of the dipole-dipole interactions}
\label{sec:dipoles}

In Section~\ref{sec:ground}, the semiclassical ground states of the effective Hamiltonian are computed first on a single
tetrahedron (Section~\ref{sec:4sbl}), then on a periodic cubic unit cell (Section~\ref{sec:16sbl}). In the former case, the bare
dipole-dipole interaction, $H_{\rm dd}$, is truncated beyond nearest neighbors. In the latter case, one should not truncate the long-range
dipole-dipole interaction. This problem has been approached in two ways. The first way is to derive the effective interactions on
a finite but large lattice. Then, to obtain the interaction between sublattices $a$ and $b$ on a periodic unit cell with sixteen
sublattices, the interactions between sublattice $a$ and all the periodic images of $b$ on the large lattice are summed up. A second way
to treat the long-range dipole is to compute the bare microscopic interactions on a periodic cubic unit cell by an Ewald summation
\cite{Ewald,Enjalran} and then to derive the effective Hamiltonian respecting the periodicity. The first approach was used in
Ref.~\onlinecite{MolavianThesis}. Here we use the latter.

\subsection{Summary}
\label{sec:summary}

We have now worked out the different types of effective interactions that
arise in the Hamiltonian $H_{\rm eff}$ obtained from the model
of Section \ref{sec:model} for \tto\ to lowest order in the virtual crystal
field excitations (VCFEs). Before describing previously published results obtained from
the effective Hamiltonian when considering a single (isolated) tetrahedron, we
briefly summarize here the results of Sections \ref{sec:classical} and
\ref{sec:quantum}.

 The effective Hamiltonian for
\tto\ has been derived to order $(\langle V\rangle/\Delta)$ which includes a classical part and also interactions coming from VCFEs
to lowest order. The classical part of the effective Hamiltonian is given by
$\mathcal{P}V\mathcal{P}$ and is nothing
other than the dipolar spin ice model (DSIM) with Ising
exchange $\mathcal{J}_{\rm classical}$ and dipole-dipole couplings that merely
differ from those of the microscopic model
(Section~\ref{sec:model}) by a constant factor related to the expectation value of the bare angular momentum in the crystal field
ground states; $\mathcal{J}_{{\rm classical}} = \mathcal{J}_{{\rm ex}}\langle
\widetilde{J}^{z} \rangle^{2}$. 

The effective Hamiltonian $H_{\rm eff}$ is
expressed in terms of spin one-half operators which have different time reversal properties (Eq.~(\ref{eqn:time}))  compared to true
angular momentum operators. A large number of different pseudospin
interactions appear in the quantum term $H_{\rm eff}^{(2)}$. These are
constrained to be invariant under time reversal and to respect lattice symmetries. If we switch off the
dipole-dipole interaction temporarily ($\mathcal{D}=0$), we find nearest neighbor Ising interactions which renormalize those from the classical
term and, also, transverse terms of the form
$\tilde{\sigma}^{\alpha}_{I_{1}}\tilde{\sigma}^{\beta}_{I_{2}}$ where $\alpha,
\beta$ can be components $x$ or $y$. Effective interactions beyond nearest
neighbor are Ising exchange interactions between second nearest neighbors and
one out of the two distinct types of third nearest
neighbors on the pyrochlore lattice (see Fig.~\ref{fig:fnnIsing}). Finally, three-body
interactions of the form $J^{\rm
  zxz}\tilde{\sigma}^{z}\tilde{\sigma}^{x}\tilde{\sigma}^{z}$ are also generated.
When dipole-dipole interactions are restored, ($\mathcal{D}\neq 0$),  $H_{\rm
  eff}$ acquires two new types of effective interaction. 
\begin{enumerate}
\item New short-range interactions acting between nearest neighbors
and beyond, decaying as $1/|\mathbf{R}|^{6}$. 
\item Long range interactions decaying as $1/|\mathbf{R}|^{3}$. 
\end{enumerate}
Both contributions, arising when the dipole-dipole interactions are
switched on, further renormalize the effective nearest neighbor Ising coupling
and contribute to the transverse effective couplings. 

For later reference we write the nearest neighbor effective couplings between ions on sites $I_{1}$ and $I_{2}$ as
\begin{equation}
J^{ zz}_{I_{1}I_{2}}\tilde{\sigma}_{I_{1}}^{z}\tilde{\sigma}_{I_{2}}^{z} +
J^{\alpha\beta}_{I_{1}I_{2}}\tilde{\sigma}_{I_{1}}^{\alpha}\tilde{\sigma}_{I_{2}}^{\beta}.
\label{eqn:couplings}
\end{equation}
where $\alpha$ and $\beta$ can each equal $x$ and $y$.

\begin{figure}
\includegraphics[width=0.5\textwidth]{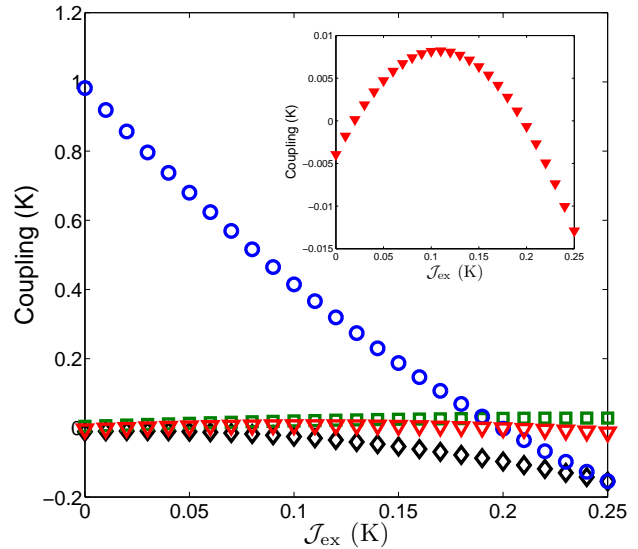}
\caption{\label{fig:Couplings} (color online). Plot showing the variation of
  various couplings in the effective Hamiltonian on a
  lattice, as a function of the bare exchange coupling $\mathcal{J}_{{\rm
  ex}}$. The bare Hamiltonian has long-range dipole-dipole
  interactions. We plot bilinear Ising $J^{ zz}$, transverse $J^{ xx}$ and
  $J^{ yy}$ couplings (Eq.~(\ref{eqn:couplings})) between nearest neighbor
  sublattices $1$ and $2$ and a three-body coupling $J^{ zxz}$ connecting
  sublattices $1$, $2$ and $3$. The difference between the $J^{ xx}$ and
  $J^{ yy}$ couplings is due to the choice of local $x$ and $y$ axes. All couplings refer to pseudospin couplings in
  the local coordinate system. The Ising coupling changes sign at about
  $\mathcal{J}_{\rm ex}\sim 0.2$ K implying a cross-over from ice-like order
  to AIAO order in the absence of other interactions. For $\mathcal{J}_{\rm
  ex}\gtrsim 0.2$ K, the Ising and transverse terms are of similar magnitude. The three-body coupling is the weakest of the three interactions; its variation is shown in the inset. 
}
\end{figure}

Now we consider the relative magnitude of some different effective couplings
on a lattice which will be of use in later sections when we interpret our ground
state phase diagrams. Fig.~\ref{fig:Couplings} shows three
different effective Hamiltonian couplings as a function of the bare exchange coupling $\mathcal{J}_{{\rm ex}}$ when the
dipole-dipole coupling $\mathcal{D}$ is fixed at the value for \tto; $\mathcal{D}=0.0315$ K. The three
couplings are for nearest neighbor interactions $J^{ 
  zz}\tilde{\sigma}^{z}\tilde{\sigma}^{z}$ and $J^{ 
  xx}\tilde{\sigma}^{x}\tilde{\sigma}^{x}$ and the three-body interaction
$J^{ zxz}\tilde{\sigma}^{z}\tilde{\sigma}^{x}\tilde{\sigma}^{z}$, all
expressed in the local coordinate system. The transverse bilinear couplings
are averaged over the bonds on a single tetrahedron to give an idea of the
scale of the interactions while the three-body
coupling is plotted for bonds connecting sublattices $1$, $2$ and $3$ on a
single tetrahedron (see Table
\ref{tab:lattice}) which is representative of the scale of these
interactions. Looking at these nearest neighbor couplings, we see that the Ising interaction changes sign at about
$\mathcal{J}_{{\rm ex}} \approx 0.2$ \nolinebreak K. For $\mathcal{J}_{{\rm ex}} < 0.2$ \nolinebreak K, the Ising interaction
favors ice-like order and, for $\mathcal{J}_{{\rm ex}} > 0.2$ \nolinebreak K, it favors all-in/all-out ordering (see
Fig.~\ref{fig:AIAO}).  The Ising coupling $J^{ zz}$ is the largest coupling over most of the range of $\mathcal{J}_{{\rm
    ex}}\lesssim 0.2$ K, followed by transverse couplings, for instance $J^{ xx}$, and then the three-body interaction $J^{
  zxz}$. For $\mathcal{J}_{{\rm ex}}\gtrsim 0.2$ K, the Ising and transverse
couplings are of similar magnitude. This is therefore a direct microscopic
derivation showing the restoration of effective pseudospin isotropy that is discussed in
Refs.~\onlinecite{Gardner2} and \onlinecite{Enjalran}.

\subsection{Relation with previous results}
\label{sec:results}

In Section~\ref{sec:classical} and so far in Section \ref{sec:quantum}, we have presented a derivation of an effective Hamiltonian for \tto\ to
lowest order in the quantum corrections to the DSIM $-$ that is, to order $\langle V\rangle/\Delta$. If this model is
to prove useful, it is important to ensure that $\langle V\rangle/\Delta$ is not so large that higher order terms contribute significantly to the low energy physics of \tto. To test the assertion that
higher order terms (those of order $(\langle V\rangle/\Delta)^{n}$ for $n\geq 2$) are not required, a direct comparison was made in Ref.~\onlinecite{Molavian1} with a microscopic model. This
microscopic model is the one presented in Section \ref{sec:model}, but with
the microscopic exchange and dipolar interactions restricted to spins on a single tetrahedron and
with the crystal field spectrum cut off beyond the four lowest energy
states. More precisely, instead of diagonalizing the full model $H_{\rm cf}+V$,
the single ion crystal field Hamiltonian is diagonalized to obtain states $|n\rangle$ and corresponding energies $E_{n}$,
whereupon all but the lowest two doublets are neglected leaving, as the new (truncated,``tr'') crystal field Hamiltonian,
\begin{multline} H_{\rm cf,tr} = \sum_{I_{p}}  E_{0}(|1_{I_{p}}\rangle \langle 1_{I_{p}}| + |2_{I_{p}}\rangle \langle 2_{I_{p}}|) \\ +
(E_{0}+\Delta)(|3_{I_{p}}\rangle \langle 3_{I_{p}}| + |4_{I_{p}}\rangle\langle 4_{I_{p}}|). \label{eqn:truncate}  \end{multline} 
This model should be a good approximation for sufficiently weak interactions given that, when interactions are switched on, the
excited crystal field levels admix into the ground state doublet with the first excited levels having the greatest contribution to
the new ground state out of all the excited levels. \cite{Admixing}  Including
interactions on a single tetrahedron within the basis of four single ion
states on each site, the model was diagonalized exactly for a range of bare exchange couplings $\mathcal{J}_{\rm ex}$. For comparison,
the effective Hamiltonian was derived on a single tetrahedron and diagonalized
computationally to obtain its spectrum. One comparison that has been made from these spectra involves
looking at the variation in the ground state degeneracy as a function of the
bare exchange coupling, $\mathcal{J}_{\rm ex}$, and the crystal field gap as shown in Fig. \ref{fig:GSdeg}. As the bare exchange
coupling $\mathcal{J}_{\rm ex}$ and the gap $\Delta$ are varied, with $\mathcal{D}$ fixed to $\mathcal{D}=0.0315$ K, there is a phase boundary between a region with a singlet ground state and a region with a doublet ground state. The boundary between these
regions is the same for both the effective Hamiltonian and the model with a truncated crystal field spectrum when the gap $\Delta$
is infinite. The boundaries move apart as $\Delta$
decreases. However, the difference between the two phase boundaries remains relatively fairly small even when the gap is about $18$ K, as for \tto;
$\mathcal{J}_{\rm ex}(\Delta =18 {\rm K})$ for the phase boundaries agree to within
ten percent. Perhaps most importantly, both calculations agree that, as the
gap decreases, the region of parameter space over which the singlet occurs
becomes larger. For the parameters estimated for \tto, \cite{Gingras,
  Mirebeau} the single tetrahedron ground state is a singlet. The ground state
for the microscopic model, for the \tto\ parameters, is mainly a superposition
of two-in/two-out states.  For this reason, this state has been called a {\it quantum spin ice}. \cite{Molavian1, Molavian2}

\begin{figure}
\subfigure[\hspace{1pt} Isotropic exchange.]{\includegraphics[width=0.5\textwidth]{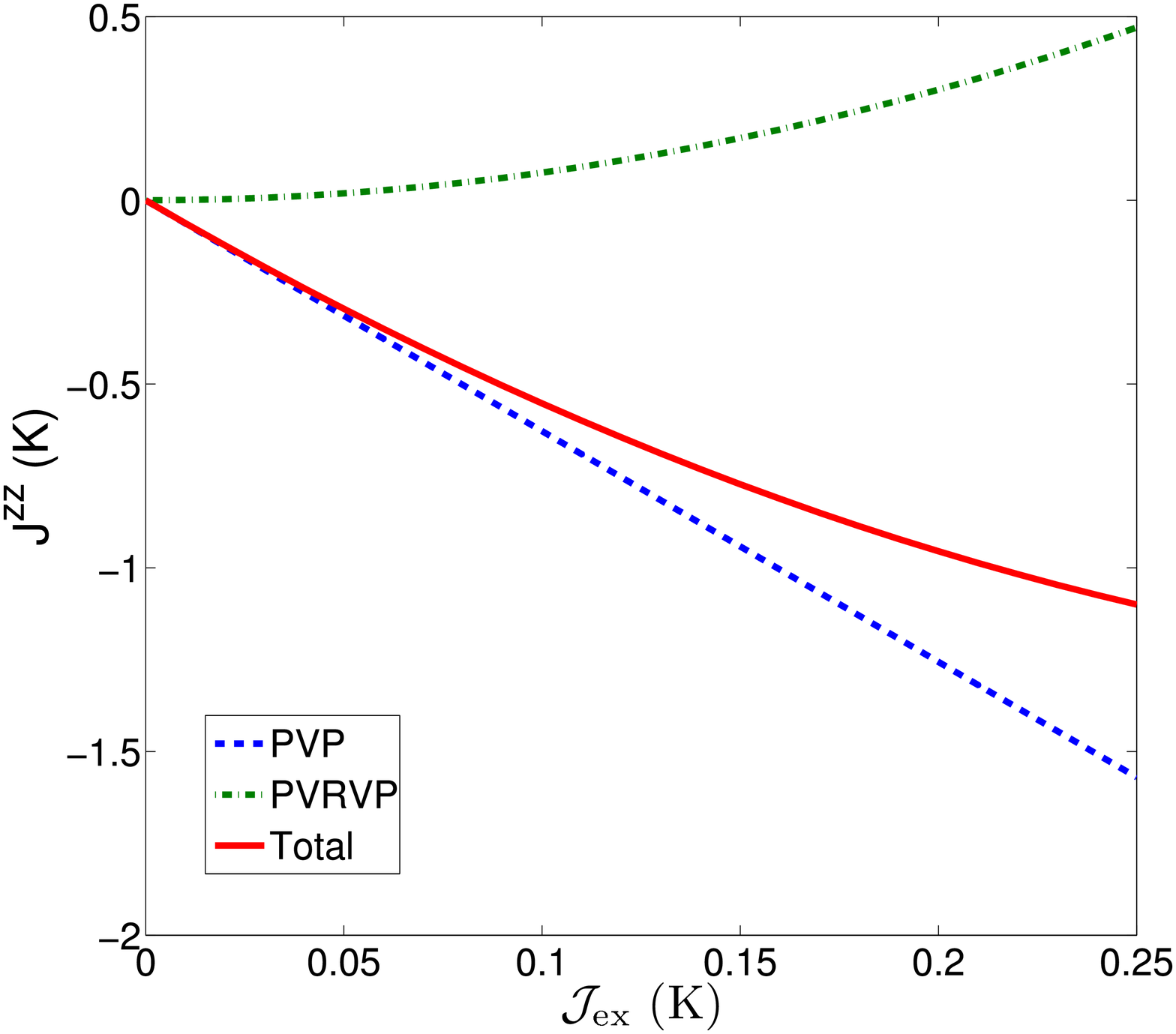}}
\subfigure[\hspace{1pt} Isotropic exchange and long-ranged dipole-dipole interaction.]{\includegraphics[width=0.5\textwidth]{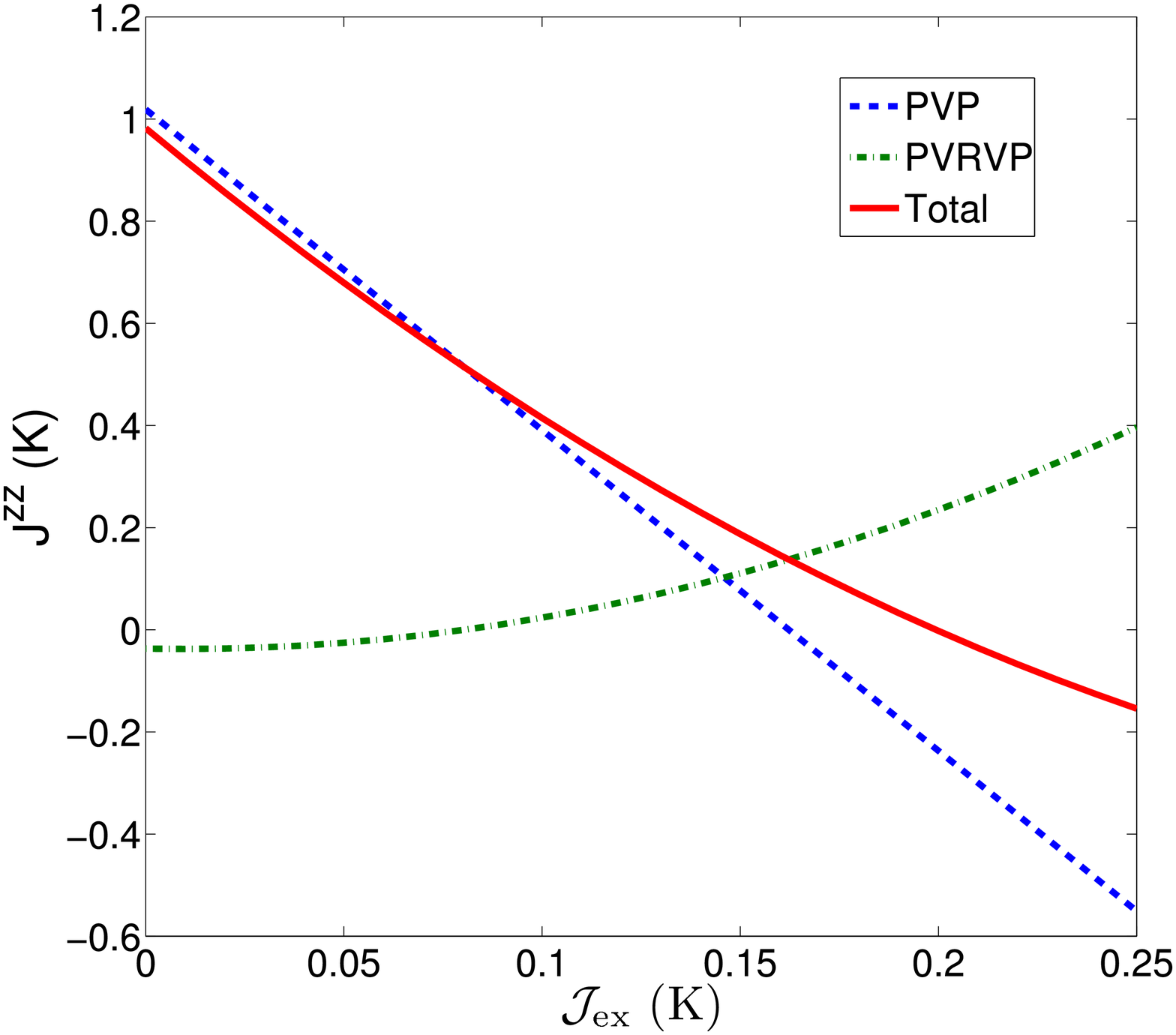}}
\caption{\label{fig:RNExchange} (color online). Plot showing how the nearest
  neighbor Ising exchange coupling $J^{ zz}$ on a lattice (Eq.~(\ref{eqn:couplings})) is renormalized by the
  quantum terms of the effective Hamiltonian. The horizontal axis is the bare exchange coupling $\mathcal{J}_{\rm ex}$. The top
  figure is for the Ising coupling when the dipole-dipole coupling is set equal to zero and the bottom figure includes the
  dipole-dipole coupling of magnetic ions in \tto: $\mathcal{D}=0.0315$ K. In both figures, the dashed line is the coupling that
  appears to lowest order $\mathcal{P}H\mathcal{P}$ in $H_{\rm eff}$ for a pair of neighboring sites. The dot-dash line is the
  correction obtained from the quantum term $H_{\rm eff}^{(2)}$ and the solid line is the sum of the two contributions to the
  Ising exchange. A positive $J^{ zz}$, in the absence of other interactions,
  favors spin ice configurations on each tetrahedron and a negative sign
  implies AIAO configurations.}
\end{figure}

That the singlet ground state of the microscopic model shows spin ice-like correlations can be understood from the effective
Hamiltonian. It is, at first sight, a peculiar result given the classical DSIM phase diagram described in Section
\ref{sec:classical} for which the \tto\ parameters lie in the AIAO phase. The explanation for this behavior lies in
the fact that, as the gap $\Delta$ is lowered, for fixed \tto\ bare parameters, the
nearest neighbor Ising exchange coupling $J^{ zz}$ (Eq.~(\ref{eqn:couplings})) is renormalized by Ising terms
coming from the quantum fluctuations $H_{\rm eff}^{(2)}\equiv\mathcal{P}V\mathcal{R}V\mathcal{P}$ (see Section \ref{sec:quantum}). The variation of Ising
exchange $J^{ zz}$ due to VCFEs as a function of the bare exchange is shown in
Fig. \ref{fig:RNExchange} when the dipole-dipole coupling $\mathcal{D}=0$. The straight line is the part from the classical term $\mathcal{P}V\mathcal{P}$ for which the
renormalized (Ising) exchange is $\mathcal{J}_{\rm classical}=\mathcal{J}_{{\rm ex}}\langle \widetilde{J}^{z}\rangle$ (where $\langle
\widetilde{J}^{z}\rangle$ is given in Eq. (\ref{eqn:ME}) and this formula in derived in Section~\ref{sec:classical}). The Ising exchange
$J^{ zz}$ from the quantum fluctuations is antiferromagnetic ($J^{ zz}>0$) so
the sum of the classical and quantum Ising couplings is less ferromagnetic
(negative) than the $J^{ zz}$ without VCFEs. Now consider the effect of
including nearest neighbor dipole-dipole coupling. With the dipole-dipole
coupling alone ($\mathcal{J}_{\rm ex}=0$), $J^{ zz}$, is antiferromagnetic ($J^{ zz}>0$), favoring spin ice correlations (see
Section~\ref{sec:convention}).  Indeed, in the \dto\ and \hto\ spin ice materials, the exchange contribution to the Ising $J^{
  zz}$ coupling is ferromagnetic but the
dipole-dipole coupling ensures that the net contribution of the bare microscopic couplings to $J^{ zz}$ is
antiferromagnetic hence frustrating a single tetrahedron and the pyrochlore lattice. \cite{denHertog} In contrast, we see in the lower part of Fig.~\ref{fig:RNExchange}, that the estimated
bare exchange coupling in \tto\ of $\mathcal{J}_{\rm ex}\sim 0.17$K, \cite{Gingras} places the
classical part of the Ising coupling (i.e. the contribution to $J^{ zz}$
from $\mathcal{P}V\mathcal{P}$) close to zero. But, VCFE
corrections lead to an effective antiferromagnetic coupling $J^{ zz}>0$ overall leading to spin ice-like correlations for \tto.

In another development, the truncated model with six crystal states per site on a tetrahedron was found to exhibit a magnetization
plateau in a $[111]$ field below $100$ mK $-$ further evidence that this model exhibits spin ice-like behavior for more
antiferromagnetic bare couplings than one would expect from the DSIM. \cite{Molavian2} 

In summary, this article is devoted to a derivation of a quantum spin-$1/2$
model with anisotropic interactions similar to the models used as starting
points in Refs.~\onlinecite{Gauge1} and \onlinecite{Gauge2}. From this
effective Hamiltonian, we establish two things. Firstly, we show that the
renormalization of the Ising exchange towards antiferromagnetic exchange ($J^{ zz}>0$) via VCFEs
occurs, not only on a single tetrahedron, but also on the full pyrochlore
lattice. That is, even if, at the classical level (which ignores exited
crystal field levels), \tto\ could be described by a non-frustrated $[111]$
pyrochlore Ising model, virtual crystal field excitations can render this
system a frustrated spin ice one, with additional transverse fluctuations. This
is the main result of this paper. Secondly, in what follows, we study the semiclassical spin
correlations that these effective interactions produce on a lattice.

\section{Semiclassical ground states}
\label{sec:ground}

\subsection{Convention and Procedure}

The effective Hamiltonian that was discussed in detail in the previous
 sections has a large number of different effective interactions arising from
virtual crystal field excitations (VCFEs) from the ground state crystal field doublet
 (see Section~\ref{sec:summary}). As we have seen, to lowest (first) order in the
perturbation expansion in $\langle V \rangle/\Delta$, the effective Hamiltonian is an Ising model. The lowest order quantum corrections to $H_{\rm eff}$ include
transverse terms between nearest neighbor spins, three-body interactions, and anisotropic interactions extending beyond nearest
neighbors. To gain some understanding of the effect of the extra terms
 generated by VCFEs on the physics of this system, it is useful to
consider a semiclassical spin model with the same interactions as in the effective quantum Hamiltonian. 

To obtain the required semiclassical model, we first observe that the effective quantum Hamiltonian should be written in terms of pseudospins one-half. That is, the
elementary quantum spins take the form $\hat{S}^{\alpha}=(1/2)\sigma^{\alpha}$ and the quantum Hamiltonian couplings derived in
the previous section are rescaled by a factor of four for the bilinear interactions and by a factor of eight for the three-body
interactions. This model is the most suitable model to consider from the point of view of a large spin $S$ expansion. Once the quantum Hamiltonian is written in terms of spins $\hat{S}^{\alpha}$, we take these
 quantum spins into classical spins which are vectors of fixed length $S=1/2$ with components parameterized by spherical
 polar angles.

To find the ground states of the semiclassical model, we start with a randomly
chosen initial spin configuration on a finite lattice and compute its energy. We then make a small random rotation of one of the pseudospins and accept this configuration only if the energy of
the new configuration is lower than that of the old configurations. This procedure is iterated until it converges, which happens within
$O(10^{4})$ steps. This zero temperature Monte Carlo may settle into a local rather than a global minimum. To alleviate this
problem, we repeat the process for a number of initial states depending on the
number of spins treated and look for the minimum energy configuration; this also allows us
to capture any ground state degeneracy.

Only a small number of independent spins are treated in this minimization procedure - we find the ground states on a single
tetrahedron (four spins) and on a cubic unit cell with periodic boundary conditions (sixteen spins). With this
number of spins, we find that only a small number of initial spin configurations
$O(10^{1})$ is necessary in the iteration scheme to find consistency between the final energies and to
capture discrete degeneracy when it arises. However, if there is a continuous degeneracy, (as
in the XY phase described below), $O(10^{2})$ initial spin configurations are necessary to confirm its existence. 
On a cubic unit cell, we capture the DSIM ground states with ordering vector $\mathbf{q}=001$  and $\mathbf{q}=0$ in the limit of
$1/\Delta=0$. When $\Delta$ is finite, VCFEs generate interactions beyond the DSIM interactions which may lead to
more complicated (modulated magnetic moment with incommensurate wavevector $\mathbf{q}$) ground states might be eliminated by (an inappropriate choice of) periodic boundary conditions. Whether
this is indeed the case for the $H_{\rm eff}$ considered below is an open question for future work. The key problem we address in computing the ground states is to establish
whether interactions generated by VCFEs beyond nearest neighbor do favor spin ice correlations over a wider range of
$\mathcal{J}_{\rm ex}$ than is observed in the absence of such terms.

\subsection{Ground states of $H_{\rm eff}$ on a single tetrahedron - 4 CF states}
\label{sec:4sbl}
 
The first results that we present are those obtained by minimizing the energy on a single tetrahedron to make a comparison with
the results of exact diagonalization of the four crystal field state
microscopic quantum model on a single tetrahedron described in
Section~\ref{sec:results}. The effective Hamiltonian required to make the comparison
includes the nearest neighbor interactions and three-body interactions on a single tetrahedron obtained by including only the
first excited crystal field doublet in the resolvent (Eq.~(\ref{eqn:resolvent}))  when computing the quantum terms, $H_{{\rm
    eff}}^{(2)}$. By truncating all the bare interactions to a single
tetrahedron, the effective Hamiltonian is derived following Section~\ref{sec:quantum} including all possible ways that the
mediating ion of Case A in Section~\ref{sec:caseone} can lie on a single tetrahedron. Cases B and C (Sections \ref{sec:casetwo}
and \ref{sec:casethree}) are treated entirely on the single
tetrahedron. At first, we present the results when the three-body terms are omitted; we include them later. When three-body
terms are omitted and the derived effective interactions are truncated beyond nearest neighbors, the ground states on a single tetrahedron
must coincide with those on the full lattice under the assumption that the lattice ground states have $\mathbf{q}=0$ ordering
wavevector. This is because the interactions for a $\mathbf{q}=0$ lattice configuration are the same as those on a single tetrahedron except for an
extra factor of two in the effective couplings on the lattice. This factor of two comes
from the fact that pairwise effective interactions couple a
spin on one sublattice, $a$, to two $b$ sublattices ($a\neq b$) whereas, on a tetrahedron, a spin with sublattice label $a$
couples to only one $b$ sublattice spin.

\begin{figure}
\begin{center}
\includegraphics[width=0.5\textwidth]{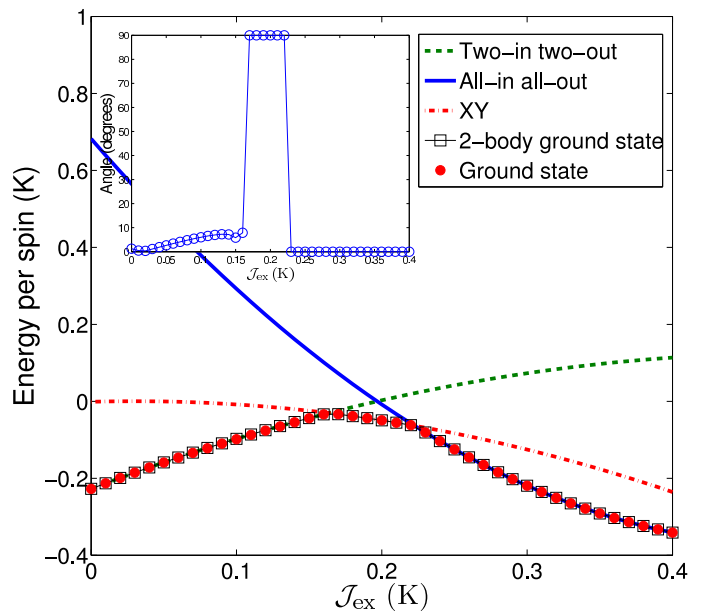}
\caption{\label{fig:Ground_Tetra_4ST} (color online). Semiclassical ground state energy of the effective Hamiltonian on a single
  tetrahedron when quantum terms are computed including VCFEs only to the first excited crystal field doublet (i.e. omitting virtual
  excitations to higher energy crystal field states). In obtaining these results, the bare microscopic exchange, $H_{\rm ex}$, and
  dipolar interaction, $H_{\rm dd}$, were truncated at the nearest neighbor distance. The resulting effective couplings generated
  in $H_{\rm eff}^{(2)}$ were also truncated beyond nearest neighbor. The parameters used for this calculation are $\Delta^{-1}=0.055$ K$^{-1}$,
  $\mathcal{D}=0.0315$ K and $\langle \widetilde{J}^{z}\rangle=3.0$. Also plotted, are energies of three different imposed spin
  configurations. The omission of three-body interactions changes the ground state energy by a few
  percent depending on the canting angle produced by these interactions. The inset shows the angle from the local $\mathbf{z}$ axes of
  Table~\ref{tab:lattice} through which the spins are canted in the ground
  states of the model - for $\mathcal{J}_{\rm ex}\lesssim 0.17$ K, the canting
  is due to the three-body
  interactions and the $90$ degree canting angles signal the onset of the
  local XY ground states which are ground states even without the three-body interactions.}
\end{center}
\end{figure}

The ground states are shown in Fig.~\ref{fig:Ground_Tetra_4ST}. This figure shows the
energies computed from the effective Hamiltonian with classical spins on a
single tetrahedron for different (imposed) specific spin configurations
and for different values of the bare exchange, $\mathcal{J}_{\rm ex}$. The energies of the ground states determined by the
minimization procedure outlined above are shown as well. 

\begin{figure}
\begin{center}
\subfigure[\label{fig:AIAO} All-in/all-out state.]{\includegraphics[width=4cm,clip]{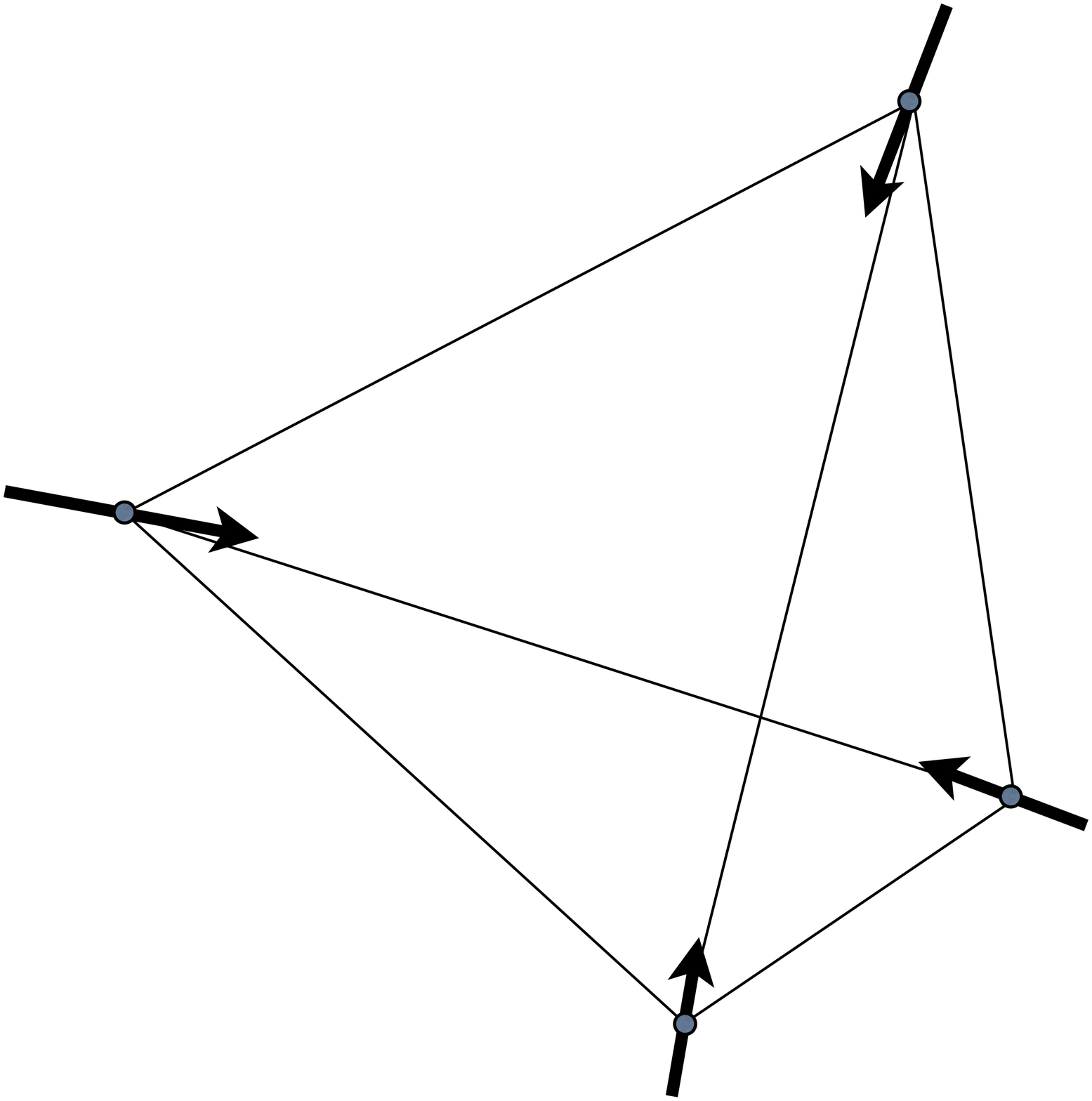}}
\subfigure[\label{fig:XY} XY configuration.]{\includegraphics[width=4cm,clip]{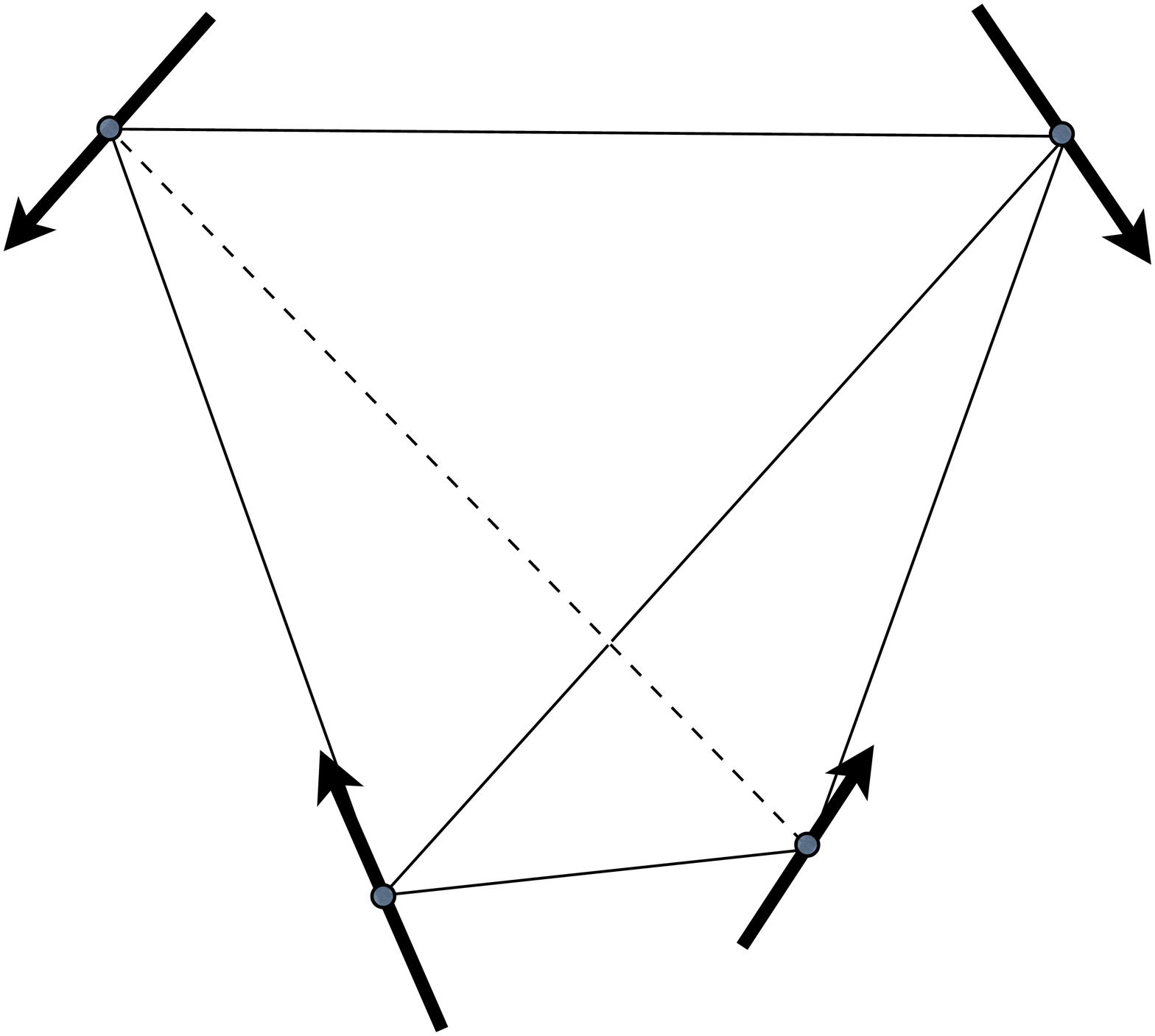}}
\end{center}
\caption{\label{fig:configs} A pair of configurations that are ground states of the four sublattice classical analog of the
  quantum effective Hamiltonian without three-body interactions. The all-in/all-out state (a) occurs for $\mathcal{J}_{\rm
  ex}>0.22$ K and the XY configuration (b) for $0.17<\mathcal{J}_{\rm ex}<0.22$ K.}
\end{figure}

One finds that as the bare exchange coupling $\mathcal{J}_{\rm ex}$  (for fixed dipole-dipole coupling
$\mathcal{D}=0.0315$ K) becomes more antiferromagnetic (i.e. positive and larger), the two-in/two-out ground state gives way to an
all-in/all-out state (Fig. \ref{fig:configs}). However, unlike the classical
model, $\Delta=\infty$, for which only the two-in/two-out and all-in/all-out states
  appear, there is a $\mathbf{q}=0$ configuration
separating the two-in/two-out state from the all-in/all-out state in which the
classical spins lie fully in their local XY planes perpendicular to the local
$[ 111 ]$ directions. 

There is a continuous degeneracy of XY configurations such that the
vector sum of the spins is zero as one should expect for sufficiently
strong antiferromagnetic exchange. A single spin configuration among the continuous
set of XY ground states is illustrated in Fig.~\ref{fig:XY}. These ground states belong to the two dimensional irreducible representation of the point group
of the tetrahedron $O_{h}$ (see, Refs.~\onlinecite{ETOPaper} and \onlinecite{Poole}) which includes the discrete ground states of the material \eto\ (see,
for example, Ref.~\onlinecite{ChampionHoldsworth}).
The onset of the XY phase
corresponds to a range of values of the bare exchange $\mathcal{J}_{\rm ex}$
where the effective Ising and transverse couplings (shown in Fig.~\ref{fig:Couplings}) are of similar magnitude
such that it is energetically favorable for the spins to lie in the local XY planes.

For the classical DSIM with dipole-dipole interactions
truncated beyond nearest neighbor, there is a phase boundary \cite{denHertog} between a spin ice state and an all-in/all-out
phase at $\mathcal{J}_{\rm ex}=5\mathcal{D}$ which is roughly $0.158$ K for
the \tto\ dipolar coupling $\mathcal{D}=0.0315$ K. Reading from the phase
diagram in Fig.~\ref{fig:Ground_Tetra_4ST}, the
spin ice to XY boundary of the nearest neighbor effective Hamiltonian is at
about $0.17$ K and the onset of the all-in/all-out (AIAO) phase is at about $\mathcal{J}_{\rm ex}=0.22$ K for the effective Hamiltonian on a single tetrahedron. A direct comparison of
these numbers with the classical model (DSIM) is possible because (i) the phase boundary of the DSIM 
depends on the ratio of the effective Ising exchange $\mathcal{J}_{\rm classical}$ (Eq.~(\ref{eqn:Jclassical})) to the effective 
dipole couplings $\mathcal{D}_{\rm classical}$ (Eq.~(\ref{eqn:Dclassical})) and (ii) the ratio of effective couplings in the classical term $\mathcal{P}V\mathcal{P}$ is simply the ratio of bare
couplings. That the phase boundary out of the two-in/two-out Ising
configuration in the semiclassical effective model appears for more positive
$\mathcal{J}_{\rm ex}$ than with the classical term
($\mathcal{P}V\mathcal{P}$) alone is because the effective Ising
exchange, $J^{zz}$, in the quantum model receives a contribution from $H_{\rm eff}^{(2)}$ that makes it more antiferromagnetic
($J^{zz}$ becomes more negative) hence making spin ice Ising configuration energetically favorable.
 We mention also that the effective Ising coupling, $J^{zz}$, between
$\mathcal{J}_{\rm ex}=0.17$ K and $\mathcal{J}_{\rm ex}=0.20$ K is
antiferromagnetic (positive) whereas $H_{\rm eff}$ calculated solely to
$\mathcal{P}V\mathcal{P}$ order has ferromagnetic effective Ising exchange (favoring all-in/all-out order) over this
range. 

\begin{figure}
\begin{center}
\includegraphics[width=0.4\textwidth]{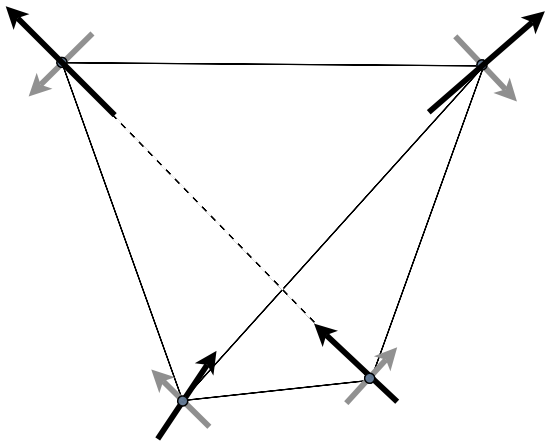}
\end{center}
\caption{\label{fig:ModelState} The semiclassical LRSI$_{000}$ ground state of the effective
Hamiltonian represented on a single tetrahedron. The black
arrows show the local Ising components of the spins and the grey arrows, the
(smaller) local XY components. The canting does not alter the moment on each
tetrahedron relative to the moment with uncanted spins.}
\end{figure}

When we include the three-body terms on a single tetrahedron, we again
find three ground state phases with the same phase boundaries
that we found in the two-body case. Whereas the AIAO phase is the same with
and without three-body terms, the other two two-body phases are modified by
the introduction of three-body couplings. Instead of ground states with spins aligned along the local Ising directions, one finds that the spins are canted away from the $[111]$ directions while retaining the spin
ice ordering in the Ising components. The canting (which is shown in
Fig.~\ref{fig:ModelState} for one of the observed ground states) is such as to preserve the
moment of the perfectly Ising spin configuration. The variation in the canting angle is shown in the inset to
Fig.~\ref{fig:Ground_Tetra_4ST}. Only in the spin ice regime, $\mathcal{J}_{\rm ex}<0.17$ K, are the ground states affected by a canting away from the Ising directions when
three-body interactions are included. The three-body terms do not produce a canting away from the two-body XY and AIAO ground states. Also, with three-body interactions included, the degenerate XY configurations cease to be the lowest
energy states in the intermediate region $0.17$ K $\lesssim \mathcal{J}_{\rm ex} \lesssim 0.22$ K -
the continuous degeneracy present without three-body terms is broken.  The transverse
(XY) components of the spins in the spin canted state are ordered into six discrete configurations. The transverse component
configurations, considered on their own, are a discrete set of configurations belonging to the aforementioned class of XY ground
states $-$ they are referred to as $\psi_{2}$ states in the literature. \cite{Poole,ChampionHoldsworth,ETOPaper} The canting angle
is zero in the AIAO phase.

Turning to the ground state energies themselves in the presence of three-body
interactions, we first note that the three-body terms make no contribution to the XY and perfectly Ising spin configurations because the three-body terms are of the form
$\tilde{\sigma}_{I_{1}}^{z}\tilde{\sigma}_{I_{2}}^{\alpha}
\tilde{\sigma}_{I_{3}}^{z}$ with $\alpha=x,y$ which vanishes in these cases. However, the perfectly Ising-like spin ice
configurations are not true ground states when three-body terms are present for $\mathcal{J}_{\rm ex}\lesssim 0.27$ K - there is a
small canting away from the Ising directions. There is a difference between the two
and three-body ground state energies that is small (about $4 \%$ at most)  arising from the relative sizes of the two and three
body couplings illustrated in Fig.~\ref{fig:Couplings}. 

Fig.~\ref{fig:Couplings} is useful in interpreting the single tetrahedron ground
state diagrams (with or without three-body interactions). In both phase
diagrams, the sign of the Ising coupling determines the correlations of the
Ising components of the classical spins (which, as we report in Section~\ref{sec:16sbl}
ceases to be true beyond a single tetrahedron), and the relative magnitude of the Ising coupling and other
couplings is correlated to the canting of the spins away from the Ising directions $-$ the spins being furthest from the Ising
directions when the transverse couplings become comparable to or greater than the Ising coupling.
  
 Now we are in a position to compare the single tetrahedron results for the quantum four crystal field state (ground and first
 excited doublet) effective Hamiltonian with the classical results. The quantum phase diagram showing the ground state
 degeneracies is plotted in Fig.~\ref{fig:GSdeg}. Focusing on the \tto\ crystal field gap of $1/\Delta=0.055$ K$^{-1}$, the boundary between the singlet and
doublet states is at about $\mathcal{J}_{\rm ex}=0.21$ K (with $\mathcal{D}=0.0315$ K). The phase boundary for the semiclassical ground
state derived from the effective Hamiltonian (Fig.~\ref{fig:Ground_Tetra_4ST}) is close to
this value, at about $\mathcal{J}_{\rm ex}=0.22$ K. The degeneracies of the semiclassical ground states and the quantum ground states
on a single tetrahedron agree for $\mathcal{J}_{\rm ex}>0.22$ K whereas for $\mathcal{J}_{\rm ex}<0.22$ K, the singlet quantum
ground state appears for the same range of couplings as both the classical six-fold degenerate canted spin ice ground state and the XY
 phase. 

\subsection{Ground states of $H_{\rm eff}$ on a single tetrahedron - 13 CF states}
\label{sec:4sbl13st}

\begin{figure}
\begin{center}
\includegraphics[width=0.5\textwidth]{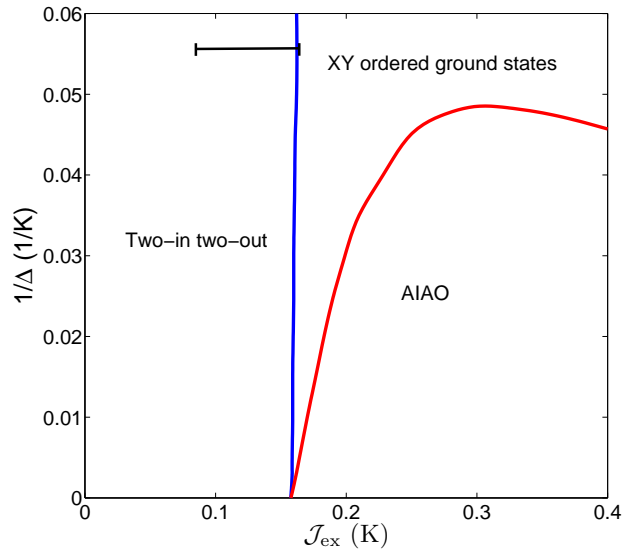}
\caption{\label{fig:NNPhaseDiagram_JDelta} (color online). Semiclassical ground
  states of $H_{\rm eff}$ on a periodic cubic unit cell with both bare and effective interactions
  truncated beyond nearest neighbor. The phase diagram shows the ground states
  that are obtained over a range of $\Delta$ and $\mathcal{J}_{\rm ex}$. Three
  body interactions are also neglected. The horizontal bar in the top left hand corner represents the uncertainty in the bare
  exchange coupling $\mathcal{J}_{\rm ex}$ in \tto\ \cite{Gingras,Mirebeau} at the value of the crystal field gap $\Delta$ appropriate to this material. }
\end{center}
\end{figure}

Before leaving the subject of the ground states on a tetrahedron, we compute the ground states on a single tetrahedron with the
full crystal field spectrum included in the resolvent operator (rather than considering a truncation of the spectrum to the ground
and first excited doublets as we have done in Section~\ref{sec:4sbl}). In this subsection, the ground states of $H_{\rm eff}$ we
present were computed as a function of both $\mathcal{J}_{\rm ex}$ and $\Delta$. Here $\Delta$ is an adjustable gap that
shifts all the excited crystal field states rigidly with respect to the ground
state doublet leaving the wavefunctions identical to those that one would
obtain by diagonalizing the \tto\ crystal field Hamiltonian. By artificially
varying $\Delta$ in this way, we can tune the system from a classical spin ice with nearest neighbor dipolar interactions to a
model in which VCFEs are significant. The results are shown in Fig.~\ref{fig:NNPhaseDiagram_JDelta}. As
one would expect based on the limiting case $1/\Delta=0$ of spin ice and the
results discussed in Section~\ref{sec:4sbl} for $1/\Delta=0.055$ K$^{-1}$, the all-in/all-out
ground states and the two-in/two-out ground states are separated by a wedge of
ordered local XY ground states (with continuous degeneracy when three-body
interactions are omitted). The range of $\mathcal{J}_{\rm ex}$ over which the
wedge extends increases as $\Delta$ decreases. For $\Delta\lesssim 20$ K, the AIAO phase is suppressed
entirely because VCFEs increase transverse couplings relative to the Ising couplings (see Fig.~\ref{fig:Couplings}). We return to
this phase diagram in Section~\ref{sec:16sbl} where we make a comparison of
Fig.~\ref{fig:NNPhaseDiagram_JDelta} with the ground states on a cubic unit
cell with periodic boundary conditions.

\subsection{Ground states with long-range interactions included - 13 CF states}
\label{sec:16sbl}

In the foregoing, we have presented the ground states for $H_{\rm eff}$ derived on a
single tetrahedron. For the case of effective nearest neighbor bilinear spin-spin
interactions, the single tetrahedron ground states are the same as the four sublattice ($\mathbf{q}=0$) ground states on the pyrochlore lattice. However,
we know that, in the DSIM, obtained from the $H_{\rm eff}$ on a lattice when the crystal field gap $\Delta$ is taken to infinity,
one of the ground states is a sixteen sublattice configuration (with ordering wavevector $(0,0,2\pi/a)$) on a conventional cubic
unit cell $-$ the LRSI$_{001}$ state shown in Fig.~\ref{fig:UnitCell}. The long-ranged nature of the dipole-dipole interaction is
responsible for the lower energy of the LRSI$_{001}$ state compared to other
ordered states that satisfy the local spin ice rules. \cite{Melko} This
observation for the DSIM tells us that we should truncate neither the bare dipole
interaction to nearest neighbor nor the effective interactions and that we should not assume
$\mathbf{q}=0$ ordering as was done implicitly in Section~\ref{sec:4sbl}. Inspired by the case of the DSIM, we investigate the
ground states on a cubic unit cell with periodic boundary conditions. The
ground states that we find in this section are for the
pyrochlore lattice, assuming that the magnetic unit cell is no bigger than
the conventional pyrochlore cubic unit cell (with $16$ sites).

For the problem of
finding ground states, the effective Hamiltonian is derived in the following
way, which differs from the approach presented above in Sections~\ref{sec:4sbl} and \ref{sec:4sbl13st}
in having to treat the long-ranged dipole-dipole interaction. The bare Hamiltonian, which
has nearest neighbor isotropic exchange and long-ranged dipole-dipole
interactions, is computed on a sixteen site cubic unit cell by summing
the dipole-dipole interaction over all periodic images by an Ewald
summation. \cite{Ewald, Enjalran} The effective Hamiltonian is then
computed numerically for this periodic model on a cubic unit cell (using identities at the end of Appendix~\ref{sec:calc} and
summing over the full $13$ state crystal field spectrum in the perturbation theory). This
procedure preserves the periodicity of the Hamiltonian. One could have instead derived the effective Hamiltonian on a lattice
and then sum the interactions over a large but finite lattice assuming periodicity in the classical spin configurations on a cubic
unit cell. We present results, in this section, for the former case, but the latter approach gives results that are quantitatively
very similar. \cite{MolavianThesis}

\begin{figure}
\begin{center}
\includegraphics[width=0.5\textwidth]{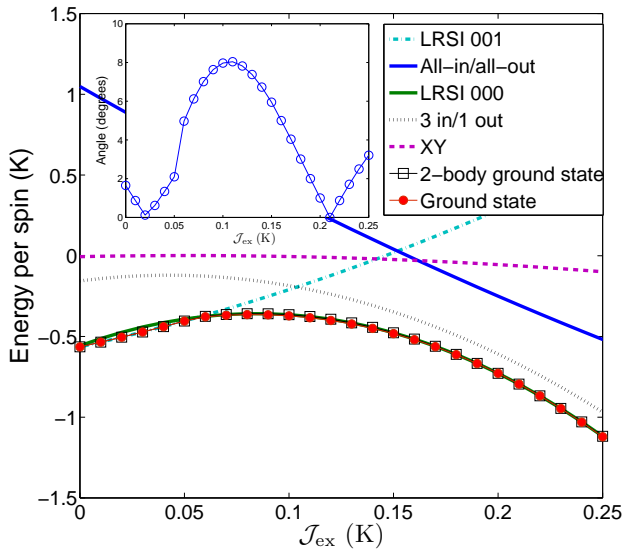}
\caption{\label{fig:Cubic_Dipole} (color online). Ground state energy for an effective Hamiltonian derived
  from a model with isotropic exchange and long ranged dipole-dipole
  interactions on a cubic unit cell with periodic boundary conditions treated by an Ewald summation with
  $\Delta^{-1}=0.055$ K$^{-1}$ and $\mathcal{D}=0.0315$ K. The ground states
  with (circles) and without (squares) including three-body interactions are
  shown as well as the energies of different (imposed) spin configurations. The two-body ground states are the LRSI$_{001}$ state for
  $\mathcal{J}_{\rm ex}\lesssim 0.06$ K which is a two-in/two-out state with
  ordering wavevector  $(0,0,2\pi/a)$ and an ordered two-in/two-out
  state with ordering wavevector $(0,0,0)$ for $\mathcal{J}_{\rm ex}\gtrsim 0.06$ K. When three-body terms are included the spins cant out of the Ising directions as
  indicated in the inset.}
\end{center}
\end{figure}

The semiclassical ground states energies of the resulting model are computed by replacing the pseudospin operators
$(1/2)\sigma^{\alpha}$ with classical spin components $S^{\alpha}$. They are
given in Fig.~\ref{fig:Cubic_Dipole} as the bare exchange coupling is varied
with $1/\Delta=0.055$ K$^{-1}$ and $\mathcal{D}=0.0315$ K; the values
appropriate to \tto. In
the same figure, we also plot, for comparison, the energies of various imposed (fixed) spin configurations. As in the single tetrahedron case, it is useful to
distinguish the ground states obtained when three-body spin interactions are
removed and the ground states for the model with all interactions included.

Without three-body spin interactions (open squares in the main panel of Fig. \ref{fig:Cubic_Dipole}) we find
that for weakly antiferromagnetic $\mathcal{J}_{{\rm ex}}\lesssim 0.06$ K, the ground state is the LRSI$_{001}$ state. But, for more
antiferromagnetic $\mathcal{J}_{{\rm ex}}$, instead of the all-in/all-out
state found for the DSIM (see inset of Fig.~\ref{fig:PhaseDiagram}), the ground
state, at least up to $\mathcal{J}_{{\rm ex}}=0.4$ K, is a state with identically ordered tetrahedra (ordering wavevector
$\mathbf{q}=0$) each obeying the ice rules with spins in the local Ising
directions $-$ we refer to this as the LRSI$_{000}$ phase.

If the dipole-dipole interaction is cut off at nearest neighbor in the microscopic bare model of Section~\ref{sec:model} before
computing the effective Hamiltonian and if effective couplings beyond nearest neighbors are removed, the LRSI$_{001}$ Ising state
has the same energy as the LRSI$_{000}$ Ising
phase for $\mathcal{J}_{{\rm ex}}\lesssim 0.06$ K. The XY phase that we found on a
single tetrahedron does not appear in the conventional cubic unit cell model
unless all effective interactions are cut off beyond nearest neighbors.

When three-body
interactions are restored (filled circles in Fig.~\ref{fig:Cubic_Dipole}), the spins
cant away from the Ising directions and the energies are lowered relative to the ground states with only two-body
interactions considered, similarly to what was found in Section~\ref{sec:4sbl} on a tetrahedron. The ordering of the Ising components of the spins is
the same regardless of whether three-body interactions are present or not. 
The canting angle of the spins away from the Ising directions is shown as
the inset in Fig.~\ref{fig:Cubic_Dipole}. There is a maximum in the angle at about $\mathcal{J}_{\rm ex} =0.11$ K and two minima
 at about $0.02$ K and $0.21$ K over the explored range of $\mathcal{J}_{\rm ex}$. The greatest angle is about $8$ degrees, compared to $90$ degrees in the
case of a single tetrahedron. Broadly, the variation in the canting angle follows the magnitude of the three-body coupling (shown
for a choice of three sublattices in Fig.~\ref{fig:Couplings}). The minimum in
the canting angle at about $\mathcal{J}_{\rm ex}\sim 0.02$ K coincides
with a minimum in the mean squared three-body coupling over all such
couplings on the cubic unit cell at this value of the
bare exchange. The minimum implies that there is little energy gain to a canting of the spins. The non-monotonic change in the
three-body couplings is allowed because the coupling has contributions both quadratic and linear in $\mathcal{J}_{\rm ex}$
coinciding with the exchange-exchange and exchange-dipole contributions to $H_{\rm eff}$ of Eq.~(\ref{eqn:decomp}). The second
minimum in the canting angle, $\mathcal{J}_{\rm ex} \approx 0.21$ K, coincides roughly with a vanishing in the three-body coupling
(shown in the inset to Fig.~\ref{fig:Couplings}) and with a change in the sign of the nearest neighbor Ising coupling. We expect
therefore, two effects to be at work - a weakening of the three-body coupling and the same effect that suppressed the canting angle in
the transition from the XY phase to the AIAO phase on a single tetrahedron (Section~\ref{sec:4sbl}). The
difference in this case is that it is a weak effect compared to that of the effective further neighbor interactions which control the Ising ordering in this range
of couplings. The ordering of the local XY components of the spins is
identical to that described in Section~\ref{sec:4sbl} - one such type of XY
ordering is shown in Fig.~\ref{fig:ModelState} in which smaller arrows
indicate the canting direction away from the Ising directions. \cite{XYorder} 

We note in passing that the material \tso\ which is, microscopically, very
similar to \tto \cite{Mirebeau}  undergoes a phase transition at about
$0.87$ K to a magnetically long-range ordered phase with ordering wavevector $\mathbf{q}=0$ and two-in/two-out spin ice
configurations on each tetrahedron. \cite{Mirebeau3} The spins in this ordered phase are canted away from the local Ising directions. In these
respects, the spin canted LRSI$_{000}$ ground state of the effective
Hamiltonian discussed above is similar to the magnetic order in \tso. But the
nature of the spin canting differs between the model and the material. The spin canting in \tso\ is such as to reduce the moment, on each
tetrahedron, compared to the moment if the spins were not canted,
\cite{Mirebeau3} whereas the canting of the effective spin $1/2$ in the LRSI$_{000}$ state (indicated in
Fig.~\ref{fig:ModelState}) produced by the three-body terms gives a moment, on each tetrahedron, that is the
same as the moment of the LRSI$_{000}$ configuration without spin canting.

The classical DSIM (which is recovered for an
infinite ground to first excited crystal field gap, $\Delta$) with long range dipole-dipole interactions has a phase boundary
between the LRSI$_{001}$ spin ice configurations and
all-in/all-out states at about $\mathcal{J}_{\rm ex}=0.14$ K (see inset of Fig.~\ref{fig:PhaseDiagram}). We have seen that,
in the effective model of \tto, the semiclassical ground states, at least for $\mathcal{J}_{{\rm  ex}}<0.25$ K, are spin
ice configurations although the bare exchange coupling $\mathcal{J}_{\rm ex}$ is antiferromagnetic so we see that
spin ice correlations are favored by VCFEs. However, Fig.~\ref{fig:Couplings} shows that the average nearest neighbor Ising
exchange $J^{zz}$ swaps sign at about $0.2$ K so the persistence of ice-like correlations in \tto, in the form of the
LRSI$_{000}$ state, up to, at least, $\mathcal{J}_{\rm ex}=0.4$ K (see
Fig.~\ref{fig:PhaseDiagram}) is not due to the renormalization of the Ising exchange described in Section~\ref{sec:results} but is
induced by further neighbor
couplings. The effective Hamiltonian to order $\Delta (\mathcal{J}_{\rm ex}/\Delta)^{2}$  is therefore a novel two-in/two-out model that does not rely on nearest neighbor
interactions to produce ice-like correlations. 

To shed some light on the fact that the all-in/all-out state,
(Fig.~\ref{fig:AIAO}), observed on a single tetrahedron and in the DSIM, \cite{denHertog, Melko, Melko2} is not seen in the
sixteen sublattice case, (for the value $1/\Delta\approx 0.055$ K$^{-1}$ as shown in Fig.~\ref{fig:PhaseDiagram} and Fig.~\ref{fig:Cubic_Dipole}), we have computed the semiclassical ground states for a range of $\Delta$ and
bare exchange couplings for the sixteen sublattice effective Hamiltonian on a cubic unit cell with periodic boundary conditions. We have omitted the three spin interactions which are not
responsible for the presence of the LRSI$_{000}$ state. The result is shown in Fig.~\ref{fig:PhaseDiagram}. This is the main
result of our paper.

 For $\Delta\gtrsim 340$ K ($1/\Delta \lesssim 0.003$), the phases are those of the DSIM with a phase boundary at about $\mathcal{J}_{\rm ex}=0.14$ K when $1/\Delta =0$. For comparison, we include an inset
showing the classical DSIM phase diagram for $\mathcal{D}=0.0315$
K. \cite{Melko, Melko2} As the gap
$\Delta$ is lowered from infinity, a $\mathbf{q}=0$ LRSI$_{000}$ phase $-$ appears at about $\Delta\sim 340$
K. \cite{LRSInote} As $\Delta$ is lowered further, the range of $\mathcal{J}_{{\rm ex}}$ over which this LRSI$_{000}$
phase is observed increases $-$ the spin-spin interactions arising from VCFEs stabilizing the LRSI$_{000}$ state. Indeed, for
$\Delta\lesssim 29$ K ($1/\Delta \sim 0.035$ K$^{-1}$ ), there is no all-in/all-out phase at least for any $\mathcal{J}_{{\rm
    ex}}<0.4$ K.  

Over the range of
$\mathcal{J}_{{\rm ex}}$ explored here, the LRSI$_{000}$ phase boundary has a dip with a minimum at about $\mathcal{J}_{{\rm ex}}=0.14$ K. The LRSI$_{000}$ is not observed in the DSIM so
the quantum terms of $H_{\rm eff}^{(2)}$ is responsible for its existence. Therefore $\mathcal{J}_{{\rm ex}}=0.14$ K is the exchange coupling at which the effect of the classical term is
minimized because the isotropic exchange and the dipole-dipole contributions
to the Ising exchange almost cancel each other. 
This accounts for the ``tail'' in Fig~\ref{fig:PhaseDiagram} where the
LRSI$_{001}$ to LRSI$_{000}$ phase boundary extends to $1/\Delta\sim 0.005$ K$^{-1}$ $-$ the
quantum terms are dominant at about $\mathcal{J}_{{\rm ex}}=0.14$ K.
 The phase diagram in
Fig.~\ref{fig:PhaseDiagram} shows that for  $\mathcal{J}_{{\rm ex}}\gtrsim 0.25$ K, the LRSI$_{000}$ spin ice appears over a larger
range of $1/\Delta$. This is because for larger values of the exchange, the quantum terms are larger for a given $\Delta$
and also because the quantum terms vary as $\mathcal{J}_{\rm ex}^{2}$ they eventually dominate over the classical terms. 

These observations lead us to two comments. Firstly, the shape of the LRSI$_{000}$ phase boundary in Fig.~\ref{fig:PhaseDiagram} is similar to the shape of the phase boundary for
nearest neighbor bare and effective interactions shown in
Fig.~\ref{fig:NNPhaseDiagram_JDelta}. The explanation we have given earlier in
this Section~\ref{sec:16sbl} (for the case
with long-range dipoles on a cubic unit cell) for the
shape of this boundary is equally applicable to the case with nearest neighbor bare and effective interactions discussed in
Section~\ref{sec:4sbl13st}. A comparison of these two figures reveals that, whereas quantum terms strongly influence the nearest
neighbor phase diagram enough to produce an XY phase, the replacement of this phase by LRSI$_{000}$ requires interactions beyond
nearest neighbor which, therefore, should not be neglected.

Secondly, as $\mathcal{J}_{{\rm ex}}/ \Delta$ increases, eventually higher order terms in powers of $\mathcal{J}_{{\rm ex}}/ \Delta$ must
become important and our effective model will break down. It is possible that the inclusion of higher order terms would lead to the
all-in/all-out phase persisting to larger values of $1/\Delta$ than we find considering only the lowest order quantum corrections
 $H_{\rm eff}^{(2)}$ to the DSIM. Fig.~\ref{fig:GSdeg} is a comparison of the exact four state model of Eq.~(\ref{eqn:truncate}) with the effective
Hamiltonian on a single tetrahedron. The singlet-doublet phase boundary indicates that the AIAO phase region should occupy
a larger range of $1/\Delta$ as $\mathcal{J}_{\rm ex}$ increases than is borne out by the semiclassical ground state calculation on
a single tetrahedron (Fig.~\ref{fig:Ground_Tetra_4ST}). On the basis of this comparison alone, however, one cannot draw any conclusions
about the effect of higher order corrections on the phase diagram of the effective Hamiltonian on a lattice.

\section{Summary and discussion}
\label{sec:discussion}

In this paper, we have introduced a low energy effective Hamiltonian for \tto\ formally derived from a minimal microscopic bare
Hamiltonian. The bare model consists of a crystal field
Hamiltonian for each magnetic ion and isotropic exchange and dipole-dipole interactions between the ${\rm J}=6$ angular momenta
(Eq. (\ref{eqn:bare}) and discussion in Section \ref{sec:model}). The low energy model is expressed in terms of effective spin
one-half operators which operate on states in the two dimensional Hilbert space spanned by the single ion ground state crystal
field doublet on each magnetic site. The effective theory is obtained as a perturbation expansion in $\langle V\rangle/\Delta$
where $\langle V\rangle$ is the
characteristic energy scale of the spin-spin interactions which incorporate exchange and dipole-dipole coupling, \cite{Vbracket} and
$\Delta$ is the energy gap between the ground and first excited levels in the crystal field spectrum. In Section \ref{sec:Heff},
we gave a detailed discussion of the terms that arise in the effective model to lowest order in the quantum corrections. To first
order in the effective Hamiltonian in powers of $\langle V\rangle/\Delta$ is the DSIM \cite{denHertog,
  Melko,Fennell} which has only interactions between the Ising components of the pseudospins. This model on its own has an antiferromagnetic
all-in/all-out (AIAO) ground state for the estimated bare exchange, dipolar and crystal field parameters for \tto\ (see vertical
dashed line in the inset to Fig.~\ref{fig:PhaseDiagram}). The next (second)
order in $\langle V\rangle/\Delta$ includes the lowest order quantum fluctuations involving virtual transitions into excited crystal field
levels. We found that the introduction of these virtual fluctuations leads to
a renormalization of the effective Ising exchange coupling in $H_{\rm eff}$ of
the lowest order (spin ice) model in such a way that two-in/two-out Ising configurations are favored on the single tetrahedron
over a wider range of $\mathcal{J}_{\rm ex}$ than one would find from the lowest order (DSIM) model. Also, to second order in
$\langle V\rangle/\Delta$, various
anisotropic transverse effective exchange couplings appear (in addition to corrections to the effective Ising exchange)
as well as some three-body interactions. Broadly speaking, the interactions between the effective spins become less Ising-like in
the presence of virtual crystal field excitations (VCFEs). This behavior is also borne out by comparisons of the diffuse neutron
scattering pattern for \tto \cite{Gardner1,Gardner2, Gardner3, Yasui} with both classical mean field
theory with classical Heisenberg spins and finite Ising-like anisotropy \cite{Enjalran} and by RPA calculations starting from the
bare microscopic model presented in Section \ref{sec:model}. \cite{Kao} In other words, the conclusion reached in
Ref.~\onlinecite{Molavian1} that, on the basis of exact diagonalization calculations and perturbation theory calculations on a single
tetrahedron, \tto\ may perhaps be described by a soft (quantum) spin ice system is upheld by the work presented in the present paper.

We studied the properties of the low energy (effective) Hamiltonian $H_{\rm
  eff}$ by finding the ground states, as a function of bare
isotropic exchange couplings $\mathcal{J}_{\rm ex}$ (from the model in Eq.~(\ref{eqn:interactions})), and for the crystal field
spectrum of \tto, under the assumption that the
pseudospins are classical (i.e. the pseudospins are vectors of fixed length $S=1/2$). Truncating the bare Hamiltonian and then the
  effective Hamiltonian to nearest neighbor interactions and assuming ground states with ordering
wavevector of $\mathbf{q}=0$ (identical spin configurations on elementary tetrahedra on the pyrochlore lattice) and omitting three
spin interactions, we found three different semiclassical ground states depending on the ratio
$\mathcal{J}_{\rm ex}/\mathcal{D}$ (see Eq.~(\ref{Eqn:Kint})). Specifically, we found (i) a
two-in/two-out state and (ii) an all-in/all-out state. In addition to these two states is one with spins lying in the local XY planes
perpendicular to the $[ 111 ]$ directions (see, for example,
  Fig. \ref{fig:XY}) with a continuous degeneracy. 

If,
instead, the original model has long-ranged dipolar interactions treated by an Ewald summation on a single cubic unit cell, then the
effective Hamiltonian has interactions with the periodicity of a cubic unit cell. For such a model, again without three spin
interactions, (and assuming that the magnetic unit cell is not larger than a single cubic unit cell), we find (for $1/\Delta =
  0.055$ K$^{-1}$ relevant to \tto) two semiclassical ground
states. For weakly antiferromagnetic bare exchange, $\mathcal{J}_{\rm ex}$ , the ground state is the LRSI$_{001}$ phase (see Fig.~\ref{fig:UnitCell}) $-$ a ground
state of the dipolar ice model $-$ and, for more antiferromagnetic $\mathcal{J}_{{\rm ex}}$, the ground state is a
two-in/two-out state with propagation (ordering) 
wavevector $\mathbf{q}=0$. The latter result $-$ the persistence of spin ice correlations with
antiferromagnetic bare coupling $-$ is partly a consequence of the renormalization of the effective Ising exchange coupling which
includes contributions from the bare dipole coupling $\mathcal{D}$ and the bare isotropic exchange coupling $\mathcal{J}_{\rm
  ex}$. It is also partly due to the presence of further neighbor interactions not present in the microscopic model. Because spin
ice-like correlations appear over a wider range of couplings than one would find in the classical model, the VCFEs
are responsible for frustrating the interactions in our simplified model (see Eqs.~(\ref{eqn:bare}),(\ref{eqn:HCF}) and
  (\ref{eqn:interactions})) of \tto. 

When the three-body interactions are incorporated, the ordering of the
Ising components of the spins is not changed from the results without three-body terms, but the effective spins then become canted out of the local
Ising directions and the local XY components are ordered into the so-called
  $\psi_{2}$ states (see
  Fig.~\ref{fig:ModelState}). \cite{Poole,ChampionHoldsworth, ETOPaper} This XY ordering is observed in the easy plane
  antiferromagnetic \eto. However, we note that the effective Hamiltonian for \eto\ has no three-body interactions (a consequence of time
  reversal within a Kramers doublet) so the effective Hamiltonian for \eto\ cannot account for the observed ordered state by means
  of three-body interactions. 

Returning to \tto, in the present work we have established that VCFEs can be included as a significant perturbation to the DSIM and that the interactions induced by VCFEs have an important effect on the physics of this material. By far the
most important problems now remaining are to establish the ground state and low energy excitations of the fully quantum effective
Hamiltonian derived in this paper beyond the single tetrahedron
approximation (see Section~\ref{sec:results}) and to assess the importance of higher order
terms in the perturbation expansion. This might be accomplished by pursuing
exact diagonalization or series expansions. \cite{SeriesExpansion}  

Further unresolved problems are to account for the long-range order in \tto\ that is induced
by a $[110]$ magnetic field \cite{Rule} and by applying pressure. \cite{Mirebeau4} There is also evidence to suggest that there
are dynamical lattice distortions away from a pyrochlore structure in zero magnetic field. \cite{Lummen, Ruff} However, the extent to
which these affect or are affected by the magnetism in the material is not
known. With the availability of an effective Hamiltonian that considers the effect of
excited crystal field levels in \tto, one can perhaps hope to supplement the model to explore the role of the
lattice on the magnetism of \tto\ and \tso.

One could extend the work in this article by including interactions in the
bare Hamiltonian besides isotropic exchange and dipole-dipole
interactions. For example, one could explore the effect of generalized
anisotropic nearest neighbor exchange interactions as was done at the mean field level in
Ref.~\onlinecite{Thompson} (for Yb$_{2}$Ti$_{2}$O$_{7}$) and
Ref.~\onlinecite{ETOPaper} (for \eto). In addition, one could include further
neighbor interactions in the bare Hamiltonian. It is already known that
further neighbor interactions are present in the related (spin ice) material
\dto. \cite{Yavorskii} If further neighbor interactions were shown to be
significant in \tto, they could be incorporated following the approach in this article.

Looking beyond the question of the ground state of \tto, we point out that an effective Hamiltonian of the type described in this
article might find a use in other problems on magnetic systems. For example,
this approach might find some use in studying the material \pso\ \hspace{1pt}
\cite{Zhou2} which has been referred to as ``dynamic spin ice'' with an ill-understood
fast dynamics compared to \hto. Two other pyrochlore magnets with Ising-like crystal field ground states are the metallic spin ice
\pio\ \hspace{1pt} \cite{Nakatsuji2} and the material \pzo \hspace{1pt} \cite{Matsuhira} both of which exhibit no long range
magnetic order at low temperature. Finally, we mention another material for which the effective Hamiltonian formalism might be
useful $-$ the langasites Nd$_{3}$Ga$_{5}$SiO$_{14}$ \cite{Simonet, Robert,
  Bordet, Zhou} and
Pr$_{3}$Ga$_{5}$SiO$_{14}$. \cite{Lumata, Bordet} These materials show no sign of
order at least down to $35$ mK although the scale of the interactions in both
compounds is much larger, as read off from the Curie-Weiss
temperatures ($-52$ K and $-2.3$ K for Nd$_{3}$Ga$_{5}$SiO$_{14}$
\cite{Robert, Bordet} and Pr$_{3}$Ga$_{5}$SiO$_{14}$ \cite{Zhou3} respectively).

We hope that the present work stimulates further theoretical investigation
into the exotic and interesting behavior displayed by these materials.

\begin{acknowledgments}
We thank Benjamin Canals, Matt Enjalran, Tom Fennell, Ludovic Jaubert and
Jacob Ruff for their critical reading of
the manuscript. This research was funded by the NSERC of Canada and the Canada
Research Chair program (M. G., Tier I), the Emerging Materials Knowledge of
Materials Manufacturing Ontario, the Canada Foundation for Innovation and the
Ontario Innovation Trust.
\end{acknowledgments}

\appendix

\section{Effective Hamiltonian}
\label{sec:general}

In order to keep this paper self-contained, and in the hope that the approach we have followed here will be of use to others, we
sketch out the main ideas behind the derivation of the effective Hamiltonian. The discussion, which we keep fairly general,
roughly follows Ref.~\onlinecite{Book} to which we refer for a broader
context. 

We consider a quantum mechanical system described by Hamiltonian
$H$ which can be split into $H_{0}$ plus a small perturbation $V$. We label
the exact eigenstates of $H$ by $|\Psi_{n}\rangle$ which
corresponds to eigenvalue $E_{n}$ for $n$ from $1$ to the dimension of the
Hilbert space $\mathcal{N}$. The eigenstates of the Hamiltonian $H_{0}$ are denoted $| n_{0} \rangle$ (where the
integer $n$ labels different eigenstates) and satisfy 
\[ H_{0}| n_{0} \rangle = E_{0,n} | n_{0} \rangle. \]
In the following, we imagine that the ground state of $H_{0}$ is $p$-fold degenerate and that eigenstates are labeled $| 1_{0}
\rangle$ to $| p_{0} \rangle$ and have energy $E_{0}$. When we introduce the perturbation $V$, to zeroth order in ordinary degenerate
perturbation theory, the ground state wavefunctions are some particular admixtures of these degenerate states $-$ in this sense
they are strongly coupled by the perturbation. 

We wish to set up a Hamiltonian that ``lives'' in the subspace spanned by the
ground state levels of $H_{0}$ and which includes
the effect of $V$ on these levels. Therefore, we introduce a projector
$\mathcal{P}$ that projects onto this subspace. Given an exact
state $|\Psi_{n}\rangle$,
\[ \mathcal{P} |\Psi_{n}\rangle \equiv |\Psi_{0,n}\rangle \]
where $| \Psi_{0,n} \rangle$ is a linear combination of $| n_{0} \rangle$ for $n =1,\ldots,p$. We refer to this subspace as the
model space $\mathfrak{M}$. Because the perturbation is assumed to be weak, the exact eigenstates $|\Psi_{n}\rangle$, for $n$ from
$1$ to $p$ lie mainly within $\mathfrak{M}$. We also introduce an operator $\Omega$ that ``undoes'' the effect of the projector
$\mathcal{P}$, \[ \Omega |\Psi_{0,n}\rangle \equiv |\Psi_{n}\rangle.  \]
It follows that $|\Psi_{0,n}\rangle= \mathcal{P}\Omega|\Psi_{0,n}\rangle$ and, because this equation is satisfied by any linear
combination of the exact states $|\Psi_{0,n}\rangle$, we find that $\mathcal{P}\Omega\mathcal{P}=\mathcal{P}$. 

The following intermediate result holds:
\begin{equation}
 [\Omega, H_{0}] \mathcal{P}= V\Omega\mathcal{P} - \Omega\mathcal{P}V\Omega\mathcal{P}. 
\label{eqn:Iden}
\end{equation}
To see this, begin with the Schr\"{o}dinger equation in the form $ (E_{n}-H_{0})|\Psi_{n}\rangle = V|\Psi_{n}\rangle $ and multiply
(from the left) by $\mathcal{P}$ to get
\[ (E_{n}-H_{0})|\Psi_{0,n}\rangle = \mathcal{P}V|\Psi_{n}\rangle \]
because the projector commutes with the Hamiltonian $H_{0}$. 

The effective Hamiltonian, $H_{\rm eff}$, is defined to be
\begin{equation} H_{\rm eff} \equiv \mathcal{P}H\Omega\mathcal{P} \label{eqn:HeffDef} \end{equation}
which has the property
\begin{equation*}  H_{\rm eff} |\Psi_{0,n}\rangle = E_{n} |\Psi_{0,n}\rangle. \end{equation*}
The effective Hamiltonian has eigenstates living in the model space $\mathfrak{M}$ and has as eigenvalues the exact eigenvalues. The
projector on the right-hand-side is there to ensure that the remaining operators $\mathcal{P}H\Omega$ operate on the model
space, $\mathfrak{M}$. Operationally, in Eq.~(\ref{eqn:HeffDef}), the $\Omega$ operator rotates the model space state into an exact
eigenstate. $H$ produces the exact eigenvalue and then the exact eigenstate is projected back into the model space.

We compute $H_{\rm eff}$ in perturbation theory by expanding $\Omega$ implicitly in powers of $V$
\begin{equation} \Omega = 1 + \Omega^{(1)} + \Omega^{(2)} + \ldots \label{eqn:omegaseries} \end{equation} 
It is then possible to eliminate $\Omega^{(k)}$ by introducing the so-called resolvent operator 
\[ \mathcal{R} \equiv (E_{0}-H_{0})^{-1} \mathcal{Q}  \]
where $\mathcal{Q}=\mathbb{I}-\mathcal{P}$. The resolvent has the spectral representation
\[ \mathcal{R} = \sum_{|\psi\rangle \notin \mathcal{M}} \frac{|\psi\rangle\langle \psi|}{E_{0}-E_{\psi}}. \]
Note that $\mathcal{R}=\mathcal{R}\mathcal{Q}$.

To eliminate $\Omega^{(k)}$, introduce the series (\ref{eqn:omegaseries}) into identity (\ref{eqn:Iden}) and use the fact that
$\mathcal{P}$ projects onto states with the same $H_{0}$ eigenvalue $E_{0}$ to obtain
\begin{align*}
\Omega^{(1)}\mathcal{P} & = \mathcal{R}V\mathcal{P}  \\
\Omega^{(2)}\mathcal{P} & = \mathcal{R}(V\Omega^{(1)}\mathcal{P} - \Omega^{(1)}\mathcal{P}V\mathcal{P})
\end{align*}
and so on. These recursion relations can be solved to get $\Omega^{(k)}$ in terms of $\mathcal{R}$, $V$ and $\mathcal{P}$.

The effective Hamiltonian (\ref{eqn:HeffDef}) then takes the form
\[ H_{\rm eff} = \mathcal{P}H\mathcal{P} + \mathcal{P}H\mathcal{R}H\mathcal{P} + \ldots \]
which, in turn, is
\begin{equation*} H_{\rm eff} = \mathcal{P}H_{0}\mathcal{P} + \mathcal{P} V \mathcal{P} + \mathcal{P}V\mathcal{R}V\mathcal{P} +
  \ldots  
\end{equation*}
because, in term $\mathcal{P}H\mathcal{R}H\mathcal{P}$, the unperturbed Hamiltonian is eliminated because it does not contain any
terms that connect the model space with the space orthogonal to it $-$ that is, terms like $\mathcal{P}H_{0}\mathcal{R}H\mathcal{P}$ 
vanish. 

\section{Calculations for Case A}
\label{sec:calc}

In this appendix, we give more details of the calculation leading to the effective pseudospin interactions from $H_{\rm
  eff}^{(2)}$ which is that part of the effective Hamiltonian that includes VCFEs to lowest order in $\langle V\rangle/\Delta$. We begin with Eq. (\ref{eqn:caseone}) which we reproduce below
\begin{multline}
\sum_{\alpha,\beta,\rho,\sigma} \sum_{{m_{p}}} \sum_{W} \mathcal{P}( m_{1},m_{2},m_{3} )  \tilde{\mathcal{K}}_{I_{1}I_{2}}^{\alpha\beta}\widetilde{J}_{I_{1}}^{\alpha}
  \widetilde{J}_{I_{2}}^{\beta} \\ \times \frac{|m_{I_{1},4},W_{I_{2}},m_{I_{3},3}\rangle\langle m_{I_{1},4}, W_{I_{2}},
  m_{I_{3},3}|}{E_{0} - E_{W}} \\ \times \tilde{\mathcal{K}}_{I_{2}I_{3}}^{\rho\sigma}\widetilde{J}_{I_{2}}^{\rho}
  \widetilde{J}_{I_{3}}^{\sigma} \mathcal{P}( m_{4},m_{5},m_{6} ).
\label{eqn:operators}
\end{multline}
In this formula, the lattice sites $I_{1}$, $I_{2}$ and $I_{3}$ have been
fixed. We observe that the matrix elements for the angular momenta on sites $I_{1}$ and $I_{3}$ are taken between states within the
ground state crystal field doublet. The nonvanishing matrix elements within this doublet are given in Eq. (\ref{eqn:ME}). From
this equation, we see that $\alpha$ and $\sigma$ must equal $z$. We consider the operators in Eq. (\ref{eqn:operators}) that act
on site $I_{1}$ 
\begin{multline*} \sum_{m_{1},m_{4}} |m_{1}\rangle\langle m_{1}| \widetilde{J}_{I_{1}}^{z} | m_{4}\rangle\langle m_{4}| \\ \longrightarrow
  |1\rangle\langle 1| \langle 1| \widetilde{J}_{I_{1}}^{z} | 1\rangle + |2\rangle\langle 2| \langle 2| \widetilde{J}_{I_{1}}^{z} |
   2\rangle \\ = - \langle \widetilde{J}_{I_{1}}^{z} \rangle  \left(|1\rangle\langle 1|- |2\rangle\langle 2| \right) \rightarrow - \langle \widetilde{J}_{I_{1}}^{z} \rangle \tilde{\sigma}_{I_{1}}^{z}.    
\end{multline*}
A similar calculation gives $- \langle \tilde{J}^{z} \rangle\tilde{\sigma}^{z}$ on site $I_{3}$. Eq. (\ref{eqn:operators}) becomes
\begin{multline}
\sum_{\beta,\rho}\tilde{\mathcal{K}}_{I_{1}I_{2}}^{z \beta} \tilde{\mathcal{K}}_{I_{2}I_{3}}^{\rho z} \langle \widetilde{J}^{z}\rangle^{2} \tilde{\sigma}_{I_{1}}^{z}
\tilde{\sigma}_{I_{3}}^{z} \\ \times  \left( \sum_{m_{2}, m_{5}} \sum_{W}   |m_{2}\rangle\langle m_{5}| \frac{ \langle m_{2} | \widetilde{J}_{I_{2}}^{\beta}|W\rangle\langle W |
\widetilde{J}_{I_{2}}^{\rho}| m_{5}\rangle }{E_{0}-E_{W}}  \right).
\label{eqn:B2} 
\end{multline}
The sum over $W$ runs over all single ion crystal field excited states. We will consider only the sum over the lowest excited
crystal field doublet states: $|3 \rangle$ and $|4 \rangle$. The denominator $E_{W}-E_{0}$ equals $\Delta$. The relevant matrix
elements are, from exact diagonalization of the crystal field Hamiltonian,
\begin{align*}
\langle 1 | \widetilde{J}^{x} | 3 \rangle & \equiv A \\ 
\langle 1 | \widetilde{J}^{y} | 3 \rangle & \equiv -iA \\ 
\langle 1 | \widetilde{J}^{z} | 4 \rangle & \equiv B \\ 
\langle 2 | \widetilde{J}^{z} | 3 \rangle & \equiv -B \\ 
\langle 2 | \widetilde{J}^{x} | 4 \rangle & \equiv -A \\ 
\langle 2 | \widetilde{J}^{y} | 4 \rangle & \equiv -iA.
\end{align*}  
All other matrix elements vanish. We shall not make any assumptions about the form of
$\tilde{\mathcal{K}}_{I_{2},I_{3}}^{\alpha\beta}$ except for symmetry under swapping both pairs of indices. The sums in brackets
in Eq. (\ref{eqn:B2}) give the operators 
\begin{align}
|1\rangle\langle 2|  \left( - \tilde{\mathcal{K}}_{I_{1}I_{2}}^{z x}\tilde{\mathcal{K}}_{I_{2}I_{3}}^{z z}
+ i\tilde{\mathcal{K}}_{I_{1}I_{2}}^{z y}\tilde{\mathcal{K}}_{I_{2}I_{3}}^{z z}  - \tilde{\mathcal{K}}_{I_{1}I_{2}}^{z
  z}\tilde{\mathcal{K}}_{I_{2}I_{3}}^{x z} \right. \nonumber \\
\left. 
+ i \tilde{\mathcal{K}}_{I_{1}I_{2}}^{z z}\tilde{\mathcal{K}}_{I_{2}I_{3}}^{y z}
\right)AB \nonumber \\
+ |2\rangle\langle 1|  \left(-\tilde{\mathcal{K}}_{I_{1}I_{2}}^{z z}\tilde{\mathcal{K}}_{I_{2}I_{3}}^{x z} 
- i \tilde{\mathcal{K}}_{I_{1}I_{2}}^{z z}\tilde{\mathcal{K}}_{I_{2}I_{3}}^{y z} - \tilde{\mathcal{K}}_{I_{1}I_{2}}^{z x}\tilde{\mathcal{K}}_{I_{2}I_{3}}^{z z}  \right. \nonumber \\
\left.  
- i \tilde{\mathcal{K}}_{I_{1}I_{2}}^{z y}\tilde{\mathcal{K}}_{I_{2}I_{3}}^{z z} 
\right)AB \nonumber \\
+ |1\rangle\langle 1|  \left( \left( \tilde{\mathcal{K}}_{I_{1}I_{2}}^{z x}\tilde{\mathcal{K}}_{I_{2}I_{3}}^{x z}
+ \tilde{\mathcal{K}}_{I_{1}I_{2}}^{z y}\tilde{\mathcal{K}}_{I_{2}I_{3}}^{y z} + i\tilde{\mathcal{K}}_{I_{1}I_{2}}^{z
  x}\tilde{\mathcal{K}}_{I_{2}I_{3}}^{y z}  \right.\right. \nonumber \\
\left.\left. 
- i\tilde{\mathcal{K}}_{I_{1}I_{2}}^{z y}\tilde{\mathcal{K}}_{I_{2}I_{3}}^{x z} \right) A^{2}
+  \tilde{\mathcal{K}}_{I_{1}I_{2}}^{z z}\tilde{\mathcal{K}}_{I_{2}I_{3}}^{z z} B^{2}
\right) \nonumber \\
+ |2\rangle\langle 2|  \left( \left( \tilde{\mathcal{K}}_{I_{1}I_{2}}^{z x}\tilde{\mathcal{K}}_{I_{2}I_{3}}^{x z}
+ \tilde{\mathcal{K}}_{I_{1}I_{2}}^{z y}\tilde{\mathcal{K}}_{I_{2}I_{3}}^{y z} - i\tilde{\mathcal{K}}_{I_{1}I_{2}}^{z
  x}\tilde{\mathcal{K}}_{I_{2}I_{3}}^{y z} \right.\right. \nonumber \\
\left. \left. 
+ i\tilde{\mathcal{K}}_{I_{1}I_{2}}^{z y}\tilde{\mathcal{K}}_{I_{2}I_{3}}^{x z} \right)A^{2}
+  \tilde{\mathcal{K}}_{I_{1}I_{2}}^{z z}\tilde{\mathcal{K}}_{I_{2}I_{3}}^{z z} B^{2}
\right).
\label{eqn:B3}
\end{align}
Of these four operators, the top two involve virtual excitations on ion $I_{2}$ that do not return the ion to its original state
but instead take it into the other crystal field ground state on ion $I_{2}$ $-$ an overall Ising spin flip. We shall see that, as
we should expect, these spin flip operations correspond to $\tilde{\sigma}^{x}$ or $\tilde{\sigma}^{y}$ effective operators. This
leads us to an important point - in order for $\tilde{\sigma}^{x}$ or $\tilde{\sigma}^{y}$ effective operators to be significant
in the effective Hamiltonian for \tto, there must be nonvanishing $\tilde{J}^{x}$, $\tilde{J}^{y}$ {\it and} $\tilde{J}^{z}$
matrix elements between the ground state doublet and first excited doublet. In order for this to be the case, the ground state and
first excited wavefunctions, which have the form
\[ | n \rangle = \sum_{M=-J}^{J} |J, M \rangle    \]
cannot have only the predominant $|J,\pm 4\rangle$ (in the ground doublet) and $|J,\pm 5\rangle$ (in the first excited doublet)
coefficients for then the $\tilde{J}^{z}$ matrix elements would vanish. Hence the conclusions of the paper are unlikely to carry
over to other materials, for example, to the spin ices. 
 
Referring to Eq. (\ref{eqn:Pauli}), we find that the above operators in Eq. (\ref{eqn:B3}) can be re-expressed in terms of Pauli
matrices. So the result of the sum of excited crystal field states in Eq. (\ref{eqn:B2}) is
\begin{align}
\left(\tilde{\mathcal{K}}_{I_{1}I_{2}}^{zz}\tilde{\mathcal{K}}_{I_{2}I_{3}}^{zz} B^{2} +
\tilde{\mathcal{K}}_{I_{1}I_{2}}^{zx}\tilde{\mathcal{K}}_{I_{2}I_{3}}^{xz} A^{2} +
\tilde{\mathcal{K}}_{I_{1}I_{2}}^{zy}\tilde{\mathcal{K}}_{I_{2}I_{3}}^{yz} A^{2}  \right) \mathbb{I}_{I_{2}} \nonumber \\
+ \left(\tilde{\mathcal{K}}_{I_{1}I_{2}}^{zx}\tilde{\mathcal{K}}_{I_{2}I_{3}}^{yz} A^{2} -
\tilde{\mathcal{K}}_{I_{1}I_{2}}^{zy}\tilde{\mathcal{K}}_{I_{2}I_{3}}^{xz} A^{2} \right) i\tilde{\sigma}^{z}_{I_{2}} \nonumber \\
- \left(\tilde{\mathcal{K}}_{I_{1}I_{2}}^{zz}\tilde{\mathcal{K}}_{I_{2}I_{3}}^{xz} +
\tilde{\mathcal{K}}_{I_{1}I_{2}}^{zx}\tilde{\mathcal{K}}_{I_{2}I_{3}}^{zz}  \right) AB \tilde{\sigma}^{x}_{I_{2}} \nonumber \\
- \left(\tilde{\mathcal{K}}_{I_{1}I_{2}}^{zy}\tilde{\mathcal{K}}_{I_{2}I_{3}}^{zz} +
\tilde{\mathcal{K}}_{I_{1}I_{2}}^{zz}\tilde{\mathcal{K}}_{I_{2}I_{3}}^{yz}  \right) AB \tilde{\sigma}^{y}_{I_{2}}. 
\label{eqn:B4}
\end{align}
 After substituting the couplings $\tilde{\mathcal{K}}$ we find that the
$\tilde{\sigma}^{z}$ terms vanish. The resulting expression is time-reversal invariant. Incorporating Eq. (\ref{eqn:B4}) into
 Eq. (\ref{eqn:B2}), we find that the overall interactions are, as we stated in the main text, Ising interactions
 $\tilde{\sigma}_{I_{1}}^{z}\tilde{\sigma}_{I_{3}}^{z}$ (arising from the unit operator in Eq. (\ref{eqn:B4})) and three-body
 interactions of the form $\tilde{\sigma}_{I_{1}}^{z}\tilde{\sigma}_{I_{2}}^{\alpha} \tilde{\sigma}_{I_{3}}^{z}$ with
 $\alpha=x,y$. Having determined the general form of the interactions and
 their couplings, we carry out a sum over all lattice sites $I_{1}$, $I_{2}$
 and $I_{3}$. The calculations for Cases B and C in the main text are carried
 out in a similar manner.

In order to organize the calculation of the terms in the effective Hamiltonian, all sums over virtual excited states 
and lattice sites are carried out numerically and the calculations described in this appendix are performed by exploiting 
the orthogonality of the Pauli matrices. As an example, suppose that the operator coefficients in Eq. 
(\ref{eqn:operators}) have been evaluated in the $|1\rangle$, $|2\rangle$ basis. We call this operator $\hat{O}$.
We want to decompose this operator into a sum of the form  
\[   \sum_{a,b,c}  A_{abc} \tilde{\sigma}^{a}\tilde{\sigma}^{b}\tilde{\sigma}^{c}     \]
where the sum runs over the Pauli operators $\tilde{\sigma}^{x}$,  $\tilde{\sigma}^{y}$, $\tilde{\sigma}^{z}$,and the unit operator.
Coefficients $A_{abc}$ are determined from
\[   A_{abc} = \frac{1}{8} {\rm Tr} [ \hat{O} \tilde{\sigma}^{a}\tilde{\sigma}^{b}\tilde{\sigma}^{c}  ] .   \]
This formula is sufficient for Case A (Section~\ref{sec:caseone}) with operators on three pyrochlore sites $I_{1}$, $I_{2}$ and $I_{3}$. 
For Cases B and C (Sections~\ref{sec:casetwo} and \ref{sec:casethree}), we decompose into a sum
\[   \sum_{a,b}  B_{ab} \tilde{\sigma}^{a}\tilde{\sigma}^{b}     \]
using
\[   B_{ab} = \frac{1}{4} {\rm Tr} [ \hat{O} \hspace{1pt} \tilde{\sigma}^{a}\tilde{\sigma}^{b} ] .   \]

\section{Crystal field parameters for T\lowercase{b}$_{2}$T\lowercase{i}$_{2}$O$_{7}$}
\label{sec:CFP}

The crystal field parameters for \tto\ are obtained from those found for \hto\ in Ref.~\onlinecite{Rosenkranz} from the formula
\ref{eqn:convert}. Ref.~\onlinecite{Rosenkranz} uses the convention 
\[ H_{\rm cf} = \sum_{l}\sum_{m=-l}^{l} \bar{B}_{l}^{m}
\left(\frac{4\pi}{2l+1}\right)^{1/2} Y_{l}^{m}  \]
for the crystal field parameters. One can convert the set of $\bar{B}_{l}^{m}$
to the $B_{l}^{m}$ using the parameters given in Ref.~\onlinecite{Kassman} and the
matrix elements of Table~\ref{tab:Rm}.

The \hto\ crystal field parameters are: 
\begin{equation}
\begin{array}{lcr}
\frac{B_{2}^{0}}{(S_{2})_{\rm Ho}} = 791 {\rm K} & \frac{B_{4}^{0}}{(S_{4})_{\rm Ho}} = 3189 {\rm K} & \frac{B_{6}^{0}}{(S_{6})_{\rm Ho}} = 1007 {\rm K} \\
\frac{B_{4}^{3}}{(S_{4})_{\rm Ho}} = 739 {\rm K} & \frac{B_{6}^{3}}{(S_{6})_{\rm Ho}} = -725 {\rm K} & \frac{B_{6}^{6}}{(S_{6})_{\rm Ho}} = 1179 {\rm K}.
\end{array}
\end{equation}
The radial expectation values $\langle r^{m}\rangle$ are given in Table~\ref{tab:Rm}
\hspace{1pt} \cite{Freeman} and the Stevens factors for \tto\ are given in
Table~\ref{tab:Stevens} \cite{Stevens}

\begin{table}
\caption{\label{tab:Rm} Table of radial expectation values, $\langle r^{m}\rangle$, for
  \tto\ and \hto. \cite{Freeman}}
\begin{ruledtabular}
\begin{tabular}{lccr}
R$^{3+}$ & $\langle \mathbf{r}^{2}\rangle$ & $\langle \mathbf{r}^{4}\rangle$ & $\langle \mathbf{r}^{6}\rangle$  \\
\hline
Ho & $0.7446$ & $1.3790$ & $5.3790$ \\
Tb & $0.8220$ & $1.6510$ & $6.8520$ \\ 
\end{tabular}
\end{ruledtabular}
\end{table}

\begin{table}
\caption{\label{tab:Stevens} Table of Stevens factors for 
  \tto\ and \hto. \cite{RareEarthMagnetism}}
\begin{ruledtabular}
\begin{tabular}{lccr}
R$^{3+}$ & $S_{2} (\times 10^{2})$ & $S_{4} (\times 10^{4})$ & $S_{6} (\times 10^{6})$  \\
\hline
Ho & $-0.2222$ & $-0.3330$ & $-1.2937$ \\
Tb & $-1.0101$ & $1.2244$ & $-1.1212$ \\ 
\end{tabular}
\end{ruledtabular}
\end{table}

\bibliography{TTOHeff}

\begin{thebibliography}{98}
\expandafter\ifx\csname natexlab\endcsname\relax\def\natexlab#1{#1}\fi
\expandafter\ifx\csname bibnamefont\endcsname\relax
  \def\bibnamefont#1{#1}\fi
\expandafter\ifx\csname bibfnamefont\endcsname\relax
  \def\bibfnamefont#1{#1}\fi
\expandafter\ifx\csname citenamefont\endcsname\relax
  \def\citenamefont#1{#1}\fi
\expandafter\ifx\csname url\endcsname\relax
  \def\url#1{\texttt{#1}}\fi
\expandafter\ifx\csname urlprefix\endcsname\relax\def\urlprefix{URL }\fi
\providecommand{\bibinfo}[2]{#2}
\providecommand{\eprint}[2][]{\url{#2}}

\bibitem[{\citenamefont{Bramwell and Gingras}(2001)}]{SpinIce}
\bibinfo{author}{\bibfnamefont{S.~T.} \bibnamefont{Bramwell}} \bibnamefont{and}
  \bibinfo{author}{\bibfnamefont{M.~J.~P.} \bibnamefont{Gingras}},
  \bibinfo{journal}{Science} \textbf{\bibinfo{volume}{294}},
  \bibinfo{pages}{1495} (\bibinfo{year}{2001}).

\bibitem[{\citenamefont{Gingras}()}]{Gingras2}
\bibinfo{author}{\bibfnamefont{M.~J.~P.} \bibnamefont{Gingras}},
  \eprint{arXiv:0903.2772}.

\bibitem[{\citenamefont{Bramwell et~al.}(2004)\citenamefont{Bramwell, Gingras,
  and Holdsworth}}]{FrusBook}
\bibinfo{author}{\bibfnamefont{S.~T.} \bibnamefont{Bramwell}},
  \bibinfo{author}{\bibfnamefont{M.~J.~P.} \bibnamefont{Gingras}},
  \bibnamefont{and} \bibinfo{author}{\bibfnamefont{P.~C.~W.}
  \bibnamefont{Holdsworth}}, \emph{\bibinfo{title}{Frustrated Spin Systems}}
  (\bibinfo{publisher}{H. T. Diep, World Scientific}, \bibinfo{year}{2004}).

\bibitem[{\citenamefont{Schrieffer and Wolff}(1966)}]{Kondo}
\bibinfo{author}{\bibfnamefont{J.~R.} \bibnamefont{Schrieffer}}
  \bibnamefont{and} \bibinfo{author}{\bibfnamefont{P.~A.} \bibnamefont{Wolff}},
  \bibinfo{journal}{Phys. Rev.} \textbf{\bibinfo{volume}{149}},
  \bibinfo{pages}{491} (\bibinfo{year}{1966}).

\bibitem[{\citenamefont{Villain}(1979)}]{Villain}
\bibinfo{author}{\bibfnamefont{J.}~\bibnamefont{Villain}}, \bibinfo{journal}{Z.
  Phys.} \textbf{\bibinfo{volume}{B33}}, \bibinfo{pages}{31}
  (\bibinfo{year}{1979}).

\bibitem[{\citenamefont{Moessner and
  Chalker}(1998{\natexlab{a}})}]{MoessnerChalker}
\bibinfo{author}{\bibfnamefont{R.}~\bibnamefont{Moessner}} \bibnamefont{and}
  \bibinfo{author}{\bibfnamefont{J.~T.} \bibnamefont{Chalker}},
  \bibinfo{journal}{Phys. Rev. Lett.} \textbf{\bibinfo{volume}{80}},
  \bibinfo{pages}{2929} (\bibinfo{year}{1998}{\natexlab{a}}).

\bibitem[{\citenamefont{Moessner and
  Chalker}(1998{\natexlab{b}})}]{MoessnerChalker2}
\bibinfo{author}{\bibfnamefont{R.}~\bibnamefont{Moessner}} \bibnamefont{and}
  \bibinfo{author}{\bibfnamefont{J.~T.} \bibnamefont{Chalker}},
  \bibinfo{journal}{Phys. Rev. B} \textbf{\bibinfo{volume}{58}},
  \bibinfo{pages}{12049} (\bibinfo{year}{1998}{\natexlab{b}}).

\bibitem[{\citenamefont{Reimers}(1992)}]{Reimers}
\bibinfo{author}{\bibfnamefont{J.~N.} \bibnamefont{Reimers}},
  \bibinfo{journal}{Phys. Rev. B} \textbf{\bibinfo{volume}{45}},
  \bibinfo{pages}{7287} (\bibinfo{year}{1992}).

\bibitem[{\citenamefont{Reimers et~al.}(1991)\citenamefont{Reimers, Berlinsky,
  and Shi}}]{Reimers2}
\bibinfo{author}{\bibfnamefont{J.~N.} \bibnamefont{Reimers}},
  \bibinfo{author}{\bibfnamefont{A.~J.} \bibnamefont{Berlinsky}},
  \bibnamefont{and} \bibinfo{author}{\bibfnamefont{A.~C.} \bibnamefont{Shi}},
  \bibinfo{journal}{Phys. Rev. B} \textbf{\bibinfo{volume}{43}},
  \bibinfo{pages}{865} (\bibinfo{year}{1991}).

\bibitem[{\citenamefont{Gardner et~al.}()\citenamefont{Gardner, Gingras, and
  Greedan}}]{GGG}
\bibinfo{author}{\bibfnamefont{J.~S.} \bibnamefont{Gardner}},
  \bibinfo{author}{\bibfnamefont{M.~J.~P.} \bibnamefont{Gingras}},
  \bibnamefont{and} \bibinfo{author}{\bibfnamefont{J.~E.}
  \bibnamefont{Greedan}}, \eprint{arXiv:0903.3661 (to appear in Rev. Mod.
  Phys.)}.

\bibitem[{\citenamefont{Palmer and Chalker}(2000)}]{PalmerChalker}
\bibinfo{author}{\bibfnamefont{S.~E.} \bibnamefont{Palmer}} \bibnamefont{and}
  \bibinfo{author}{\bibfnamefont{J.~T.} \bibnamefont{Chalker}},
  \bibinfo{journal}{Phys. Rev. B} \textbf{\bibinfo{volume}{62}},
  \bibinfo{pages}{488} (\bibinfo{year}{2000}).

\bibitem[{\citenamefont{Elhajal et~al.}(2005)\citenamefont{Elhajal, Canals,
  Sunyer, and Lacroix}}]{Elhajal}
\bibinfo{author}{\bibfnamefont{M.}~\bibnamefont{Elhajal}},
  \bibinfo{author}{\bibfnamefont{B.}~\bibnamefont{Canals}},
  \bibinfo{author}{\bibfnamefont{R.}~\bibnamefont{Sunyer}}, \bibnamefont{and}
  \bibinfo{author}{\bibfnamefont{C.}~\bibnamefont{Lacroix}},
  \bibinfo{journal}{Phys. Rev. B} \textbf{\bibinfo{volume}{71}},
  \bibinfo{pages}{094420} (\bibinfo{year}{2005}).

\bibitem[{\citenamefont{Champion et~al.}(2003)\citenamefont{Champion, Harris,
  Holdsworth, Wills, Balakrishnan, Bramwell, Cizmar, Fennell, Gardner, Lago
  et~al.}}]{Champion}
\bibinfo{author}{\bibfnamefont{J.~D.~M.} \bibnamefont{Champion}},
  \bibinfo{author}{\bibfnamefont{M.~J.} \bibnamefont{Harris}},
  \bibinfo{author}{\bibfnamefont{P.~C.~W.} \bibnamefont{Holdsworth}},
  \bibinfo{author}{\bibfnamefont{A.~S.} \bibnamefont{Wills}},
  \bibinfo{author}{\bibfnamefont{G.}~\bibnamefont{Balakrishnan}},
  \bibinfo{author}{\bibfnamefont{S.~T.} \bibnamefont{Bramwell}},
  \bibinfo{author}{\bibfnamefont{E.}~\bibnamefont{Cizmar}},
  \bibinfo{author}{\bibfnamefont{T.}~\bibnamefont{Fennell}},
  \bibinfo{author}{\bibfnamefont{J.~S.} \bibnamefont{Gardner}},
  \bibinfo{author}{\bibfnamefont{J.}~\bibnamefont{Lago}}, \bibnamefont{et~al.},
  \bibinfo{journal}{Phys. Rev. B} \textbf{\bibinfo{volume}{68}},
  \bibinfo{pages}{020401} (\bibinfo{year}{2003}).

\bibitem[{\citenamefont{Champion and Holdsworth}(2004)}]{ChampionHoldsworth}
\bibinfo{author}{\bibfnamefont{J.~D.~M.} \bibnamefont{Champion}}
  \bibnamefont{and} \bibinfo{author}{\bibfnamefont{P.~C.~W.}
  \bibnamefont{Holdsworth}}, \bibinfo{journal}{J. Phys.: Condens. Matter}
  \textbf{\bibinfo{volume}{16}}, \bibinfo{pages}{S665} (\bibinfo{year}{2004}).

\bibitem[{\citenamefont{Champion et~al.}(2001)\citenamefont{Champion, Wills,
  Fennell, Bramwell, Gardner, and Green}}]{GTO1}
\bibinfo{author}{\bibfnamefont{J.~D.~M.} \bibnamefont{Champion}},
  \bibinfo{author}{\bibfnamefont{A.~S.} \bibnamefont{Wills}},
  \bibinfo{author}{\bibfnamefont{T.}~\bibnamefont{Fennell}},
  \bibinfo{author}{\bibfnamefont{S.~T.} \bibnamefont{Bramwell}},
  \bibinfo{author}{\bibfnamefont{J.~S.} \bibnamefont{Gardner}},
  \bibnamefont{and} \bibinfo{author}{\bibfnamefont{M.~A.} \bibnamefont{Green}},
  \bibinfo{journal}{Phys. Rev. B} \textbf{\bibinfo{volume}{64}},
  \bibinfo{pages}{140407(R)} (\bibinfo{year}{2001}).

\bibitem[{\citenamefont{Stewart et~al.}(2004)\citenamefont{Stewart, Ehlers,
  Wills, Bramwell, and Gardner}}]{GTO2}
\bibinfo{author}{\bibfnamefont{J.~R.} \bibnamefont{Stewart}},
  \bibinfo{author}{\bibfnamefont{G.}~\bibnamefont{Ehlers}},
  \bibinfo{author}{\bibfnamefont{A.~S.} \bibnamefont{Wills}},
  \bibinfo{author}{\bibfnamefont{S.~T.} \bibnamefont{Bramwell}},
  \bibnamefont{and} \bibinfo{author}{\bibfnamefont{J.~S.}
  \bibnamefont{Gardner}}, \bibinfo{journal}{J. Phys.: Condens. Matter}
  \textbf{\bibinfo{volume}{16}}, \bibinfo{pages}{L321} (\bibinfo{year}{2004}).

\bibitem[{\citenamefont{Wills et~al.}(2006{\natexlab{a}})\citenamefont{Wills,
  Zhitomirsky, Canals, Sanchez, Bonville, de~Reotier, and Yaouanc}}]{GSO}
\bibinfo{author}{\bibfnamefont{A.~S.} \bibnamefont{Wills}},
  \bibinfo{author}{\bibfnamefont{M.~E.} \bibnamefont{Zhitomirsky}},
  \bibinfo{author}{\bibfnamefont{B.}~\bibnamefont{Canals}},
  \bibinfo{author}{\bibfnamefont{J.~P.} \bibnamefont{Sanchez}},
  \bibinfo{author}{\bibfnamefont{P.}~\bibnamefont{Bonville}},
  \bibinfo{author}{\bibfnamefont{P.~D.} \bibnamefont{de~Reotier}},
  \bibnamefont{and} \bibinfo{author}{\bibfnamefont{A.}~\bibnamefont{Yaouanc}},
  \bibinfo{journal}{J. Phys.: Condens. Matter} \textbf{\bibinfo{volume}{18}},
  \bibinfo{pages}{L37} (\bibinfo{year}{2006}{\natexlab{a}}).

\bibitem[{\citenamefont{Gingras et~al.}(1997)\citenamefont{Gingras, Stager,
  Raju, Gaulin, and Greedan}}]{Gingras3}
\bibinfo{author}{\bibfnamefont{M.~J.~P.} \bibnamefont{Gingras}},
  \bibinfo{author}{\bibfnamefont{C.~V.} \bibnamefont{Stager}},
  \bibinfo{author}{\bibfnamefont{N.~P.} \bibnamefont{Raju}},
  \bibinfo{author}{\bibfnamefont{B.~D.} \bibnamefont{Gaulin}},
  \bibnamefont{and} \bibinfo{author}{\bibfnamefont{J.~E.}
  \bibnamefont{Greedan}}, \bibinfo{journal}{Phys.\ Rev.\ Lett.}
  \textbf{\bibinfo{volume}{78}}, \bibinfo{pages}{947} (\bibinfo{year}{1997}).

\bibitem[{\citenamefont{Gardner
  et~al.}(1999{\natexlab{a}})\citenamefont{Gardner, Gaulin, Lee, Broholm, Raju,
  and Greedan}}]{Gardner4}
\bibinfo{author}{\bibfnamefont{J.~S.} \bibnamefont{Gardner}},
  \bibinfo{author}{\bibfnamefont{B.~D.} \bibnamefont{Gaulin}},
  \bibinfo{author}{\bibfnamefont{S.-H.} \bibnamefont{Lee}},
  \bibinfo{author}{\bibfnamefont{C.}~\bibnamefont{Broholm}},
  \bibinfo{author}{\bibfnamefont{N.~P.} \bibnamefont{Raju}}, \bibnamefont{and}
  \bibinfo{author}{\bibfnamefont{J.~E.} \bibnamefont{Greedan}},
  \bibinfo{journal}{Phys. Rev. Lett.} \textbf{\bibinfo{volume}{83}},
  \bibinfo{pages}{211} (\bibinfo{year}{1999}{\natexlab{a}}).

\bibitem[{\citenamefont{Lee}(2008)}]{PALee}
\bibinfo{author}{\bibfnamefont{P.~A.} \bibnamefont{Lee}},
  \bibinfo{journal}{Science} \textbf{\bibinfo{volume}{321}},
  \bibinfo{pages}{1306} (\bibinfo{year}{2008}).

\bibitem[{\citenamefont{Levi}(2007)}]{Levi}
\bibinfo{author}{\bibfnamefont{B.~G.} \bibnamefont{Levi}},
  \bibinfo{journal}{Physics Today} \textbf{\bibinfo{volume}{60}},
  \bibinfo{pages}{16} (\bibinfo{year}{2007}).

\bibitem[{\citenamefont{Nakatsuji et~al.}(2005)\citenamefont{Nakatsuji, Nambu,
  Tonomura, Sakai, Jonas, Broholm, Tsunetsugu, Qiu, and Maeno}}]{Nakatsuji}
\bibinfo{author}{\bibfnamefont{S.}~\bibnamefont{Nakatsuji}},
  \bibinfo{author}{\bibfnamefont{Y.}~\bibnamefont{Nambu}},
  \bibinfo{author}{\bibfnamefont{H.}~\bibnamefont{Tonomura}},
  \bibinfo{author}{\bibfnamefont{O.}~\bibnamefont{Sakai}},
  \bibinfo{author}{\bibfnamefont{S.}~\bibnamefont{Jonas}},
  \bibinfo{author}{\bibfnamefont{C.}~\bibnamefont{Broholm}},
  \bibinfo{author}{\bibfnamefont{H.}~\bibnamefont{Tsunetsugu}},
  \bibinfo{author}{\bibfnamefont{Y.}~\bibnamefont{Qiu}}, \bibnamefont{and}
  \bibinfo{author}{\bibfnamefont{Y.}~\bibnamefont{Maeno}},
  \bibinfo{journal}{Science} \textbf{\bibinfo{volume}{309}},
  \bibinfo{pages}{1697} (\bibinfo{year}{2005}).

\bibitem[{\citenamefont{Mendels et~al.}(2007)\citenamefont{Mendels, Bert,
  de~Vries, Olariu, Harrison, Duc, Trombe, Lord, Amato, and Baines}}]{Mendels}
\bibinfo{author}{\bibfnamefont{P.}~\bibnamefont{Mendels}},
  \bibinfo{author}{\bibfnamefont{F.}~\bibnamefont{Bert}},
  \bibinfo{author}{\bibfnamefont{M.~A.} \bibnamefont{de~Vries}},
  \bibinfo{author}{\bibfnamefont{A.}~\bibnamefont{Olariu}},
  \bibinfo{author}{\bibfnamefont{A.}~\bibnamefont{Harrison}},
  \bibinfo{author}{\bibfnamefont{F.}~\bibnamefont{Duc}},
  \bibinfo{author}{\bibfnamefont{J.~C.} \bibnamefont{Trombe}},
  \bibinfo{author}{\bibfnamefont{J.~S.} \bibnamefont{Lord}},
  \bibinfo{author}{\bibfnamefont{A.}~\bibnamefont{Amato}}, \bibnamefont{and}
  \bibinfo{author}{\bibfnamefont{C.}~\bibnamefont{Baines}},
  \bibinfo{journal}{Phys. Rev. Lett.} \textbf{\bibinfo{volume}{98}},
  \bibinfo{pages}{077204} (\bibinfo{year}{2007}).

\bibitem[{\citenamefont{Simonet et~al.}(2008)\citenamefont{Simonet, Ballou,
  Robert, Canals, Hippert, Bordet, Lejay, Fouquet, Ollivier, and
  Braithwaite}}]{Simonet}
\bibinfo{author}{\bibfnamefont{V.}~\bibnamefont{Simonet}},
  \bibinfo{author}{\bibfnamefont{R.}~\bibnamefont{Ballou}},
  \bibinfo{author}{\bibfnamefont{J.}~\bibnamefont{Robert}},
  \bibinfo{author}{\bibfnamefont{B.}~\bibnamefont{Canals}},
  \bibinfo{author}{\bibfnamefont{F.}~\bibnamefont{Hippert}},
  \bibinfo{author}{\bibfnamefont{P.}~\bibnamefont{Bordet}},
  \bibinfo{author}{\bibfnamefont{P.}~\bibnamefont{Lejay}},
  \bibinfo{author}{\bibfnamefont{P.}~\bibnamefont{Fouquet}},
  \bibinfo{author}{\bibfnamefont{J.}~\bibnamefont{Ollivier}}, \bibnamefont{and}
  \bibinfo{author}{\bibfnamefont{D.}~\bibnamefont{Braithwaite}},
  \bibinfo{journal}{Phys. Rev. Lett.} \textbf{\bibinfo{volume}{100}},
  \bibinfo{pages}{237204} (\bibinfo{year}{2008}).

\bibitem[{\citenamefont{Okamoto et~al.}(2007)\citenamefont{Okamoto, Nohara,
  Aruga-Katori, and Takagi}}]{Okamoto}
\bibinfo{author}{\bibfnamefont{Y.}~\bibnamefont{Okamoto}},
  \bibinfo{author}{\bibfnamefont{M.}~\bibnamefont{Nohara}},
  \bibinfo{author}{\bibfnamefont{H.}~\bibnamefont{Aruga-Katori}},
  \bibnamefont{and} \bibinfo{author}{\bibfnamefont{H.}~\bibnamefont{Takagi}},
  \bibinfo{journal}{Phys. Rev. Lett.} \textbf{\bibinfo{volume}{99}},
  \bibinfo{pages}{137207} (\bibinfo{year}{2007}).

\bibitem[{\citenamefont{Gardner
  et~al.}(1999{\natexlab{b}})\citenamefont{Gardner, Dunsiger, Gaulin, Gingras,
  Greedan, Kiefl, Lumsden, MacFarlane, Raju, Sonier et~al.}}]{Gardner1}
\bibinfo{author}{\bibfnamefont{J.~S.} \bibnamefont{Gardner}},
  \bibinfo{author}{\bibfnamefont{S.~R.} \bibnamefont{Dunsiger}},
  \bibinfo{author}{\bibfnamefont{B.~D.} \bibnamefont{Gaulin}},
  \bibinfo{author}{\bibfnamefont{M.~J.~P.} \bibnamefont{Gingras}},
  \bibinfo{author}{\bibfnamefont{J.~E.} \bibnamefont{Greedan}},
  \bibinfo{author}{\bibfnamefont{R.~F.} \bibnamefont{Kiefl}},
  \bibinfo{author}{\bibfnamefont{M.~D.} \bibnamefont{Lumsden}},
  \bibinfo{author}{\bibfnamefont{W.~A.} \bibnamefont{MacFarlane}},
  \bibinfo{author}{\bibfnamefont{N.~P.} \bibnamefont{Raju}},
  \bibinfo{author}{\bibfnamefont{J.~E.} \bibnamefont{Sonier}},
  \bibnamefont{et~al.}, \bibinfo{journal}{Phys. Rev. Lett.}
  \textbf{\bibinfo{volume}{82}}, \bibinfo{pages}{1012}
  (\bibinfo{year}{1999}{\natexlab{b}}).

\bibitem[{\citenamefont{Gardner et~al.}(2001)\citenamefont{Gardner, Gaulin,
  Berlinsky, Waldron, Dunsiger, Raju, and Greedan}}]{Gardner2}
\bibinfo{author}{\bibfnamefont{J.~S.} \bibnamefont{Gardner}},
  \bibinfo{author}{\bibfnamefont{B.~D.} \bibnamefont{Gaulin}},
  \bibinfo{author}{\bibfnamefont{A.~J.} \bibnamefont{Berlinsky}},
  \bibinfo{author}{\bibfnamefont{P.}~\bibnamefont{Waldron}},
  \bibinfo{author}{\bibfnamefont{S.~R.} \bibnamefont{Dunsiger}},
  \bibinfo{author}{\bibfnamefont{N.~P.} \bibnamefont{Raju}}, \bibnamefont{and}
  \bibinfo{author}{\bibfnamefont{J.~D.} \bibnamefont{Greedan}},
  \bibinfo{journal}{Phys. Rev. B} \textbf{\bibinfo{volume}{64}},
  \bibinfo{pages}{224416} (\bibinfo{year}{2001}).

\bibitem[{\citenamefont{Gardner et~al.}(2003)\citenamefont{Gardner, Keren,
  Ehlers, Stock, Segal, Roper, F{\aa}k, Stone, Hammar, Reich
  et~al.}}]{Gardner3}
\bibinfo{author}{\bibfnamefont{J.~S.} \bibnamefont{Gardner}},
  \bibinfo{author}{\bibfnamefont{A.}~\bibnamefont{Keren}},
  \bibinfo{author}{\bibfnamefont{G.}~\bibnamefont{Ehlers}},
  \bibinfo{author}{\bibfnamefont{C.}~\bibnamefont{Stock}},
  \bibinfo{author}{\bibfnamefont{E.}~\bibnamefont{Segal}},
  \bibinfo{author}{\bibfnamefont{J.~M.} \bibnamefont{Roper}},
  \bibinfo{author}{\bibfnamefont{B.}~\bibnamefont{F{\aa}k}},
  \bibinfo{author}{\bibfnamefont{M.~B.} \bibnamefont{Stone}},
  \bibinfo{author}{\bibfnamefont{P.~R.} \bibnamefont{Hammar}},
  \bibinfo{author}{\bibfnamefont{D.~H.} \bibnamefont{Reich}},
  \bibnamefont{et~al.}, \bibinfo{journal}{Phys. Rev. B}
  \textbf{\bibinfo{volume}{68}}, \bibinfo{pages}{180401(R)}
  (\bibinfo{year}{2003}).

\bibitem[{Gla()}]{Glass}
\eprint{However, some studies have reported glassy behavior. See
  Refs.~\onlinecite{Hamaguchi}, \onlinecite{Yasui} and \onlinecite{Luo}.}

\bibitem[{\citenamefont{Gingras et~al.}(2000)\citenamefont{Gingras, den Hertog,
  Faucher, Gardner, Dunsiger, Chang, Gaulin, Raju, and Greedan}}]{Gingras}
\bibinfo{author}{\bibfnamefont{M.~J.~P.} \bibnamefont{Gingras}},
  \bibinfo{author}{\bibfnamefont{B.~C.} \bibnamefont{den Hertog}},
  \bibinfo{author}{\bibfnamefont{M.}~\bibnamefont{Faucher}},
  \bibinfo{author}{\bibfnamefont{J.~S.} \bibnamefont{Gardner}},
  \bibinfo{author}{\bibfnamefont{S.~R.} \bibnamefont{Dunsiger}},
  \bibinfo{author}{\bibfnamefont{L.~J.} \bibnamefont{Chang}},
  \bibinfo{author}{\bibfnamefont{B.~D.} \bibnamefont{Gaulin}},
  \bibinfo{author}{\bibfnamefont{N.~P.} \bibnamefont{Raju}}, \bibnamefont{and}
  \bibinfo{author}{\bibfnamefont{J.~E.} \bibnamefont{Greedan}},
  \bibinfo{journal}{Phys.\ Rev.\ B.} \textbf{\bibinfo{volume}{62}},
  \bibinfo{pages}{6496} (\bibinfo{year}{2000}).

\bibitem[{\citenamefont{Mirebeau et~al.}(2006)\citenamefont{Mirebeau, Apetrei,
  Goncharenko, and Moessner}}]{Mirebeau2}
\bibinfo{author}{\bibfnamefont{I.}~\bibnamefont{Mirebeau}},
  \bibinfo{author}{\bibfnamefont{A.}~\bibnamefont{Apetrei}},
  \bibinfo{author}{\bibfnamefont{I.~N.} \bibnamefont{Goncharenko}},
  \bibnamefont{and} \bibinfo{author}{\bibfnamefont{R.}~\bibnamefont{Moessner}},
  \bibinfo{journal}{Physica} \textbf{\bibinfo{volume}{B385}},
  \bibinfo{pages}{307} (\bibinfo{year}{2006}).

\bibitem[{\citenamefont{Enjalran et~al.}(2004)\citenamefont{Enjalran, Gingras,
  Kao, Maestro, and Molavian}}]{Enjalran2}
\bibinfo{author}{\bibfnamefont{M.}~\bibnamefont{Enjalran}},
  \bibinfo{author}{\bibfnamefont{M.~J.~P.} \bibnamefont{Gingras}},
  \bibinfo{author}{\bibfnamefont{Y.-J.} \bibnamefont{Kao}},
  \bibinfo{author}{\bibfnamefont{A.~D.} \bibnamefont{Maestro}},
  \bibnamefont{and} \bibinfo{author}{\bibfnamefont{H.~R.}
  \bibnamefont{Molavian}}, \bibinfo{journal}{J. Phys.: Condens Matter}
  \textbf{\bibinfo{volume}{16}}, \bibinfo{pages}{5673} (\bibinfo{year}{2004}).

\bibitem[{\citenamefont{Molavian et~al.}(2007)\citenamefont{Molavian, Gingras,
  and Canals}}]{Molavian1}
\bibinfo{author}{\bibfnamefont{H.~R.} \bibnamefont{Molavian}},
  \bibinfo{author}{\bibfnamefont{M.~J.~P.} \bibnamefont{Gingras}},
  \bibnamefont{and} \bibinfo{author}{\bibfnamefont{B.}~\bibnamefont{Canals}},
  \bibinfo{journal}{Phys.\ Rev.\ Lett.} \textbf{\bibinfo{volume}{98}},
  \bibinfo{pages}{157204} (\bibinfo{year}{2007}).

\bibitem[{\citenamefont{Chernyshev et~al.}(2004)\citenamefont{Chernyshev,
  Galanakis, Phillips, Rozhkov, and Tremblay}}]{Hubbard}
\bibinfo{author}{\bibfnamefont{A.~L.} \bibnamefont{Chernyshev}},
  \bibinfo{author}{\bibfnamefont{D.}~\bibnamefont{Galanakis}},
  \bibinfo{author}{\bibfnamefont{P.}~\bibnamefont{Phillips}},
  \bibinfo{author}{\bibfnamefont{A.~V.} \bibnamefont{Rozhkov}},
  \bibnamefont{and} \bibinfo{author}{\bibfnamefont{A.-M.~S.}
  \bibnamefont{Tremblay}}, \bibinfo{journal}{Phys. Rev. B}
  \textbf{\bibinfo{volume}{70}}, \bibinfo{pages}{235111}
  (\bibinfo{year}{2004}).

\bibitem[{\citenamefont{Delannoy et~al.}(2005)\citenamefont{Delannoy, Gingras,
  Holdsworth, and Tremblay}}]{Delannoy}
\bibinfo{author}{\bibfnamefont{J.-Y.~P.} \bibnamefont{Delannoy}},
  \bibinfo{author}{\bibfnamefont{M.~J.~P.} \bibnamefont{Gingras}},
  \bibinfo{author}{\bibfnamefont{P.~C.~W.} \bibnamefont{Holdsworth}},
  \bibnamefont{and} \bibinfo{author}{\bibfnamefont{A.-M.~S.}
  \bibnamefont{Tremblay}}, \bibinfo{journal}{Phys. Rev. B}
  \textbf{\bibinfo{volume}{72}}, \bibinfo{pages}{115114}
  (\bibinfo{year}{2005}).

\bibitem[{\citenamefont{Delannoy et~al.}(2009)\citenamefont{Delannoy, Gingras,
  Holdsworth, and Tremblay}}]{Delannoy2}
\bibinfo{author}{\bibfnamefont{J.-Y.~P.} \bibnamefont{Delannoy}},
  \bibinfo{author}{\bibfnamefont{M.~J.~P.} \bibnamefont{Gingras}},
  \bibinfo{author}{\bibfnamefont{P.~C.~W.} \bibnamefont{Holdsworth}},
  \bibnamefont{and} \bibinfo{author}{\bibfnamefont{A.-M.~S.}
  \bibnamefont{Tremblay}}, \bibinfo{journal}{Phys. Rev. B}
  \textbf{\bibinfo{volume}{79}}, \bibinfo{pages}{235130}
  (\bibinfo{year}{2009}).

\bibitem[{\citenamefont{Leigh et~al.}(2006)\citenamefont{Leigh, Phillips, and
  Choy}}]{Phillips}
\bibinfo{author}{\bibfnamefont{R.~G.} \bibnamefont{Leigh}},
  \bibinfo{author}{\bibfnamefont{P.}~\bibnamefont{Phillips}}, \bibnamefont{and}
  \bibinfo{author}{\bibfnamefont{T.-P.} \bibnamefont{Choy}},
  \bibinfo{journal}{Phys. Rev. Lett.} \textbf{\bibinfo{volume}{99}},
  \bibinfo{pages}{046404} (\bibinfo{year}{2006}).

\bibitem[{\citenamefont{Hermele et~al.}(2004)\citenamefont{Hermele, Fisher, and
  Balents}}]{Gauge1}
\bibinfo{author}{\bibfnamefont{M.}~\bibnamefont{Hermele}},
  \bibinfo{author}{\bibfnamefont{M.~P.~A.} \bibnamefont{Fisher}},
  \bibnamefont{and} \bibinfo{author}{\bibfnamefont{L.}~\bibnamefont{Balents}},
  \bibinfo{journal}{Phys. Rev. B} \textbf{\bibinfo{volume}{69}},
  \bibinfo{pages}{064404} (\bibinfo{year}{2004}).

\bibitem[{\citenamefont{Neto et~al.}(2006)\citenamefont{Neto, Pujol, and
  Fradkin}}]{Gauge2}
\bibinfo{author}{\bibfnamefont{A.~H.~C.} \bibnamefont{Neto}},
  \bibinfo{author}{\bibfnamefont{P.}~\bibnamefont{Pujol}}, \bibnamefont{and}
  \bibinfo{author}{\bibfnamefont{E.}~\bibnamefont{Fradkin}},
  \bibinfo{journal}{Phys. Rev. B} \textbf{\bibinfo{volume}{74}},
  \bibinfo{pages}{024302} (\bibinfo{year}{2006}).

\bibitem[{\citenamefont{Goremychkin et~al.}(2008)\citenamefont{Goremychkin,
  Osborn, Rainford, Macaluso, Adroja, and Koza}}]{DynamicFrustration}
\bibinfo{author}{\bibfnamefont{E.~A.} \bibnamefont{Goremychkin}},
  \bibinfo{author}{\bibfnamefont{R.}~\bibnamefont{Osborn}},
  \bibinfo{author}{\bibfnamefont{B.~D.} \bibnamefont{Rainford}},
  \bibinfo{author}{\bibfnamefont{R.~T.} \bibnamefont{Macaluso}},
  \bibinfo{author}{\bibfnamefont{D.~T.} \bibnamefont{Adroja}},
  \bibnamefont{and} \bibinfo{author}{\bibfnamefont{M.}~\bibnamefont{Koza}},
  \bibinfo{journal}{Nature Physics} \textbf{\bibinfo{volume}{4}},
  \bibinfo{pages}{766} (\bibinfo{year}{2008}).

\bibitem[{\citenamefont{Melko et~al.}(2001)\citenamefont{Melko, den Hertog, and
  Gingras}}]{Melko}
\bibinfo{author}{\bibfnamefont{R.~G.} \bibnamefont{Melko}},
  \bibinfo{author}{\bibfnamefont{B.~C.} \bibnamefont{den Hertog}},
  \bibnamefont{and} \bibinfo{author}{\bibfnamefont{M.~J.~P.}
  \bibnamefont{Gingras}}, \bibinfo{journal}{Phys.\ Rev.\ Lett.}
  \textbf{\bibinfo{volume}{87}}, \bibinfo{pages}{067203}
  (\bibinfo{year}{2001}).

\bibitem[{\citenamefont{Rosenkranz et~al.}(2000)\citenamefont{Rosenkranz,
  Ramirez, Hayashi, Cava, Siddharthan, and Shastry}}]{Rosenkranz}
\bibinfo{author}{\bibfnamefont{S.}~\bibnamefont{Rosenkranz}},
  \bibinfo{author}{\bibfnamefont{A.~P.} \bibnamefont{Ramirez}},
  \bibinfo{author}{\bibfnamefont{A.}~\bibnamefont{Hayashi}},
  \bibinfo{author}{\bibfnamefont{R.~J.} \bibnamefont{Cava}},
  \bibinfo{author}{\bibfnamefont{R.}~\bibnamefont{Siddharthan}},
  \bibnamefont{and} \bibinfo{author}{\bibfnamefont{B.~S.}
  \bibnamefont{Shastry}}, \bibinfo{journal}{J. Appl. Phys.}
  \textbf{\bibinfo{volume}{87}}, \bibinfo{pages}{5914} (\bibinfo{year}{2000}).

\bibitem[{\citenamefont{Mirebeau et~al.}(2007)\citenamefont{Mirebeau, Bonville,
  and Hennion}}]{Mirebeau}
\bibinfo{author}{\bibfnamefont{I.}~\bibnamefont{Mirebeau}},
  \bibinfo{author}{\bibfnamefont{P.}~\bibnamefont{Bonville}}, \bibnamefont{and}
  \bibinfo{author}{\bibfnamefont{M.}~\bibnamefont{Hennion}},
  \bibinfo{journal}{Phys. Rev. B} \textbf{\bibinfo{volume}{76}},
  \bibinfo{pages}{184436} (\bibinfo{year}{2007}).

\bibitem[{Vbr()}]{Vbracket}
\eprint{$\langle V\rangle$ denotes symbolically the energy scale of the
  interactions $V$, defined as the maximum of the exchange coupling and the
  dipole-dipole coupling.}

\bibitem[{\citenamefont{den Hertog and Gingras}(2000)}]{denHertog}
\bibinfo{author}{\bibfnamefont{B.~C.} \bibnamefont{den Hertog}}
  \bibnamefont{and} \bibinfo{author}{\bibfnamefont{M.~J.~P.}
  \bibnamefont{Gingras}}, \bibinfo{journal}{Phys.\ Rev.\ Lett.}
  \textbf{\bibinfo{volume}{84}}, \bibinfo{pages}{3430} (\bibinfo{year}{2000}).

\bibitem[{\citenamefont{Bramwell et~al.}(2001)\citenamefont{Bramwell, Harris,
  den Hertog, Gingras, Gardner, McMorrow, Wildes, Cornelius, Champion, Melko
  et~al.}}]{Bramwell}
\bibinfo{author}{\bibfnamefont{S.~T.} \bibnamefont{Bramwell}},
  \bibinfo{author}{\bibfnamefont{M.~J.} \bibnamefont{Harris}},
  \bibinfo{author}{\bibfnamefont{B.~C.} \bibnamefont{den Hertog}},
  \bibinfo{author}{\bibfnamefont{M.~J.~P.} \bibnamefont{Gingras}},
  \bibinfo{author}{\bibfnamefont{J.~S.} \bibnamefont{Gardner}},
  \bibinfo{author}{\bibfnamefont{D.~F.} \bibnamefont{McMorrow}},
  \bibinfo{author}{\bibfnamefont{A.~R.} \bibnamefont{Wildes}},
  \bibinfo{author}{\bibfnamefont{A.~L.} \bibnamefont{Cornelius}},
  \bibinfo{author}{\bibfnamefont{J.~D.~M.} \bibnamefont{Champion}},
  \bibinfo{author}{\bibfnamefont{R.~G.} \bibnamefont{Melko}},
  \bibnamefont{et~al.}, \bibinfo{journal}{Phys.\ Rev.\ Lett.}
  \textbf{\bibinfo{volume}{87}}, \bibinfo{pages}{047205}
  (\bibinfo{year}{2001}).

\bibitem[{\citenamefont{Yavors'kii et~al.}(2008)\citenamefont{Yavors'kii,
  Fennell, Gingras, and Bramwell}}]{Yavorskii}
\bibinfo{author}{\bibfnamefont{T.}~\bibnamefont{Yavors'kii}},
  \bibinfo{author}{\bibfnamefont{T.}~\bibnamefont{Fennell}},
  \bibinfo{author}{\bibfnamefont{M.~J.~P.} \bibnamefont{Gingras}},
  \bibnamefont{and} \bibinfo{author}{\bibfnamefont{S.~T.}
  \bibnamefont{Bramwell}}, \bibinfo{journal}{Phys. Rev. Lett.}
  \textbf{\bibinfo{volume}{101}}, \bibinfo{pages}{037204}
  (\bibinfo{year}{2008}).

\bibitem[{\citenamefont{Melko and Gingras}(2004)}]{Melko2}
\bibinfo{author}{\bibfnamefont{R.~G.} \bibnamefont{Melko}} \bibnamefont{and}
  \bibinfo{author}{\bibfnamefont{M.~J.~P.} \bibnamefont{Gingras}},
  \bibinfo{journal}{J. Phys.:Condens. Matter} \textbf{\bibinfo{volume}{16}},
  \bibinfo{pages}{R1277} (\bibinfo{year}{2004}).

\bibitem[{\citenamefont{Ramirez et~al.}(1999)\citenamefont{Ramirez, Hayashi,
  Cava, Siddharthan, and Shastry}}]{Ramirez}
\bibinfo{author}{\bibfnamefont{A.~P.} \bibnamefont{Ramirez}},
  \bibinfo{author}{\bibfnamefont{A.}~\bibnamefont{Hayashi}},
  \bibinfo{author}{\bibfnamefont{R.~J.} \bibnamefont{Cava}},
  \bibinfo{author}{\bibfnamefont{R.}~\bibnamefont{Siddharthan}},
  \bibnamefont{and} \bibinfo{author}{\bibfnamefont{B.~S.}
  \bibnamefont{Shastry}}, \bibinfo{journal}{Nature}
  \textbf{\bibinfo{volume}{399}}, \bibinfo{pages}{333} (\bibinfo{year}{1999}).

\bibitem[{\citenamefont{Hamaguchi et~al.}(2004)\citenamefont{Hamaguchi,
  Matsushita, Wada, Yasui, and Sato}}]{Hamaguchi}
\bibinfo{author}{\bibfnamefont{N.}~\bibnamefont{Hamaguchi}},
  \bibinfo{author}{\bibfnamefont{T.}~\bibnamefont{Matsushita}},
  \bibinfo{author}{\bibfnamefont{N.}~\bibnamefont{Wada}},
  \bibinfo{author}{\bibfnamefont{Y.}~\bibnamefont{Yasui}}, \bibnamefont{and}
  \bibinfo{author}{\bibfnamefont{M.}~\bibnamefont{Sato}},
  \bibinfo{journal}{Phys.\ Rev.\ B.} \textbf{\bibinfo{volume}{69}},
  \bibinfo{pages}{132413} (\bibinfo{year}{2004}).

\bibitem[{\citenamefont{Ke et~al.}(2009)\citenamefont{Ke, West, Cava, and
  Schiffer}}]{Ke}
\bibinfo{author}{\bibfnamefont{X.}~\bibnamefont{Ke}},
  \bibinfo{author}{\bibfnamefont{D.~V.} \bibnamefont{West}},
  \bibinfo{author}{\bibfnamefont{R.~J.} \bibnamefont{Cava}}, \bibnamefont{and}
  \bibinfo{author}{\bibfnamefont{P.}~\bibnamefont{Schiffer}},
  \bibinfo{journal}{Phys. Rev. B} \textbf{\bibinfo{volume}{80}},
  \bibinfo{pages}{144426} (\bibinfo{year}{2009}).

\bibitem[{\citenamefont{Luo et~al.}(2001)\citenamefont{Luo, Hess, and
  Corruccini}}]{Luo}
\bibinfo{author}{\bibfnamefont{G.}~\bibnamefont{Luo}},
  \bibinfo{author}{\bibfnamefont{S.~T.} \bibnamefont{Hess}}, \bibnamefont{and}
  \bibinfo{author}{\bibfnamefont{L.~R.} \bibnamefont{Corruccini}},
  \bibinfo{journal}{Phys.\ Lett.\ A.} \textbf{\bibinfo{volume}{291}},
  \bibinfo{pages}{306} (\bibinfo{year}{2001}).

\bibitem[{\citenamefont{Yasui et~al.}(2002)\citenamefont{Yasui, Kanada, Ito,
  Harashina, Sato, Okumura, Kakura, and Kadawski}}]{Yasui}
\bibinfo{author}{\bibfnamefont{Y.}~\bibnamefont{Yasui}},
  \bibinfo{author}{\bibfnamefont{M.}~\bibnamefont{Kanada}},
  \bibinfo{author}{\bibfnamefont{M.}~\bibnamefont{Ito}},
  \bibinfo{author}{\bibfnamefont{H.}~\bibnamefont{Harashina}},
  \bibinfo{author}{\bibfnamefont{M.}~\bibnamefont{Sato}},
  \bibinfo{author}{\bibfnamefont{H.}~\bibnamefont{Okumura}},
  \bibinfo{author}{\bibfnamefont{K.}~\bibnamefont{Kakura}}, \bibnamefont{and}
  \bibinfo{author}{\bibfnamefont{H.}~\bibnamefont{Kadawski}},
  \bibinfo{journal}{J. Phys. Soc. Jpn.} \textbf{\bibinfo{volume}{71}},
  \bibinfo{pages}{599} (\bibinfo{year}{2002}).

\bibitem[{\citenamefont{Enjalran and Gingras}(2004)}]{Enjalran}
\bibinfo{author}{\bibfnamefont{M.}~\bibnamefont{Enjalran}} \bibnamefont{and}
  \bibinfo{author}{\bibfnamefont{M.~J.~P.} \bibnamefont{Gingras}},
  \bibinfo{journal}{Phys.\ Rev.\ B.} \textbf{\bibinfo{volume}{70}},
  \bibinfo{pages}{174426} (\bibinfo{year}{2004}).

\bibitem[{\citenamefont{Kao et~al.}(2003)\citenamefont{Kao, Enjalran, Maestro,
  Molavian, and Gingras}}]{Kao}
\bibinfo{author}{\bibfnamefont{Y.-J.} \bibnamefont{Kao}},
  \bibinfo{author}{\bibfnamefont{M.}~\bibnamefont{Enjalran}},
  \bibinfo{author}{\bibfnamefont{A.~D.} \bibnamefont{Maestro}},
  \bibinfo{author}{\bibfnamefont{H.~R.} \bibnamefont{Molavian}},
  \bibnamefont{and} \bibinfo{author}{\bibfnamefont{M.~J.~P.}
  \bibnamefont{Gingras}}, \bibinfo{journal}{Phys.\ Rev.\ B.}
  \textbf{\bibinfo{volume}{68}}, \bibinfo{pages}{172407}
  (\bibinfo{year}{2003}).

\bibitem[{\citenamefont{Oitmaa et~al.}(2006)\citenamefont{Oitmaa, Hamer, and
  Zheng}}]{SeriesExpansion}
\bibinfo{author}{\bibfnamefont{J.}~\bibnamefont{Oitmaa}},
  \bibinfo{author}{\bibfnamefont{C.}~\bibnamefont{Hamer}}, \bibnamefont{and}
  \bibinfo{author}{\bibfnamefont{W.}~\bibnamefont{Zheng}},
  \emph{\bibinfo{title}{Series Expansion Methods For Strongly Interacting \\
  Lattice Models}} (\bibinfo{publisher}{Cambridge University Press},
  \bibinfo{year}{2006}).

\bibitem[{\citenamefont{Hutchings}(1964)}]{Hutchings}
\bibinfo{author}{\bibfnamefont{M.~J.} \bibnamefont{Hutchings}},
  \bibinfo{journal}{Solid State Phys.} \textbf{\bibinfo{volume}{16}},
  \bibinfo{pages}{227} (\bibinfo{year}{1964}).

\bibitem[{\citenamefont{Stevens}(1952)}]{Stevens}
\bibinfo{author}{\bibfnamefont{K.~W.~H.} \bibnamefont{Stevens}},
  \bibinfo{journal}{Proc. Phys. Soc., London} \textbf{\bibinfo{volume}{A65}},
  \bibinfo{pages}{209} (\bibinfo{year}{1952}).

\bibitem[{\citenamefont{Freeman and Desclaux}(1979)}]{Freeman}
\bibinfo{author}{\bibfnamefont{A.~J.} \bibnamefont{Freeman}} \bibnamefont{and}
  \bibinfo{author}{\bibfnamefont{J.~P.} \bibnamefont{Desclaux}},
  \bibinfo{journal}{J. Mag. and Mag. Mat.} \textbf{\bibinfo{volume}{12}},
  \bibinfo{pages}{11} (\bibinfo{year}{1979}).

\bibitem[{Tim()}]{TimeReversal}
\eprint{This relation is consistent with the sign change of angular momentum
  operators under time reversal operation $\mathfrak{T}$ associated with
  antiunitary operator $\theta$ - the anticommutation of $J$ and $\theta$. For
  then $J^{z}(\theta|J,M\rangle) = -M(\theta|J,M\rangle)$ and
  $J^{\pm}(\theta|J,M\rangle) =\theta J^{\mp}|J,M\rangle = -\sqrt{(J\mp
  M+1)(J\pm 1)}(\theta|J,M\mp 1\rangle)$ from which one has the relation
  $\theta|J,M\rangle = (-)^{J-M}|J,M\rangle$. This ensures that a wavefunction
  is invariant under time reversal provided $c^{M}=(-)^{J-M}c^{-M}$.}

\bibitem[{Est()}]{Estimate}
\eprint{Ref.~\onlinecite{Mirebeau} estimates the exchange by fitting the
  experimental Curie-Weiss temperature to a model with crystal field and an
  isotropic exchange coupling that is treated within mean field theory.}

\bibitem[{Int()}]{Interactions}
\eprint{In retaining $V=H_{\rm ex}+H_{\rm dd}$ in Eq.~(\ref{eqn:interactions})
  as our microscopic model of \tto, we were guided by the observation that this
  model does a good job in describing the inelastic and diffuse neutron
  scattering of \tto\ in the paramagnetic regime (Ref.~\cite{Kao}).}

\bibitem[{\citenamefont{McClarty et~al.}(2009)\citenamefont{McClarty, Curnoe,
  and Gingras}}]{ETOPaper}
\bibinfo{author}{\bibfnamefont{P.~A.} \bibnamefont{McClarty}},
  \bibinfo{author}{\bibfnamefont{S.~H.} \bibnamefont{Curnoe}},
  \bibnamefont{and} \bibinfo{author}{\bibfnamefont{M.~J.~P.}
  \bibnamefont{Gingras}}, \bibinfo{journal}{J. Phys.: Conference Series}
  \textbf{\bibinfo{volume}{145}}, \bibinfo{pages}{012032}
  (\bibinfo{year}{2009}).

\bibitem[{\citenamefont{Santini et~al.}(2009)\citenamefont{Santini, Carretta,
  Amoretti, Caciuffo, Magnani, and Lander}}]{Multipole}
\bibinfo{author}{\bibfnamefont{P.}~\bibnamefont{Santini}},
  \bibinfo{author}{\bibfnamefont{S.}~\bibnamefont{Carretta}},
  \bibinfo{author}{\bibfnamefont{G.}~\bibnamefont{Amoretti}},
  \bibinfo{author}{\bibfnamefont{R.}~\bibnamefont{Caciuffo}},
  \bibinfo{author}{\bibfnamefont{N.}~\bibnamefont{Magnani}}, \bibnamefont{and}
  \bibinfo{author}{\bibfnamefont{G.~H.} \bibnamefont{Lander}},
  \bibinfo{journal}{Rev. Mod. Phys.} \textbf{\bibinfo{volume}{81}},
  \bibinfo{pages}{807} (\bibinfo{year}{2009}).

\bibitem[{\citenamefont{Lindgren and Morrison}(1982)}]{Book}
\bibinfo{author}{\bibfnamefont{I.}~\bibnamefont{Lindgren}} \bibnamefont{and}
  \bibinfo{author}{\bibfnamefont{J.}~\bibnamefont{Morrison}},
  \emph{\bibinfo{title}{Atomic Many-Body Theory}}
  (\bibinfo{publisher}{Springer-Verlag}, \bibinfo{year}{1982}).

\bibitem[{\citenamefont{Bergman et~al.}(2007)\citenamefont{Bergman, Shindou,
  Fiete, and Balents}}]{Bergman}
\bibinfo{author}{\bibfnamefont{D.~L.} \bibnamefont{Bergman}},
  \bibinfo{author}{\bibfnamefont{R.}~\bibnamefont{Shindou}},
  \bibinfo{author}{\bibfnamefont{G.~A.} \bibnamefont{Fiete}}, \bibnamefont{and}
  \bibinfo{author}{\bibfnamefont{L.}~\bibnamefont{Balents}},
  \bibinfo{journal}{Phys. Rev. B} \textbf{\bibinfo{volume}{75}},
  \bibinfo{pages}{094403} (\bibinfo{year}{2007}).

\bibitem[{\citenamefont{Molavian and Gingras}(2009)}]{Molavian2}
\bibinfo{author}{\bibfnamefont{H.~R.} \bibnamefont{Molavian}} \bibnamefont{and}
  \bibinfo{author}{\bibfnamefont{M.~J.~P.} \bibnamefont{Gingras}},
  \bibinfo{journal}{J. Phys.: Condens. Matter} \textbf{\bibinfo{volume}{21}},
  \bibinfo{pages}{172201} (\bibinfo{year}{2009}).

\bibitem[{\citenamefont{Fennell et~al.}(2004)\citenamefont{Fennell, Petrenko,
  Fak, Bramwell, Enjalran, Yavors'kii, Gingras, Melko, and
  Balakrishnan}}]{Fennell}
\bibinfo{author}{\bibfnamefont{T.}~\bibnamefont{Fennell}},
  \bibinfo{author}{\bibfnamefont{O.~A.} \bibnamefont{Petrenko}},
  \bibinfo{author}{\bibfnamefont{B.}~\bibnamefont{Fak}},
  \bibinfo{author}{\bibfnamefont{S.~T.} \bibnamefont{Bramwell}},
  \bibinfo{author}{\bibfnamefont{M.}~\bibnamefont{Enjalran}},
  \bibinfo{author}{\bibfnamefont{T.}~\bibnamefont{Yavors'kii}},
  \bibinfo{author}{\bibfnamefont{M.~J.~P.} \bibnamefont{Gingras}},
  \bibinfo{author}{\bibfnamefont{R.~G.} \bibnamefont{Melko}}, \bibnamefont{and}
  \bibinfo{author}{\bibfnamefont{G.}~\bibnamefont{Balakrishnan}},
  \bibinfo{journal}{Phys. Rev. B} \textbf{\bibinfo{volume}{70}},
  \bibinfo{pages}{134408} (\bibinfo{year}{2004}).

\bibitem[{Tra()}]{Transition}
\eprint{This zero temperature transition at $\mathcal{J}_{\rm
  ex}/\mathcal{D}=4.525$ corresponds to the boundary at $J_{\rm nn}/D_{\rm
  nn}=-0.905$ in Ref.~\onlinecite{Melko} since (i) here we have switched the
  sign convention for antiferromagnetic $\mathcal{J}_{\rm ex}$ compared to
  Ref.~\onlinecite{Melko} and (ii) since we have $J_{\rm nn}$ and $D_{\rm nn}$
  here given by $J_{\rm nn}=\mathcal{J}_{\rm ex}\langle
  \tilde{J}^{z}\rangle^{2} /3$ and $D_{\rm nn}=5 \mathcal{D}\langle
  \tilde{J}^{z}\rangle^{2} /3$.}

\bibitem[{Spi({\natexlab{a}})}]{SpinIceFNN}
\eprint{It is also because exchange couplings beyond nearest neighbors are weak
  in comparison to the Ising interactions such that there exists a prevailing
  (classical) Ising energy scale set by $\mathcal{P}H\mathcal{P}$. See
  Refs.~\onlinecite{Yavorskii} and \onlinecite{Fennell}.}

\bibitem[{Spi({\natexlab{b}})}]{SpinIceFNN2}
\eprint{The bare nearest neighbor isotropic exchange $\mathcal{J}_{\rm ex}$ is
  not known with good precision. There are currently differing estimates of
  $\mathcal{J}_{\rm ex}$ in the literature. \cite{Gingras, Mirebeau} Also,
  further neighbor exchange is measurable, at least in the spin ice \dto,
  \cite{Yavorskii} (see also Ref.~\onlinecite{Ruff2}) so it is not unlikely
  that further neighbor interactions also play a role in the physics of other
  rare earth titanates including \tto.}

\bibitem[{\citenamefont{Wills et~al.}(2006{\natexlab{b}})\citenamefont{Wills,
  Zhitomirsky, Canals, Sanchez, Bonville, de~R\'{e}otier, and Yaouanc}}]{Wills}
\bibinfo{author}{\bibfnamefont{A.~S.} \bibnamefont{Wills}},
  \bibinfo{author}{\bibfnamefont{M.~E.} \bibnamefont{Zhitomirsky}},
  \bibinfo{author}{\bibfnamefont{B.}~\bibnamefont{Canals}},
  \bibinfo{author}{\bibfnamefont{J.~P.} \bibnamefont{Sanchez}},
  \bibinfo{author}{\bibfnamefont{P.}~\bibnamefont{Bonville}},
  \bibinfo{author}{\bibfnamefont{P.~D.} \bibnamefont{de~R\'{e}otier}},
  \bibnamefont{and} \bibinfo{author}{\bibfnamefont{A.}~\bibnamefont{Yaouanc}},
  \bibinfo{journal}{J. Phys.: Condens. Matter} \textbf{\bibinfo{volume}{18}},
  \bibinfo{pages}{L37} (\bibinfo{year}{2006}{\natexlab{b}}).

\bibitem[{\citenamefont{Maestro and Gingras}(2007)}]{DelMaestro}
\bibinfo{author}{\bibfnamefont{A.~D.} \bibnamefont{Maestro}} \bibnamefont{and}
  \bibinfo{author}{\bibfnamefont{M.~J.~P.} \bibnamefont{Gingras}},
  \bibinfo{journal}{Phys. Rev. B} \textbf{\bibinfo{volume}{76}},
  \bibinfo{pages}{064418} (\bibinfo{year}{2007}).

\bibitem[{\citenamefont{Keren et~al.}(2004)\citenamefont{Keren, Gardner,
  Ehlers, Fukaya, Segal, and Uemura}}]{Keren}
\bibinfo{author}{\bibfnamefont{A.}~\bibnamefont{Keren}},
  \bibinfo{author}{\bibfnamefont{J.~S.} \bibnamefont{Gardner}},
  \bibinfo{author}{\bibfnamefont{G.}~\bibnamefont{Ehlers}},
  \bibinfo{author}{\bibfnamefont{A.}~\bibnamefont{Fukaya}},
  \bibinfo{author}{\bibfnamefont{E.}~\bibnamefont{Segal}}, \bibnamefont{and}
  \bibinfo{author}{\bibfnamefont{Y.~J.} \bibnamefont{Uemura}},
  \bibinfo{journal}{Phys. Rev. Lett.} \textbf{\bibinfo{volume}{92}},
  \bibinfo{pages}{107204} (\bibinfo{year}{2004}).

\bibitem[{\citenamefont{van Duijn et~al.}(2005)\citenamefont{van Duijn, Kim,
  Hur, Adroja, Adams, Huang, Jaime, Cheong, Broholm, and Perring}}]{Duijn}
\bibinfo{author}{\bibfnamefont{J.}~\bibnamefont{van Duijn}},
  \bibinfo{author}{\bibfnamefont{K.~H.} \bibnamefont{Kim}},
  \bibinfo{author}{\bibfnamefont{N.}~\bibnamefont{Hur}},
  \bibinfo{author}{\bibfnamefont{D.}~\bibnamefont{Adroja}},
  \bibinfo{author}{\bibfnamefont{M.~A.} \bibnamefont{Adams}},
  \bibinfo{author}{\bibfnamefont{Q.~Z.} \bibnamefont{Huang}},
  \bibinfo{author}{\bibfnamefont{M.}~\bibnamefont{Jaime}},
  \bibinfo{author}{\bibfnamefont{S.-W.} \bibnamefont{Cheong}},
  \bibinfo{author}{\bibfnamefont{C.}~\bibnamefont{Broholm}}, \bibnamefont{and}
  \bibinfo{author}{\bibfnamefont{T.~G.} \bibnamefont{Perring}},
  \bibinfo{journal}{Phys. Rev. Lett.} \textbf{\bibinfo{volume}{94}},
  \bibinfo{pages}{177201} (\bibinfo{year}{2005}).

\bibitem[{\citenamefont{de~Leeuw et~al.}(1980)\citenamefont{de~Leeuw, Perram,
  and Smith}}]{Ewald}
\bibinfo{author}{\bibfnamefont{S.~W.} \bibnamefont{de~Leeuw}},
  \bibinfo{author}{\bibfnamefont{J.~W.} \bibnamefont{Perram}},
  \bibnamefont{and} \bibinfo{author}{\bibfnamefont{E.~R.} \bibnamefont{Smith}},
  \bibinfo{journal}{Proc. Roy. Soc. London} \textbf{\bibinfo{volume}{373}},
  \bibinfo{pages}{27} (\bibinfo{year}{1980}).

\bibitem[{\citenamefont{Molavian}()}]{MolavianThesis}
\bibinfo{author}{\bibfnamefont{H.~R.} \bibnamefont{Molavian}}, \eprint{Ph. D.
  thesis, U. of Waterloo (2007)}.

\bibitem[{Adm()}]{Admixing}
\eprint{Admixing between the excited crystal field states is neglected in this
  model, (a consequence of studying the effective Hamiltonian to second order
  in perturbation theory and neglecting higher order terms), which should be a
  good approximation because the splitting between the first excited doublet
  and the next excited singlet is about $120$ K. Note that in the remainder of
  the article all excited crystal field states are included in our derivation
  of $H_{\rm eff}$.}

\bibitem[{\citenamefont{Poole et~al.}(2007)\citenamefont{Poole, Wills, and
  Leli\`{e}vre-Berna}}]{Poole}
\bibinfo{author}{\bibfnamefont{A.}~\bibnamefont{Poole}},
  \bibinfo{author}{\bibfnamefont{A.~S.} \bibnamefont{Wills}}, \bibnamefont{and}
  \bibinfo{author}{\bibfnamefont{E.}~\bibnamefont{Leli\`{e}vre-Berna}},
  \bibinfo{journal}{J. Phys.:Condens. Matter} \textbf{\bibinfo{volume}{19}},
  \bibinfo{pages}{452201} (\bibinfo{year}{2007}).

\bibitem[{XYo()}]{XYorder}
\eprint{It is worth noting here that since the ordering of the XY components is
  the same ordering found in Er$_{2}$Ti$_{2}$O$_{7}$ which has an almost
  perfect XY anisotropy. Given that the ordering mechanism
  Er$_{2}$Ti$_{2}$O$_{7}$ is currently not understood, \cite{ETOPaper} this
  result might lead one to suspect that three spin interactions produced by
  VCFEs are responsible. This turns out not to be the case: if one computes the
  effective Hamiltonian for Er$_{2}$Ti$_{2}$O$_{7}$, one finds that the time
  reversal properties of the effective spins are the same as for ordinary
  angular momenta (in contrast to \tto\ effective spins) so that effective
  three spin interactions cannot appear in this case.}

\bibitem[{\citenamefont{Mirebeau et~al.}(2005)\citenamefont{Mirebeau, Apetrei,
  Rodríguez-Carvajal, Bonville, Forget, Colson, Glazkov, Sanchez, Isnard, and
  Suard}}]{Mirebeau3}
\bibinfo{author}{\bibfnamefont{I.}~\bibnamefont{Mirebeau}},
  \bibinfo{author}{\bibfnamefont{A.}~\bibnamefont{Apetrei}},
  \bibinfo{author}{\bibfnamefont{J.}~\bibnamefont{Rodríguez-Carvajal}},
  \bibinfo{author}{\bibfnamefont{P.}~\bibnamefont{Bonville}},
  \bibinfo{author}{\bibfnamefont{A.}~\bibnamefont{Forget}},
  \bibinfo{author}{\bibfnamefont{D.}~\bibnamefont{Colson}},
  \bibinfo{author}{\bibfnamefont{V.}~\bibnamefont{Glazkov}},
  \bibinfo{author}{\bibfnamefont{J.~P.} \bibnamefont{Sanchez}},
  \bibinfo{author}{\bibfnamefont{O.}~\bibnamefont{Isnard}}, \bibnamefont{and}
  \bibinfo{author}{\bibfnamefont{E.}~\bibnamefont{Suard}},
  \bibinfo{journal}{Phys. Rev. Lett.} \textbf{\bibinfo{volume}{94}},
  \bibinfo{pages}{246402} (\bibinfo{year}{2005}).

\bibitem[{LRS()}]{LRSInote}
\eprint{Note that the ground to first excited crystal field gap in spin ice
  \hto\ is about $250$ K, which might suggest that VCFEs could be significant
  based on the fact that the LRSI$_{000}$ phase appears at about this value of
  $\Delta$. However, $\mathcal{J}_{\rm ex}/\mathcal{D}$ in \hto\ places this
  material in the LRSI$_{001}$ part of the phase diagram as one would expect
  from the DSIM.}

\bibitem[{\citenamefont{Rule et~al.}(2006)\citenamefont{Rule, Ruff, Gaulin,
  Dunsiger, Gardner, Clancy, Lewis, Dabkowska, Mirebeau, Manuel et~al.}}]{Rule}
\bibinfo{author}{\bibfnamefont{K.~C.} \bibnamefont{Rule}},
  \bibinfo{author}{\bibfnamefont{J.~P.~C.} \bibnamefont{Ruff}},
  \bibinfo{author}{\bibfnamefont{B.~D.} \bibnamefont{Gaulin}},
  \bibinfo{author}{\bibfnamefont{S.~R.} \bibnamefont{Dunsiger}},
  \bibinfo{author}{\bibfnamefont{J.~S.} \bibnamefont{Gardner}},
  \bibinfo{author}{\bibfnamefont{J.~P.} \bibnamefont{Clancy}},
  \bibinfo{author}{\bibfnamefont{M.~J.} \bibnamefont{Lewis}},
  \bibinfo{author}{\bibfnamefont{H.~A.} \bibnamefont{Dabkowska}},
  \bibinfo{author}{\bibfnamefont{I.}~\bibnamefont{Mirebeau}},
  \bibinfo{author}{\bibfnamefont{P.}~\bibnamefont{Manuel}},
  \bibnamefont{et~al.}, \bibinfo{journal}{Phys.\ Rev.\ Lett.}
  \textbf{\bibinfo{volume}{96}}, \bibinfo{pages}{177201}
  (\bibinfo{year}{2006}).

\bibitem[{\citenamefont{Mirebeau et~al.}(2002)\citenamefont{Mirebeau,
  Goncharenko, Cadavez-Peres, Bramwell, Gingras, and Gardner}}]{Mirebeau4}
\bibinfo{author}{\bibfnamefont{I.}~\bibnamefont{Mirebeau}},
  \bibinfo{author}{\bibfnamefont{I.~N.} \bibnamefont{Goncharenko}},
  \bibinfo{author}{\bibfnamefont{P.}~\bibnamefont{Cadavez-Peres}},
  \bibinfo{author}{\bibfnamefont{S.~T.} \bibnamefont{Bramwell}},
  \bibinfo{author}{\bibfnamefont{M.~J.~P.} \bibnamefont{Gingras}},
  \bibnamefont{and} \bibinfo{author}{\bibfnamefont{J.~S.}
  \bibnamefont{Gardner}}, \bibinfo{journal}{Nature}
  \textbf{\bibinfo{volume}{420}}, \bibinfo{pages}{54} (\bibinfo{year}{2002}).

\bibitem[{\citenamefont{Lummen et~al.}(2008)\citenamefont{Lummen, Handayani,
  Donker, Fausti, Dhalenne, Berthet, Revcolevschi, and van
  Loosdrecht}}]{Lummen}
\bibinfo{author}{\bibfnamefont{T.~T.~A.} \bibnamefont{Lummen}},
  \bibinfo{author}{\bibfnamefont{I.~P.} \bibnamefont{Handayani}},
  \bibinfo{author}{\bibfnamefont{M.~C.} \bibnamefont{Donker}},
  \bibinfo{author}{\bibfnamefont{D.}~\bibnamefont{Fausti}},
  \bibinfo{author}{\bibfnamefont{G.}~\bibnamefont{Dhalenne}},
  \bibinfo{author}{\bibfnamefont{P.}~\bibnamefont{Berthet}},
  \bibinfo{author}{\bibfnamefont{A.}~\bibnamefont{Revcolevschi}},
  \bibnamefont{and} \bibinfo{author}{\bibfnamefont{P.~H.~M.} \bibnamefont{van
  Loosdrecht}}, \bibinfo{journal}{Phys.\ Rev.\ B.}
  \textbf{\bibinfo{volume}{78}}, \bibinfo{pages}{094418}
  (\bibinfo{year}{2008}).

\bibitem[{\citenamefont{Ruff et~al.}(2007)\citenamefont{Ruff, Gaulin,
  Castellan, Rule, Clancy, and Rodriguez}}]{Ruff}
\bibinfo{author}{\bibfnamefont{J.~P.~C.} \bibnamefont{Ruff}},
  \bibinfo{author}{\bibfnamefont{B.~D.} \bibnamefont{Gaulin}},
  \bibinfo{author}{\bibfnamefont{J.~P.} \bibnamefont{Castellan}},
  \bibinfo{author}{\bibfnamefont{K.~C.} \bibnamefont{Rule}},
  \bibinfo{author}{\bibfnamefont{J.~P.} \bibnamefont{Clancy}},
  \bibnamefont{and}
  \bibinfo{author}{\bibfnamefont{J.}~\bibnamefont{Rodriguez}},
  \bibinfo{journal}{Phys.\ Rev.\ Lett.} \textbf{\bibinfo{volume}{99}},
  \bibinfo{pages}{237202} (\bibinfo{year}{2007}).

\bibitem[{\citenamefont{Thompson et~al.}()\citenamefont{Thompson, McClarty,
  R\o{}nnow, Regnault, Sorge, and Gingras}}]{Thompson}
\bibinfo{author}{\bibfnamefont{J.~D.} \bibnamefont{Thompson}},
  \bibinfo{author}{\bibfnamefont{P.~A.} \bibnamefont{McClarty}},
  \bibinfo{author}{\bibfnamefont{H.~M.} \bibnamefont{R\o{}nnow}},
  \bibinfo{author}{\bibfnamefont{L.~P.} \bibnamefont{Regnault}},
  \bibinfo{author}{\bibfnamefont{A.}~\bibnamefont{Sorge}}, \bibnamefont{and}
  \bibinfo{author}{\bibfnamefont{M.~J.~P.} \bibnamefont{Gingras}},
  \eprint{(unpublished)}.

\bibitem[{\citenamefont{Zhou et~al.}(2008)\citenamefont{Zhou, Wiebe, Janik,
  Balicas, Yo, Qiu, Copley, and Gardner}}]{Zhou2}
\bibinfo{author}{\bibfnamefont{H.~D.} \bibnamefont{Zhou}},
  \bibinfo{author}{\bibfnamefont{C.~R.} \bibnamefont{Wiebe}},
  \bibinfo{author}{\bibfnamefont{J.~A.} \bibnamefont{Janik}},
  \bibinfo{author}{\bibfnamefont{L.}~\bibnamefont{Balicas}},
  \bibinfo{author}{\bibfnamefont{Y.~J.} \bibnamefont{Yo}},
  \bibinfo{author}{\bibfnamefont{Y.}~\bibnamefont{Qiu}},
  \bibinfo{author}{\bibfnamefont{J.~R.~D.} \bibnamefont{Copley}},
  \bibnamefont{and} \bibinfo{author}{\bibfnamefont{J.~S.}
  \bibnamefont{Gardner}}, \bibinfo{journal}{Phys. Rev. Lett.}
  \textbf{\bibinfo{volume}{101}}, \bibinfo{pages}{227204}
  (\bibinfo{year}{2008}).

\bibitem[{\citenamefont{Nakatsuji et~al.}(2009)\citenamefont{Nakatsuji,
  Machida, Maeno, Tayama, Sakakibara, van Duijn, Balicas, Millican, Macaluso,
  and Chan}}]{Nakatsuji2}
\bibinfo{author}{\bibfnamefont{S.}~\bibnamefont{Nakatsuji}},
  \bibinfo{author}{\bibfnamefont{Y.}~\bibnamefont{Machida}},
  \bibinfo{author}{\bibfnamefont{Y.}~\bibnamefont{Maeno}},
  \bibinfo{author}{\bibfnamefont{T.}~\bibnamefont{Tayama}},
  \bibinfo{author}{\bibfnamefont{T.}~\bibnamefont{Sakakibara}},
  \bibinfo{author}{\bibfnamefont{J.}~\bibnamefont{van Duijn}},
  \bibinfo{author}{\bibfnamefont{L.}~\bibnamefont{Balicas}},
  \bibinfo{author}{\bibfnamefont{J.~N.} \bibnamefont{Millican}},
  \bibinfo{author}{\bibfnamefont{R.~T.} \bibnamefont{Macaluso}},
  \bibnamefont{and} \bibinfo{author}{\bibfnamefont{J.~Y.} \bibnamefont{Chan}},
  \bibinfo{journal}{Phys. Rev. Lett.} \textbf{\bibinfo{volume}{96}},
  \bibinfo{pages}{087204} (\bibinfo{year}{2009}).

\bibitem[{\citenamefont{Matsuhira et~al.}(2009)\citenamefont{Matsuhira, Sekine,
  Paulsen, Wakeshima, Hinatsu, Kitazawa, Kiuchi, Hiroi, and
  Takagi}}]{Matsuhira}
\bibinfo{author}{\bibfnamefont{K.}~\bibnamefont{Matsuhira}},
  \bibinfo{author}{\bibfnamefont{C.}~\bibnamefont{Sekine}},
  \bibinfo{author}{\bibfnamefont{C.}~\bibnamefont{Paulsen}},
  \bibinfo{author}{\bibfnamefont{M.}~\bibnamefont{Wakeshima}},
  \bibinfo{author}{\bibfnamefont{Y.}~\bibnamefont{Hinatsu}},
  \bibinfo{author}{\bibfnamefont{T.}~\bibnamefont{Kitazawa}},
  \bibinfo{author}{\bibfnamefont{Y.}~\bibnamefont{Kiuchi}},
  \bibinfo{author}{\bibfnamefont{Z.}~\bibnamefont{Hiroi}}, \bibnamefont{and}
  \bibinfo{author}{\bibfnamefont{S.}~\bibnamefont{Takagi}},
  \bibinfo{journal}{J. Phys.: Conference Series}
  \textbf{\bibinfo{volume}{145}}, \bibinfo{pages}{012031}
  (\bibinfo{year}{2009}).

\bibitem[{\citenamefont{Robert et~al.}(2006)\citenamefont{Robert, Simonet,
  Canals, Ballou, Bordet, Lejay, and Stunault}}]{Robert}
\bibinfo{author}{\bibfnamefont{J.}~\bibnamefont{Robert}},
  \bibinfo{author}{\bibfnamefont{V.}~\bibnamefont{Simonet}},
  \bibinfo{author}{\bibfnamefont{B.}~\bibnamefont{Canals}},
  \bibinfo{author}{\bibfnamefont{R.}~\bibnamefont{Ballou}},
  \bibinfo{author}{\bibfnamefont{P.}~\bibnamefont{Bordet}},
  \bibinfo{author}{\bibfnamefont{P.}~\bibnamefont{Lejay}}, \bibnamefont{and}
  \bibinfo{author}{\bibfnamefont{A.}~\bibnamefont{Stunault}},
  \bibinfo{journal}{Phys. Rev. Lett.} \textbf{\bibinfo{volume}{96}},
  \bibinfo{pages}{197205} (\bibinfo{year}{2006}).

\bibitem[{\citenamefont{Bordet et~al.}(2006)\citenamefont{Bordet, Gelard,
  Marty, Ibanez, Robert, Simonet, Canals, Ballou, and Lejay}}]{Bordet}
\bibinfo{author}{\bibfnamefont{P.}~\bibnamefont{Bordet}},
  \bibinfo{author}{\bibfnamefont{I.}~\bibnamefont{Gelard}},
  \bibinfo{author}{\bibfnamefont{K.}~\bibnamefont{Marty}},
  \bibinfo{author}{\bibfnamefont{A.}~\bibnamefont{Ibanez}},
  \bibinfo{author}{\bibfnamefont{J.}~\bibnamefont{Robert}},
  \bibinfo{author}{\bibfnamefont{V.}~\bibnamefont{Simonet}},
  \bibinfo{author}{\bibfnamefont{B.}~\bibnamefont{Canals}},
  \bibinfo{author}{\bibfnamefont{R.}~\bibnamefont{Ballou}}, \bibnamefont{and}
  \bibinfo{author}{\bibfnamefont{P.}~\bibnamefont{Lejay}}, \bibinfo{journal}{J.
  Phys.: Condens. Matter} \textbf{\bibinfo{volume}{18}}, \bibinfo{pages}{5147}
  (\bibinfo{year}{2006}).

\bibitem[{\citenamefont{Zhou et~al.}(2007)\citenamefont{Zhou, Vogt, Janik, Jo,
  Balicas, Qiu, Copley, Gardner, and Wiebe}}]{Zhou}
\bibinfo{author}{\bibfnamefont{H.~D.} \bibnamefont{Zhou}},
  \bibinfo{author}{\bibfnamefont{B.~W.} \bibnamefont{Vogt}},
  \bibinfo{author}{\bibfnamefont{J.~A.} \bibnamefont{Janik}},
  \bibinfo{author}{\bibfnamefont{Y.-J.} \bibnamefont{Jo}},
  \bibinfo{author}{\bibfnamefont{L.}~\bibnamefont{Balicas}},
  \bibinfo{author}{\bibfnamefont{Y.}~\bibnamefont{Qiu}},
  \bibinfo{author}{\bibfnamefont{J.~R.~D.} \bibnamefont{Copley}},
  \bibinfo{author}{\bibfnamefont{J.~S.} \bibnamefont{Gardner}},
  \bibnamefont{and} \bibinfo{author}{\bibfnamefont{C.~R.} \bibnamefont{Wiebe}},
  \bibinfo{journal}{Phys. Rev. Lett.} \textbf{\bibinfo{volume}{99}},
  \bibinfo{pages}{236401} (\bibinfo{year}{2007}).

\bibitem[{\citenamefont{Lumata et~al.}()\citenamefont{Lumata, Choi, Besara,
  Hoch, Zhou, Brooks, Kuhns, Reyes, Dalal, and Wiebe}}]{Lumata}
\bibinfo{author}{\bibfnamefont{L.~L.} \bibnamefont{Lumata}},
  \bibinfo{author}{\bibfnamefont{K.~Y.} \bibnamefont{Choi}},
  \bibinfo{author}{\bibfnamefont{T.}~\bibnamefont{Besara}},
  \bibinfo{author}{\bibfnamefont{M.~J.~R.} \bibnamefont{Hoch}},
  \bibinfo{author}{\bibfnamefont{H.~D.} \bibnamefont{Zhou}},
  \bibinfo{author}{\bibfnamefont{J.~S.} \bibnamefont{Brooks}},
  \bibinfo{author}{\bibfnamefont{P.~L.} \bibnamefont{Kuhns}},
  \bibinfo{author}{\bibfnamefont{A.~P.} \bibnamefont{Reyes}},
  \bibinfo{author}{\bibfnamefont{N.~S.} \bibnamefont{Dalal}}, \bibnamefont{and}
  \bibinfo{author}{\bibfnamefont{C.~R.} \bibnamefont{Wiebe}},
  \eprint{arXiv:0811.3367}.

\bibitem[{\citenamefont{Zhou et~al.}(2009)\citenamefont{Zhou, Wiebe, Balicas,
  Jo, Takano, Case, Qiu, Copley, and Gardner}}]{Zhou3}
\bibinfo{author}{\bibfnamefont{H.~D.} \bibnamefont{Zhou}},
  \bibinfo{author}{\bibfnamefont{C.~R.} \bibnamefont{Wiebe}},
  \bibinfo{author}{\bibfnamefont{L.}~\bibnamefont{Balicas}},
  \bibinfo{author}{\bibfnamefont{Y.-J.} \bibnamefont{Jo}},
  \bibinfo{author}{\bibfnamefont{Y.}~\bibnamefont{Takano}},
  \bibinfo{author}{\bibfnamefont{M.~J.} \bibnamefont{Case}},
  \bibinfo{author}{\bibfnamefont{Y.}~\bibnamefont{Qiu}},
  \bibinfo{author}{\bibfnamefont{J.~R.~D.} \bibnamefont{Copley}},
  \bibnamefont{and} \bibinfo{author}{\bibfnamefont{J.~S.}
  \bibnamefont{Gardner}}, \bibinfo{journal}{Phys. Rev. Lett.}
  \textbf{\bibinfo{volume}{102}}, \bibinfo{pages}{067203}
  (\bibinfo{year}{2009}).

\bibitem[{\citenamefont{Kassman}(1970)}]{Kassman}
\bibinfo{author}{\bibfnamefont{A.}~\bibnamefont{Kassman}}, \bibinfo{journal}{J.
  Chem. Phys.} \textbf{\bibinfo{volume}{53}}, \bibinfo{pages}{4118}
  (\bibinfo{year}{1970}).

\bibitem[{\citenamefont{Jensen and Mackintosh}(1991)}]{RareEarthMagnetism}
\bibinfo{author}{\bibfnamefont{J.}~\bibnamefont{Jensen}} \bibnamefont{and}
  \bibinfo{author}{\bibfnamefont{A.}~\bibnamefont{Mackintosh}},
  \emph{\bibinfo{title}{Rare Earth Magnetism}} (\bibinfo{publisher}{Oxford
  University Press}, \bibinfo{year}{1991}).

\bibitem[{\citenamefont{Ruff et~al.}(2005)\citenamefont{Ruff, Melko, and
  Gingras}}]{Ruff2}
\bibinfo{author}{\bibfnamefont{J.~P.~C.} \bibnamefont{Ruff}},
  \bibinfo{author}{\bibfnamefont{R.~G.} \bibnamefont{Melko}}, \bibnamefont{and}
  \bibinfo{author}{\bibfnamefont{M.~J.~P.} \bibnamefont{Gingras}},
  \bibinfo{journal}{Phys.\ Rev.\ Lett.} \textbf{\bibinfo{volume}{95}},
  \bibinfo{pages}{097202} (\bibinfo{year}{2005}).

\end{thebibliography}
   
\end{document}